\documentclass[iop]{emulateapj2}
\usepackage{morefloats}
\usepackage{lscape}
\newcommand{\myemail}{jcurtis@psu.edu}
\newcommand{\caii}{Ca {\sc ii} H \& K}
\newcommand{\mgi}{Mg {\sc i}}
\newcommand{\fei}{Fe{\sc i}}
\newcommand{\feii}{Fe{\sc ii}}
\newcommand{\feih}{\ensuremath{[\mbox{Fe{\sc i}/H}]}}
\newcommand{\feiih}{\ensuremath{[\mbox{Fe{\sc ii}/H}]}}

\newcommand{\teff}{\ensuremath{T_{\mbox{\scriptsize eff}}}}
\newcommand{\logg}{\ensuremath{\log g}}
\newcommand{\persec}{\ensuremath{\mbox{s}^{-1}}}
\newcommand{\griz}{\ensuremath{g^\prime r^\prime i^\prime z^\prime}}
\newcommand{\degree}{\ensuremath{^\circ}}
\newcommand{\astrosun}{\odot}
\newcommand{\msun}{\ensuremath{\mbox{M}_{\astrosun}}}
\newcommand{\vsini}{\ensuremath{v \sin i}}
\newcommand{\gabor}{Gabor F{\"u}r{\'e}sz}
\newcommand{\andy}{Andrew Szentgyorgyi}

\newcommand{\kms}{\ensuremath{\mbox{km s}^{-1}}}
\newcommand{\mps}{\mbox{m} \persec}

\submitted{{\sc Accepted to AJ:} December 7, 2012}

\begin{document}
\title{Ruprecht 147:  The Oldest Nearby Open Cluster as \\ a New Benchmark for Stellar Astrophysics}
\author{Jason L. Curtis \altaffilmark{1,2,3}}
\author{Angie Wolfgang \altaffilmark{3,4}}
\author{Jason T. Wright  \altaffilmark{1}}
\author{John M. Brewer \altaffilmark{5}}
\author{John Asher Johnson  \altaffilmark{6}}

\altaffiltext{1}{Department of Astronomy \& Astrophysics, The Pennsylvania State University, University Park, PA 16802, USA}
\altaffiltext{2}{\myemail}
\altaffiltext{3}{NSF Graduate Research Fellow}
\altaffiltext{4}{Department of Astronomy \& Astrophysics, University of California, Santa Cruz, CA 95064, USA}
\altaffiltext{5}{Department of Astronomy, Yale University, New Haven, CT 06511, USA}
\altaffiltext{6}{Department of Astrophysics, California Institute of Technology, Pasadena, CA 91125, USA}

\begin{abstract}
Ruprecht 147 is a hitherto unappreciated open cluster that holds great promise 
as a standard in fundamental stellar astrophysics. 
We have conducted a radial velocity survey of astrometric candidates with 
Lick, Palomar, and MMT observatories and have identified over 100 members, 
including 5 blue stragglers, 11 red giants, 
and 5 double-lined spectroscopic binaries (SB2s).
We estimate the cluster metallicity from spectroscopic analysis, 
using Spectroscopy Made Easy (SME), 
and find it to be [M/H] = $+0.07 \pm 0.03$. 
We have obtained deep CFHT/MegaCam \griz\ photometry
and fit Padova isochrones to the 
($g' - i'$) and 2MASS ($J - K_S$) CMDs, 
using the $\tau^2$ maximum-likelihood procedure of \citet{naylor2009}, 
and an alternative method using 2D cross-correlations developed in this work.
We find best fits for isochrones at 
age $t = 2.5 \pm 0.25$ Gyr, 
$m - M = 7.35 \pm 0.1$, and $A_V = 0.25 \pm 0.05$, 
with additional uncertainty from the unresolved binary population and 
possibility of differential extinction across this large cluster. 
The inferred age is heavily dependent by our choice of stellar evolution model:   
fitting Dartmouth and PARSEC models yield age parameters of 3 Gyr and 3.25 Gyr respectively.
At $\sim$300 pc and $\sim$3 Gyr, Ruprecht 147 is by far the oldest nearby star cluster. 
\end{abstract}


\section{Introduction}
The observational foundations of stellar astrophysics are studies of the Sun and
stellar clusters.  A few ``benchmark'' clusters form the basis of our 
understanding of stellar evolution, and the effects of abundance, age, and
mass on stars.   When fully characterized with precise ages, distances
and metallicities, these clusters become touchstones for similar stars
in the field and test models of stellar evolution and structure. 

Galactic gravitational tidal forces are effective at disrupting most
Galactic clusters on a time scale of a few hundred Myr
\citep{soderblom2010}, so most clusters tend to be relatively
young.  This is fortunate for studies of early stellar evolution and
massive stars: for such work stellar astrophysicists
have access to several nearby young clusters (e.g. Pleiades
$\sim$100 -- 200 Myr; Hyades and Praesepe $\sim$700 Myr). 

Studies of the older cool stars (age $\gtrsim 1$ Gyr) that typify 
the field must rely on rarer and thus more distant clusters.  
Studies of the typical rotation, activity level, and photometry of G, K and
M dwarfs as a function of age and mass, such as the WIYN Open Cluster Survey
(WOCS\footnote{http://www.astro.wisc.edu/wocs/}), 
the Palomar Transient Factory \citep[PTF,][]{PTFpraesepe} and 
the {\it Kepler} Cluster Study \citep{KeplerCluster}, 
investigate clusters with distances of 1 -- 4 kpc.  
These larger distance moduli can make spectroscopic study of 
their low mass members extremely difficult.  

\begin{figure}\begin{center}
\plotone{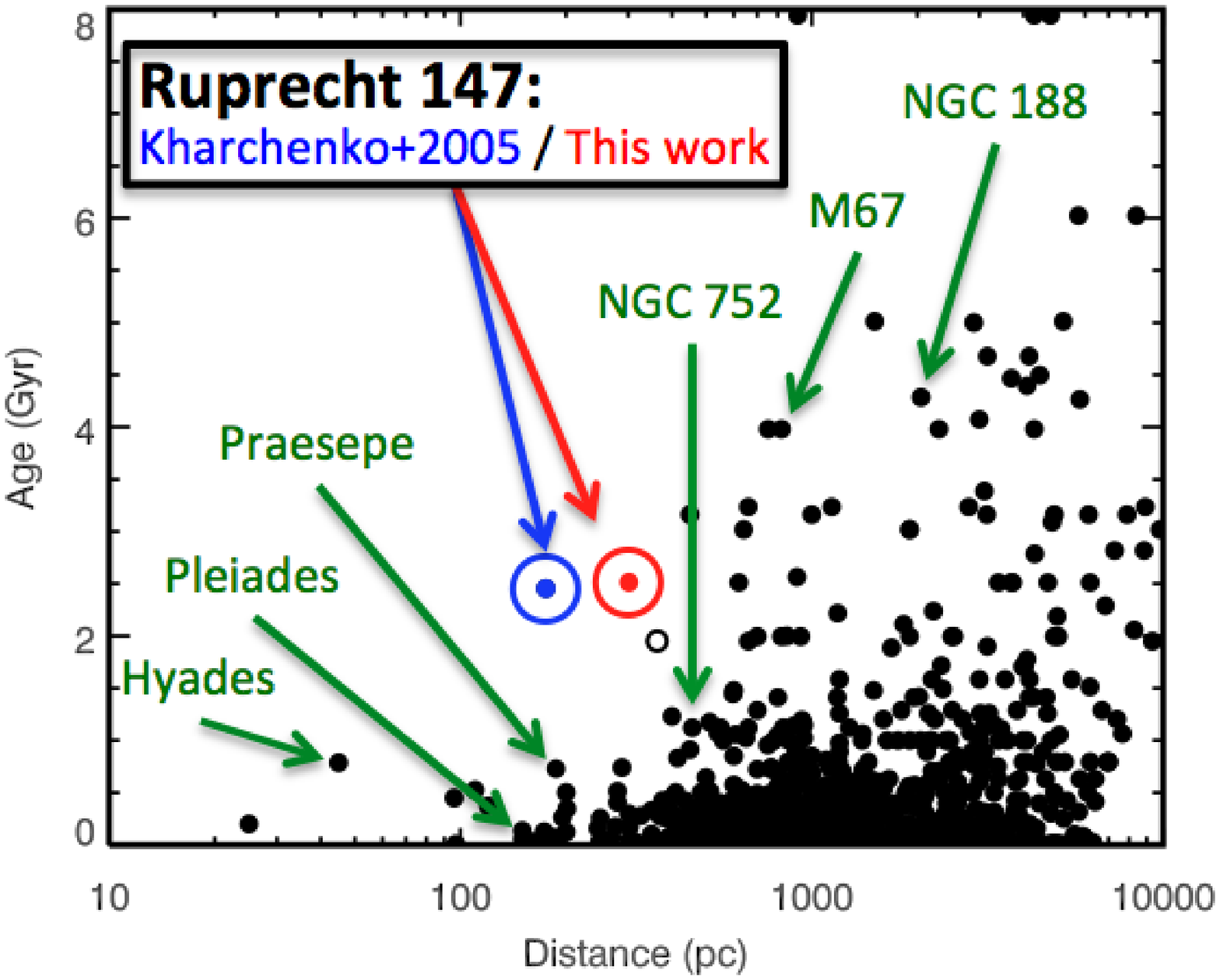}
	\caption{Data from the WEBDA database \citep{webda} showing all known
	clusters in distance -- age space. R147 is by far the oldest nearby
	cluster, and holds great promise as a standard in fundamental stellar
	astrophysics: \citet{khar2005} values shown in blue, ours in red. 
	WEBDA lists an age for M67 at 2.5 Gyr, but we plot it at 4 Gyr according 
	to \citet{sunM67} and references therein. 
	The open circle next to R147 denotes Loden 1, 
	an unconfirmed grouping of stars with unreliable properties -- see footnote 8 for a discussion.
	\label{f:webda}}
\end{center}\end{figure}

Fortunately, \citet{dias2001} and \citet{khar2005} used catalog data to
identify Ruprecht 147 (R147 = NGC 6774), and estimated its age to 
be $\sim$2.5 Gyr at a distance of 175 -- 270 pc, making R147 by far the 
oldest nearby cluster\footnote{
One WEBDA cluster, Loden 1, is plotted as an open circle in Figure 1, and has 
properties that are apparently similar to Ruprecht 147. 
The membership and properties of Loden 1 were determined by \citet{khar2005}. 
They identify only nine 1-$\sigma$ members from proper motions and photometry, 
and none have measured radial velocities. 
The Loden 1 grouping has not been confirmed as a real open cluster, 
and the properties derived by the automated search of \citet{khar2005} 
are thus unreliable.}
(e.g. WEBDA lists NGC 752 at 1.1 Gyr and 457 pc, 
Figure \ref{f:webda}).

\subsection{Pre-2000 literature}
Despite its promising scientific potential due to the 
unique combination of its age and distance, 
and despite having a similar distance and size to Praesepe, 
R147 was completely overlooked by stellar astronomers until 
the works by \citeauthor{dias2006} and \citeauthor{khar2005}.
This is likely because its proximity makes R147 a very sparse 
cluster on the sky:  there are only $\sim$50 members with 
$V<11$ and only $\sim$10 with $V < 9$ spread over 5 square degrees. 
Its presence is also obscured by its location in the 
Galactic plane ($-14\degree < b < -12\degree$, in Sagittarius), 
and the fact that due to its age, it lacks the many 
bright A stars that made similarly nearby clusters so 
obvious, even to the astronomers of antiquity. 

\begin{figure}[!ht]\begin{center}
\plotone{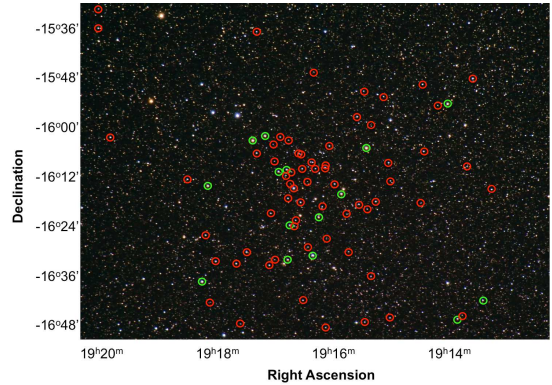}
		\caption{This astrophotograph of a portion of the Ruprecht 147 field 
        was taken and kindly provided by Chris Beckett and Stefano Meneguolo 
        of the Royal Astronomical Society of Canada. 
        We have attached an approximate coordinate system 
        solution (we have not solved for the field distortions)
        and circled the 80 high-confidence members in red. 
		There are 52 stars in this image with $V < 10$, and only 
		17 are members of R147 and are circled green. 
		Of the 47 stars with $V < 9$ within $\approx 2$\degree\ 
		of the cluster center (extending beyond this image), only 11 are members.
		It is remarkable that Herschel correctly identified this 
		as an open cluster in 1830.
          \label{f:imgid}}
\end{center}\end{figure}

In fact, a complete pre-2000 bibliography of R147 consists almost 
exclusively of entries in various catalogues. 
R147 was originally discovered in 1830 by John Herschel, 
who described it as 
``a very large straggling space full of loose stars" 
\citep{firstr147ref},
and labelled it GC 4481 \citep{GC}.
Since then it has appeared with numerous designations
including NGC 6774, OCL 65, Lund 883 
\citep{NGC, Alte58, Lyng87, mermcat1995}.
Some star charts have even designated R147 as 
an asterism, and not a true cluster 
(e.g. ``Burnham's Celestial Handbook: An Observer's Guide to the Universe 
Beyond the Solar System" lists NGC 6774 as ``possibly not a true cluster" 
\citep{burnham}). 
The name we use here originates from \citet{Rupr66}, 
who classified R147 as a III-2-m cluster in the Trumpler system
\citep[160]{trumpler}.
According to \citet{kenjanes}, Brian Skiff realized that 
NGC 6774 and R147 are likely the same star cluster. 
\citet{kenjanes} describe R147 as a ``45$'$ sized V-shaped group 
of bright stars" that is ``a sparse possible open cluster", 
and estimate the cluster center as the location of HD 180228 
(while this star's photometry apparently places it 
on the R147 red giant branch, the Tycho-2 proper motions,  
-1.6 and -6.3 mas/yr in right ascension and declination, 
are inconsistent with cluster membership, see Figure \ref{f:dias}). 
Figure \ref{f:imgid} highlights our high-confidence members on 
an optical image. Herschel's cluster identification is truly 
amazing, given the lack of a well defined cluster core. 
But those arguing for the asterism status were not entirely 
wrong either: of the 51 stars with $V < 9$ within $\approx 2$\degree\ 
of the cluster center, we confirm only 11 as members.

\subsection{Recent work in the literature}
\label{s:litrev}
Only in the last decade has R147 received any individual attention 
in studies of open clusters.
\citet{dias2001} first identified R147's membership based on the
stellar population's common proper motion:
selecting stars in the Tycho-2 Catalogue \citep{tycho2}
that were spatially coincident with BDA clusters 
\citep[The Open Cluster DataBase,][]{mermcat1995},
they determined cluster membership with the Tycho-2 proper motions
using the statistical method of \citet{sanders1971}
and found 33 stars with mean proper motion of 
$\mu_\alpha = -0.8 \pm 2.3$ and $\mu_\delta = -28.5 \pm 2.3$ mas/yr.
\citet{dias2001} also provided the first distance estimate
based on only two Hipparcos parallax measurements
\footnote{Actually, three R147 members appear in the Hipparcos catalog, see \S \ref{s:distance}}
\citep[HIP1,][]{HIPPARCOS}: 
$\pi = 3.57 \pm 1.01$ mas ($280 \pm 79$ pc) for HIP 94635 (CWW 1)
\footnote{Throughout this paper, we will refer to individual stars with the designation 
``CWW \#" (CWW = Curtis, Wolfgang and Wright). 
Our membership list provides 2MASS IDs, astrometry, photometry, 
radial velocities, and membership probabilities for 108 stars. 
The CWW ID numbers sort these stars according to $V$ magnitude
(see Table 6).} , 
and $3.75 \pm 1.04$ mas ($267 \pm 74$ pc) for HIP 94803 (CWW 2), 
which they average to $3.7 \pm 0.2$ mas, 
estimating the distance to R147 to be 
250 pc\footnote{although this is a numerical error as 
1000 / 3.7 = 270, not 250.}.
Since then, \citet{newHIPPARCOS} (HIP2) has performed 
a new data reduction and issued an updated catalog 
with parallaxes of 
$5.48 \pm 0.65$ mas ($182 \pm 22$ pc) for HIP 94635, and
$4.92 \pm 0.79$ mas ($203 \pm 33$ pc) for HIP 94803
\footnote{\citet[][\S 3.3.1, 3.3.2]{vanLeeuwenbook} cautions against deriving 
distances and distance moduli from parallaxes when the relative error is 
greater than 10\%. The Lutz-Kelker bias can also introduce a 0.1 magnitude 
systematic offset at 10\% relative error}.

\citet{dias2002} compiled all available data for 2095 galactic clusters
(\textit{The New Catalogue of Optically Visible Open Clusters and Candidates}, or DAML02)
and published an updated membership list and cluster properties for R147:
25 members, proper motion $\mu_\alpha = -0.9 \pm 0.3$ and
$\mu_\delta = -29.3 \pm 0.3$ mas/yr, 
RV = $41$ \kms\ 
(from the single published measurement in \citet{wilsonRV}, see \S \ref{s:rv}),
distance = 200 pc, color excess E($B-V$) = 0.2 mag., 
and an age 3.2 Myr
(presumably from misidentifying blue stragglers as main sequence turnoff stars). 
\citeauthor{dias2002} re-classified R147 as IV-2-p (Trumpler system).

\begin{figure}[h!]\begin{center}
\plotone{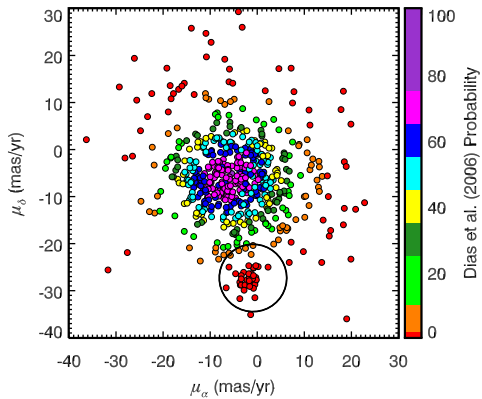}
	\caption{Proper motion diagram of stars in the R147 field,
	color shaded by membership probability as derived by \citet{dias2006}.
	The black circle highlights the proper motion 
	of R147 \citep[][and confirmed here]{khar2005}. 
	The Dias membership probablities are clearly in
	error. \label{f:dias}}
\end{center}\end{figure}

Following their 2002 work, \citet{dias2006} selected all 
clusters in their DAML02 catalogue with known distances
and queried the UCAC-2 catalogue \citep{UCAC2} for all stars within 
the measured cluster radii, plus $2^\prime$, 
of their tabulated cluster centers.
Employing similar methods as \citet{dias2001}, 
they derived a mean proper motion for R147 of
$\mu_\alpha = -4.6 \pm 0.4$ 
and $\mu_\delta = -5.6 \pm 0.4$,
and identified 200 cluster members. 
Figure \ref{f:dias} shows the proper motions for stars 
in the R147 field, 
color-shaded by membership probability as derived by \citet{dias2006}. 
The black circle highlights the proper motion of R147 according to 
\citet{khar2005} and confirmed in this work, and shows that the Dias algorithm
missed the cluster, locating the field stars instead. 
The \citet{dias2006} membership list and cluster parameters are thus unreliable.
\citet{dias2006} attribute their algorithm's failure to the 
large angular size of R147. 

A similar automated effort has been undertaken by \citet{khar2001},
who assembled the All-Sky Compiled Catalogue of 2.5 Million Stars
(ASCC-2.5), including 
proper motions from the Tycho-2 catalog \citep{tycho2},
Johnson $BV$ photometry, and radial velocities and spectral types 
when they are available.

\begin{figure*}[h!]\begin{center}
\plotone{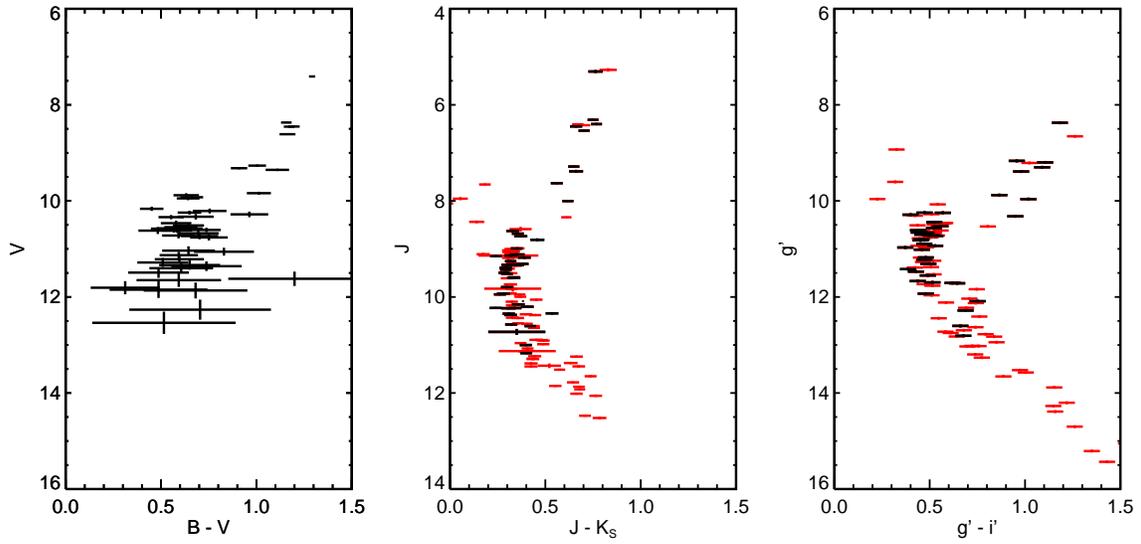}
	\caption{Ruprecht 147 color -- magnitude diagrams. 
	The left panel shows the ($B - V$) photometry used by \citet{khar2005} 
	to estimate age and distance by isochrone fitting. 
	These data are magnitude limited at the main sequence turnoff. 
	The central and right panels plot the same stars in black, 
	along with our additional members in red. 
	The main sequence is better defined in the 2MASS NIR ($J - K_S$) and 
	our optical ($g' - i'$) CMDs, 
	which explains the $\approx$80\% discrepancy between 
	the \citeauthor{khar2005} distance of 175 pc and our value of $\approx$300 pc.
	The $g'$ and $i'$ error bars are set at 0.03 magnitude.
	The color errors are the magnitude errors added in quadrature. 
	\label{f:khar}}
\end{center}\end{figure*}

\citet{khar2005} searched this catalog and identified
520 Galactic open clusters, including R147.
Their algorithm determined the core and cluster angular radii,
and the distances, mean space motions (proper motion and radial velocity), 
and ages of the clusters.
Three important differences exist between the \citet{dias2006} membership and 
properties and those of \citet{khar2005}: 
(1) \citeauthor{khar2005} correctly identify the cluster, 
cataloging 41 1-$\sigma$ members;
(2) they provide the first reliable age estimate of 
2.45 Gyr from their isochrone fitting; 
and (3) they claim a new distance of only 175 pc, 
75 pc closer than that inferred from the original 
Hipparcos parallaxes, 
but similar to the distances derived in HIP2.
While we determine a similar age of $\sim$2.5 Gyr, 
we derive a distance $d \approx 300$ pc 
(\S \ref{s:isosme}, \ref{s:isofit}) by fitting isochrones 
to a spectroscopically derived \teff\ -- \logg\ diagram, and
2MASS ($J - K_S$) and CFHT/MegaCam ($g' - i'$) color -- magnitude diagrams.
Figure \ref{f:khar} plots the CMD used by 
\citet{khar2005} to derive age and distance.
The Tycho-2 $BV$ photometry is magnitude limited at $V \sim 11$, 
near the R147 main sequence turnoff. 
ASCC-2.5 is supplemented with various ground based photometry for 
fainter magnitudes, which Figure \ref{f:khar} demonstrates is 
insufficient for main sequence fitting. 
While the MSTO provides a strong constraint on the age, 
the descrepancy between our derived distance and that of \citeauthor{khar2005} 
can be explained by the ill-defined ($B - V$) main sequence. 
Their analysis was also hindered by a lack of a spectroscopically determined 
composition, and they assumed Solar metallicity. 
Without an accurate metallicity, 
and with a main sequence dominated by photometric error, 
it is difficult to disentangle visual extinction, age, composition and distance. 
Instead, \citet{khar2005} assumed $A_V = 0.465$ from the 
\citet{dustmap} dust map at their location for the cluster center, 
even though according to this dust map $A_V$ varies from 0.3 to 0.6 mag. across the cluster 
(see \S \ref{s:av}). 
Although \citet{dias2001} were the first to determine the distance, 
\citet{kharseg} were the first to discuss Ruprecht 147 specifically as an old nearby 
cluster in a peer-reviewed publication.

Despite these issues, the works of \citeauthor{dias2001} and \citeauthor{khar2005} 
are significant because they essentially re-discoverd Ruprecht 147 and 
provided the first good evidence that R147 is in fact the oldest nearby star cluster.

Most recently, \citet{redgiants} observed three cluster red giants and 
spectroscopically measured radial velocities and stellar properties 
(discussed in \S \ref{s:rv}, \ref{s:sme}).
They determined the cluster metallicity to be super-Solar, 
thereby decreasing the estimated age, from a fit to 
an enriched Padova isochrone \citep{padova1,padova2}\footnote{The Padova isochrones are available at http://stev.oapd.inaf.it/cgi-bin/cmd. 
We primarily use these stellar evolution models because the Padova group provides synthetic 
photometry in a large number of systems, including the CFHT/MegaCam \griz\ filter set 
allowing us to analyze our optical photometry.}
to $\sim$1.25 Gyr,
and derived a distance of $280 \pm 100$ pc, 
along with a color excess of E($B-V$) = 0.11 
(or $A_V$ = 0.34, assuming $R_V$ = 3.1).

Ruprecht 147 has also appeared in the open cluster 
luminosity function study of \citet{r147lf} and 
a paper on Galactic kinematics and structure as 
defined by open clusters by \citet{zhi2009}, 
but these works undoubtedly suffer from a 
poorly determined membership
and uncertain cluster properties.

We have begun an observational campagin to characterize R147, 
catalog its members, and prove its benchmark status.  
Here we present our initial efforts, 
detailing in particular our R147 membership search 
that more than doubles the number of known cluster members 
(\S \ref{s:membership}),
and our derivation of the cluster's age, distance, and metallicity
(\S \ref{s:properties}).
We begin with an overview of our photometric and spectroscopic datasets.

\section{Observational datasets}\label{s:data}
Cluster members are identified by their common space motion, 
determined from proper motions and radial velocities, 
and by their placement on a color -- magnitude diagram (CMD). 
We utilize the NOMAD, UCAC-3 and PPMXL astrometric catalogs for proper motions. 
We have high resolution, single order echelle spectra from MMT Observatory;  
and high-resolution, cross-dispersed echelle spectra from 
Lick, Palomar, and Keck Observatories. 
We acquired deep \griz\ photometry 
of a 4 square degree field with CFHT/MegaCam, 
and utilize NIR $JHK_S$ photometry from the 2MASS Point Source Catalog.
Other observing projects are underway, 
including deep NIR imaging with UKIRT/WFCAM (PI Adam Kraus), 
a 250 ks exposure of the cluster core 
with Chandra/ACIS (PI Steve Saar), 
and an RV survey for K and M dwarf members with 
Magellan/MIKE+MagE (PI Steve Saar). 

\subsection{Astrometric catalogs}
\label{s:pm}
    Our initial list of candidate members was drawn from 
    the NOMAD and UCAC-3 catalogs.
    The Naval Observatory Merged Astrometric Dataset \citep[NOMAD,][]{nomad}
    combines data (positions, proper motions, and
    $BVR/JHK$ photometry) for over 1 billion stars from the Hipparcos \citep{HIPPARCOS},
    Tycho-2 \citep{tycho2}, UCAC-2 \citep{UCAC2}, USNO-B1.0 \citep{usnob1}, and 2MASS \citep{2MASS} catalogs.
    The Third USNO CCD Astrograph Catalog \citep[UCAC-3,][]{UCAC3} expands 
    on NOMAD by improving UCAC-2 in many ways, including
    complete sky coverage, reduced systematic errors for CCD observations,
    deeper photometry ($R \approx$ 8 -- 16) 
    for $\sim$80 million stars, and 
    improved astrometry (resolved double stars,
    inclusion of several new catalogs,
    and re-reduction of early epoch photographic plates
    to derive proper motions). 
	
	In this paper, we use proper motions from the PPMXL catalog \citep{ppmxl}, 
	and provide these values in Table 6. 
	PPMXL utilizes astrometry from the USNO-B1.0 and 2MASS catalogs to 
	calculate proper motions in the ICRS system for 
	approximately 900 million objects, 
	including $\sim$410 million with 2MASS photometry. 
	The catalog covers the entire sky down to $V \approx 20$. 
	PPMXL was released in 2010, 
	after we had derived our initial membership catalog. 
	Some stars have NOMAD and/or UCAC-3 proper motions consistent with cluster 
	membership, but are discrepant according to the PPMXL values (and vice versa). 
	We include these stars in our membership list despite this, 
	and we evaluate their probability of membership based on 
	all available kinematic and photometric data (\S \ref{s:membership}). 

\subsection{Lick 3-m and Palomar 200-in spectra and radial velocities}
\label{s:lp}
	We performed initial radial velocity confirmation of suspected members to 
	verify the existence of the cluster with the Hamilton echelle spectrometer 
	on the 120 inch Shane telescope at Lick Observatory \citep[$R \sim 50,000$;][]{Hamilton}.  
	Our objectives were to obtain RVs of known and suspected members, 
	to identify new members, and to obtain high resolution spectra of the brightest 
	members at high signal-to-noise ratios (SNRs) for more detailed analysis of 
	abundances and chromospheric activity.

\begin{figure}[!ht]\begin{center}
	\plotone{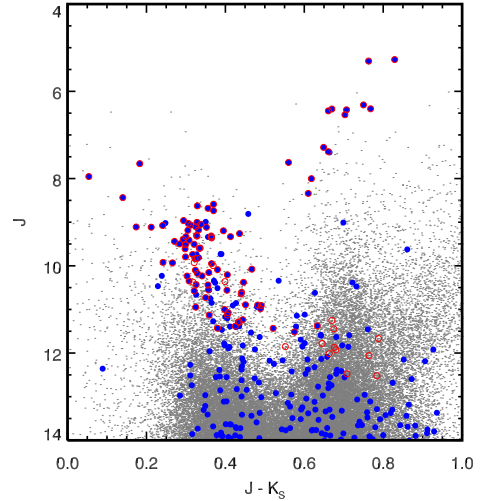}
    	\caption{Proper motion cuts made to NOMAD stars to identify candidate R147 members. 
    	Of the over 750,000 NOMAD stars in the 1\degree .5 radius region centered on R147, 
    	38,623 have $0 < J - K_S < 1$ and $4 < J < 14$ which are shown in gray. 
    	There are 1348 stars with NOMAD proper motions within 5 mas/yr of the cluster value, 
    	280 of which have $J < 14$ and are plotted in blue. 
		Our final membership list of 108 stars are circled in red.
		This plot demonstrates that there is really only one obvious sequence 
		at bright magnitudes where the proper motions are reliable, 
		including a well-defined red giant branch.
          \label{f:nomad}}
\end{center}\end{figure}

	We observed candidate cluster members on UT 2007 July 31 -- August 1 and 
	2007 August 22 -- 23, including the members identified by \citet{khar2005}.  
	To locate additional candidate members, we selected stars from the
	NOMAD catalog that were within $1^\circ.25$ of the published cluster center
    and that had UCAC-2 and TYCHO-2 proper motions within 5 mas/yr of the
    \citeauthor{khar2005} value. 
    Although there are over 750,000 NOMAD stars in the field due to its large size and 
    location in the Galactic plane, the cluster is separated enough from the field in 
    proper motion space that this yielded a list of 1348 stars, 
    illustrated in Figures \ref{f:nomad}.

  	To further vet this list, we used NOMAD ($B - V$) and 
  	2MASS ($J - K_S$) color - magnitude diagrams to identify
    stars consistent with an assumed distance of 230 pc, 
    a compromise between the HIP1 distance to the cluster (270 pc) and
    the value of \citeauthor{khar2005} (175 pc).  
    We combined the CMDs and proper motion information to estimate crude 
    membership probabilities based on the Hipparcos main sequence 
    with no reddening corrections, calculated generously to
    account for uncertainties in the cluster parameters and for the poor
    quality of some of the NOMAD proper motion entries, 
    and we favored brighter targets to improve the efficiency of vetting 
    candidate members at the telescope.

	We drew from this list, sorted by membership probability, to choose targets for 
	spectroscopic study at Lick Observatory. We used these spectra to measure 
	radial velocities for the stars and determine the space motion of the cluster. 

	\subsubsection{Data acquisition and raw reduction}
	We adopted the spectrograph setup procedure of the California and Carnegie Planet
	Search, which placed bright emission lines from a thorium-argon (ThAr)
	lamp on specific pixels to approximately reproduce a known wavelength
	solution to a fraction of a pixel.  
	Our prior experience using the Hamilton spectrograph 
	at coud\'e focus
	revealed that the wavelength solution is reliable to a pixel or two 
	over the course of the night. 
	This was sufficient for our purposes of measuring radial velocities to $<5$ \kms, 
	a precision which allows most interloping field stars to be identified, 
	so we did not attempt any further wavelength calibration throughout the night. 
	In practice, our radial velocity accuracy proved to be much better than 5 \kms.  
	
	Observing conditions were good, and we obtained several high SNR spectra of 
	radial velocity standard stars of various spectral types throughout the nights, 
	chosen from the catalog of \citet{RVstandard}.
	For candidate members we used exposure times of 60 -- 90 s, 
	depending on the magnitude of the star. 
	For fainter stars we obtained SNRs as low as 1 per pixel, 
	which is sufficient for our velocity work because of the 
	broad spectral coverage of the Hamilton spectrograph.
	This strategy allowed a large number of stars to be observed in our 
	allocated time for this low-declination cluster,  
	which only spent a few hours per night at sufficiently low airmass to be useful.  

	The raw spectra were processed with the standard Hamilton Spectrograph data reduction
	pipeline used for precise radial velocity work by the California and
 	Carnegie Planet Search, which includes bias subtraction and flat fielding of each
	frame and which results in a one-dimensional spectrum for each of 92 orders.  
	
	We calculated an empirical blaze function for each order by fitting a
	polynomial to the spectra of several rapidly rotating B stars that 
	we observed for this purpose.
	These stars show no high resolution spectral features, 
	and we corrected orders contaminated by the effects of the broad Balmer lines
	by averaging the polynomial blaze function of the neighboring orders. 
	Variations in slit illumination from target to target 
	created apparent continuum variations that were not perfectly removed by this process, 
	and the nature of the polynomial fitting process caused the fit to diverge 
	from the actual spectrum significantly at the edges of orders. 
	The resulting spectra were nonetheless sufficiently flat that the cross correlations
	required for our data analysis (\S \ref{s:angrv}) could be confidently performed.

	\subsubsection{Palomar spectra}
	We followed a similar procedure at Palomar Observatory to determine membership 
	probabilities and activity measurements of fainter candidate members.  
	Target stars were drawn from the same sorted list that was compiled for the 
	Lick observing run the previous year, including 25 targets that were chosen for follow-up 
	observations based on qualitative examination of the radial velocity measurements 
	derived from the Lick data, either because the signal-to-noise ratio of the 
	Lick data was too low for a definitive velocity measurement or the star showed 
	evidence of binarity, necessitating a second epoch.

	We observed on 2008 August 5 and 18 with the East-Arm Echelle 
	\citep[$R \sim 33,000$;][]{EastArm} on the Hale 200-inch at Palomar, 
	following our earlier procedure of short integrations at very low signal-to-noise ratio 
	(the additional aperture of the 200-in over the 3-m,  
	somewhat mitigated by the low throughput of the East Arm Echelle, 
	allowed us to explore fainter targets, or brighter targets at better SNR).  

	\subsubsection{Radial velocity determination}
	\label{s:angrv}
	Although we adopted the rough wavelength calibration used for planet search work, 
	we did not attempt to use this calibration to measure our radial velocities. 
	Rather, we extracted radial velocities in pixel space by cross-correlating the 
	spectra of our candidate cluster members with those of our observed RV standard stars 
	(a similar pixel-space cross-correlation method was employed by \citet{Norris2011}).
	To reduce the errors introduced by comparing two stars of different spectral types, 
	we paired each candidate member to an RV standard star that minimized the difference 
	between their $V-J$ colors ($\Delta(V - J)$), with $V - J = 0.8$ for the bluest standard star 
	and $V - J = 2.4$ for the reddest. 
	
	Imperfect flat fielding produced a sharp spike at exactly zero shift 
	in the cross-correlation functions (CCFs), 
	and the presence of telluric lines created a narrow peak there, 
	complicating the radial velocity measurements derived from these CCFs.  
	This justified our use of velocity standard stars as cross-correlation templates rather 
	than high SNR spectra of actual cluster members, 
	since the standards have different radial velocities than the cluster 
	and so the true CCF peak is far removed from the spurious peaks at zero shift. 
	To further address the problem of telluric lines, 
	we empirically rejected those portions of the spectrum 
	where these lines dominated the CCF: after dividing 
	each of the 92 orders into three segments, 
	we discarded from all spectra those segments that 
	showed strong telluric peaks near zero shift.  

	Computing these CCFs for the different combinations of RV standard stars 
	with $\Delta(V-J) < 0.5$ allowed us to calibrate the conversion from 
	pixel space shifts to radial velocities, 
	after applying a barycentric radial velocity correction.  
	This calibration step thus obviates the need for a transformation 
	into wavelength space.  
	Specifically, we fit a linear function with zero intercept to the 
	measured RV standard stars' CCF pixel shifts as a function of 
	the difference in their radial velocities, giving us the 
	velocity shift per pixel in each spectrum segment.
	The root mean squared of the residuals of this fit is less than 0.6 \kms; 
	this provides our best measure of the 
	systematic velocity precision we expect at high SNRs. 

	Comparison of this calibration constant among the segments confirmed that 
	the velocity shift per pixel of the Hamilton spectrograph is nearly constant 
	for each of the 92 orders. 
	This is not surprising given that both the resolution, 
	$R$ ($= \lambda / \Delta \lambda$ per resolution element), 
	and the sampling, $s$ (pixels/resolution element), 
	are nearly constant across an echellogram, 
	and that our calibration constant, having units of \kms /pixel, 
	is essentially $c/Rs$, where $c$ is the speed of light.
	This allowed us to add the CCFs of the remaining segments together 
	to improve the SNR of the stellar signal, 
	and enabled the clear identification of a peak 
	and its associated pixel shift in the combined CCF.
	To be conservative, we divided each spectrum into three sets of segments, 
	corresponding to the left, middle and right sides of each order. 
	After separately summing the CCFs in each set, we required 
	that the location of the tallest peak in the summed CCFs to be 
	identical in all three sets; 
	however, when the side segments produced noisy CCFs, as was the case for 
	stars with SNR $\sim 10$, we used the location of the tallest peak in 
	the middle segments' summed CCF, as long as the peak met our high quality 
	classification. 
		
	We visually inspected each CCF produced by our procedure and 
	classified the stars into two categories based on the quality 
	of their CCFs:  either the combined CCF had one clear peak, 
	which corresponded to high SNR spectra, or the combined CCF 
	had multiple peaks of approximately equal height, which 
	indicated that the CCF was dominated by noise, and which
	usually corresponded to spectra with 
	SNR $\lesssim 2$ per pixel 
	(some of these discarded stars were later revisited in the Palomar observing run 
	in order to acquire higher SNR spectra). 

	We found a clear clustering of barycentric velocities near 43 \kms, 
	which is within 2 \kms\ of the cluster RV quoted by \citet{dias2002} 
	that was based on a single measurement of a single putative member (\S \ref{s:rv}).  
	Upon closer inspection of the Lick velocities for signs of systematics, 
	we found that these apparent 
	cluster members' velocities exhibited a slight correlation with time 
	from the beginning of the observing run to the end, 
	with magnitude 2 -- 4 \kms.
	We fit this trend to a linear function and removed it.  
	We have also observed 49 of these stars with MMT/Hectochelle 
	(with very high signal-to-noise ratio, 
	we expect all to have RV precision $\approx$0.5 \kms, \S \ref{s:mmt}),
	and measure differences between the Lick/Palomar and Hectochelle RVs 
	as large as 5 \kms, 
	except for a possible single-lined spectroscopic binary (SB1) with a difference of 20 \kms.
	We interpret this offset between telescopes and spectrographs as a measure 
	of systematic error in our absolute barycentric radial velocities.
	
	Figure \ref{f:rvhist} illustrates the resulting RVs and shows a 
	clear clustering around the cluster velocity. 
	We tentatively identified as cluster members any stars 
	with a measured radial velocity between 32 and 54 \kms, 
	or roughly twice the typical systematic error.

\subsection{Spectra from Keck/HIRES}
\label{s:keck}
    Spectra of four cluster members were obtained on 2008 September 12 and 18
    and for two members on 2011 October 17 
    with the High Resolution Echelle Spectrometer \citep[HIRES,][]{HIRES} 
    on the 10-m telescope at Keck Observatory.  
    The stars were kindly observed by the 
    California Planet Survey team (CPS\footnote{http://exoplanets.org/cps.html})
    without an iodine cell, 
    and with the B5 decker (slit of 3.5$''$ length and 0.861$''$ width),
    giving a typical resolution $R \sim 50,000$ in the 3360 -- 8100 \AA\ bandpass.
    Exposure times were monitored with a photomultiplier tube exposure meter to ensure 
    high signal-to-noise ($S/N \sim 50-100$).
    These observations are summarized in Table \ref{t:keckobs}, 
    and were reduced by the standard CPS pipeline.
	\citet{chubak2012}
	measured absolute radial velocities (results discussed in \S \ref{s:rv}) 
	and we derive stellar properties in Section \ref{s:sme}.
	
\begin{deluxetable*}{cccccccl}
\tabletypesize{\scriptsize}
\tablecaption{Keck/HIRES observations. \label{t:keckobs}}
\tablewidth{0pt}
\tablehead{
\colhead{CWW ID} & \colhead{2MASS ID} & \colhead{Obs Date} & \colhead{Exposure Time} & 
\colhead{Airmass} & \colhead{$V$\tablenotemark{a}} & 
\colhead{S/N\tablenotemark{b}} & \colhead{Notes\tablenotemark{c}} \\
&  & \colhead{JD} & \colhead{seconds} & \colhead{} & \colhead{mag.} & 
 & 
}
\startdata
72	& 	19165800-1614277	& 2008-09-12 & 210 & 1.43 & 11.52 & 50   & G dwarf / SB2 \\
78	&	19160879-1524279	& 2008-09-12 & 170 & 1.44 & 11.82 & 60   & late F dwarf \\
21	&	19132220-1645096	& 2008-09-18 & 90  & 2.17 &  9.98 & 80   & Subgiant  \\
22	&	19172382-1612488	& 2008-09-18 & 93  & 2.12 & 10.04 & 70   & mid-F MSTO / SB1 \\
44	&	19164495-1717074	& 2011-10-17  & 167   & 1.39 & 10.61 & 80 & mid-F dwarf MSTO \\
91	&	19164725-1604093	& 2011-10-17  & 822   & 1.32 & 12.39 & 50 & early G dwarf \\
\enddata
\tablenotetext{a}{$V$ magnitudes are drawn from the NOMAD catalog.}
\tablenotetext{b}{Signal-to-noise ratio measured in the spectral order 
	encompassing the Mg b triplet, in the 5034 -- 5036 \AA\ continuum.} 
\tablenotetext{c}{Approximate classification, performed by matching spectroscopic and 
photometric properties to isochrone masses (\S \ref{s:cmd}).
SB1 status suggested by inconsistent RVs from multiple epochs; MSTO = main sequence turnoff}
\end{deluxetable*}

\subsection{Spectra from MMT/Hectochelle}
\label{s:mmt}
	We obtained high-resolution spectra with MMT/Hectochelle
    in the vicinity of the \caii\ lines for 48 members 
    (as determined from Lick/Palomar RVs), 
    10 candidate members (from astrometry and photometry alone),
    and 23 potential astrometric reference stars.  
    These data provide radial velocities, 
    chromospheric activity indicators (e.g. \citep{wright2004}),
    and gravity diagnostics \citep[via the Wilson-Bappu effect,][]{WilsonBappu} 
    useful for identifying background giants as astrometric references.
	The remaining fibers not allocated for sky subtraction were 
	assigned to proper motion candidates with inconsistent photometry to 
	assess how R147 members are kinematically distinct from the field.
	
    MMT is a 6.5-m telescope located at the 
    Fred Lawrence Whipple Observatory on Mt. Hopkins, AZ \citep{mmt}.
    Hectochelle is a high-resolution ($R \sim$ 32,000 -- 40,000) fiber-feb spectrograph,
    which provides simultaneous observations for 240 targets in a one square degree field 
    \citep{hectochelle, gabor}.     
    We observed the central square degree with the `Ca41' \caii\ filter 
    with 1x1 on-chip binning. Eight total hours were obtained to ensure
    sufficient signal-to-noise for a future chromospheric activity study.
    All observed targets have $g' = 9 - 15.5$.

    Twelve 40-minute exposures were obtained over the nights of UT 2010 July 4 -- 5.
    These data were reduced with an IRAF\footnote{IRAF is distributed by the National Optical Astronomy Observatories,
    which are operated by the Association of Universities for Research
    in Astronomy, Inc., under cooperative agreement with the National
    Science Foundation.}-based automated pipeline developed 
    at the Harvard-Smithsonian Center for Astrophysics,
    provided and run by \gabor\ and \andy, 
    which flat-fielded, cosmic-ray removed, and wavelength calibrated
    our targets and sky flats.
    The wavelength solution was determined from Thorium-Argon (ThAr) 
    lamp comparison spectra, with an RMS precison of $0.2 - 0.5$ \kms\ 
    \citep[for reduction details, see][]{hectopipe}.

    Radial velocities were measured by cross-correlating the target spectrum 
    with Solar spectra obtained from the sky flat exposures, 
    then corrected for Earth's heliocentric motion.
    We checked the fiber-to-fiber and day-to-day stability of the spectrograph by 
    measuring velocity shifts determined by cross-correlating matched 
    ThAr, Solar, and target spectra. 
    The fiber-to-fiber velocity shift on the first night
    was $12 \pm 35$ \mps\ (for the ThAr spectra) and 
    $200 \pm 200$ \mps\ (for the Solar spectra).
	We also measure a night-to-night variation between each fiber of 
	$31 \pm 41$ \mps\ (ThAr) and 
	$200 \pm 180$ \mps (Solar).
	The RVs measured each day for R147 stars show a mean 
	absolute difference of 0.23 \kms.
    In summary, fiber-to-fiber and day-to-day offsets and errors are well under or 
    comparable to the precision set by the wavelength solution. 
	The RV distribution for 49 member stars is shown in Figure \ref{f:rvhist}, 
	and is discussed in \S \ref{s:rv}.

\subsection{Preliminary optical photometry}
\label{s:griz}
\begin{figure}[h!]\begin{center}
\plotone{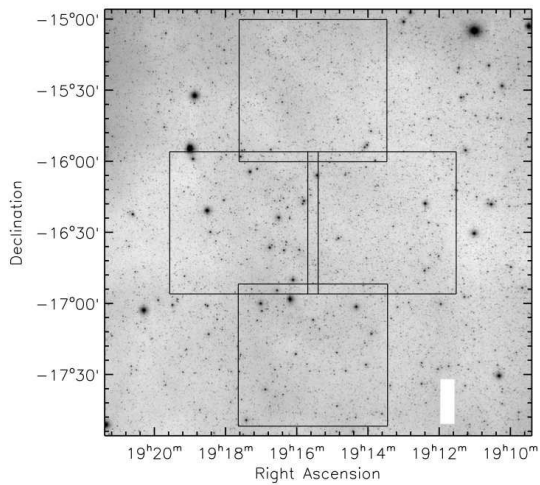}
	\caption{The four fields we imaged in \griz\ with CFHT/MegaCam, 
	overlaid on a 2MASS $J$ band mosaic image generated with 
	Montage (http://montage.ipac.caltech.edu). 
	We log-scaled and smoothed the image with a 3$''$ boxcar.
	\label{f:2mass}}
\end{center}\end{figure}

	We imaged a 4 square degree field in the optical 
    \griz\ bands in 2008 April and May with CFHT/MegaCam \citep{MegaCam}
    \footnote{http://www.cfht.hawaii.edu/Instruments/Imaging/MegaPrime/}.
    With MegaCam's one square degree field of view, four fields were 
    required to cover the majority of the known cluster. 
    The fields are outlined over a 2MASS $J$ band mosaic image 
    in Figure \ref{f:2mass}.
    Six additional surrounding fields were imaged solely in $i'$ band, 
    for the purpose of first-epoch astrometry for the entire cluster, 
    including any extended halo. 
	We obtained both a series of short (1 second) and long (few minutes) exposures 
	in queued service observing mode.
	Typically, 5 dithered exposures were obtained for each field and exposure time.
	
	These observations were pre-processed at CFHT with the Elixir pipeline
	\citep{elixir}\footnote{http://www.cfht.hawaii.edu/Instruments/Elixir/home.html}. 
	Elixir creates master bias, dark and flat images, 
	which are used to detrend the observation frames.
	SExtractor identifies sources and determines their pixel coordinates and raw flux. 
	The astrometric calibration is performed by comparison to the USNO-B1.0 catalog.

Photometric magnitudes are calibrated from the instrumental magnitude 
with the application of a zero point, an airmass term and a color term. 
The coefficients are derived from observations of standard stars. 
Every night, one Landolt (1992) field was observed (SA-101, SA-107 and SA-113), 
along with two spectrophotometric standards 
(i.e. an O star or white dwarf: Feige 110, GD 153, HZ 43, BD+28 4211) 
and at least one CFHTLS Deep field.
The zero points for each frame were determined from 13 to 26 standards observed in 
4 to 9 separate images during a run. 
Standards were not necessarily observed in all the filters utilized on a given night, 
but the zero point scatter across an observing run for each filter ranged from 0.0073 to 0.0180, 
and is therefore quite stable. 
Frames obtained on photometric nights provide the means to calibrate observations taken 
under less transparent conditions (the image scaling is done by TERAPIX, see below).
The zero points were determeind after the application of the superflat, 
and therefore are valid for all 36 CCDs.
The photometric and astrometric calibration data are stored in the FITS image headers, 
and the data transferred to the Canadian Astronomy Data Centre 
(CADC) in Victoria\footnote{http://www1.cadc-ccda.hia-iha.nrc-cnrc.gc.ca/cadc/}.

TERAPIX performed the final photometric and astrometric reduction of 
our MegaCam imagery\footnote{TERAPIX is a data reduction center located at the 
Institut d'Astrophysique in Paris, France:  http://terapix.iap.fr/}. 
The TERAPIX pipeline \citep{terapix} takes the detrended images and 
the preliminary calibration from Elixir,
and completes a final photometric and astrometric calibration and 
provides source merged catalogs. 
First the images are re-scaled:  
the photometry is analyzed in each overlapping frame, and the 
frames are re-scaled to the photometry in the image with highest flux per source, 
which is considered to be the least extinguished. 
The overlapping images are stacked and the sources are re-extracted. 
The TERAPIX pipeline co-addition and astrometric calibration modules 
are now maintained at AstrOmatic.net, and documentation for each 
can be found at the 
SWarp\footnote{http://www.astromatic.net/software/swarp}
and SCAMP \citep[][Software for Calibrating AstroMetry and Photometry,]{scamp}\footnote{http://www.astromatic.net/software/scamp} 
webpages.

TERAPIX handles the CFHT Legacy Survey reduction. 
The CFHTLS\footnote{http://www.cfht.hawaii.edu/Science/CFHTLS/}
and TERAPIX's CFHTLS reduction\footnote{http://terapix.iap.fr/article.php?id\_article=383}
webpages provide additional details relevant to the final 
photometric and astrometric calibrations.
	
TERAPIX kindly provided us merged \griz\ source catalogs for each field and exposure duration.
The short exposures saturate at $g' \approx 9.5$ and 
the long exposures saturate at $g' \sim 16$. 
Sources with $g' \sim 17 - 18$ have consistent photometric magnitudes 
in each catalog, and sources are detected down to $g' \sim 24$ in 
the long exposure catalog. 
We therefore have photometry covering $10 \lesssim g' \lesssim 24$.
	
Our faintest red giant branch member is $g' = 10.53$ and $i' = 9.73$. 
Given the saturation limit in each band at approximately 9.5, 
the majority of the red giant branch stars are saturated in $g'$ and $i'$.
We will include the optical RGB in our figures in this work, but 
with stars plotted with open circles to distinguish these data 
from the more reliable optical photometry for the rest of the membership.

\begin{figure}[!ht]\begin{center}
    \plotone{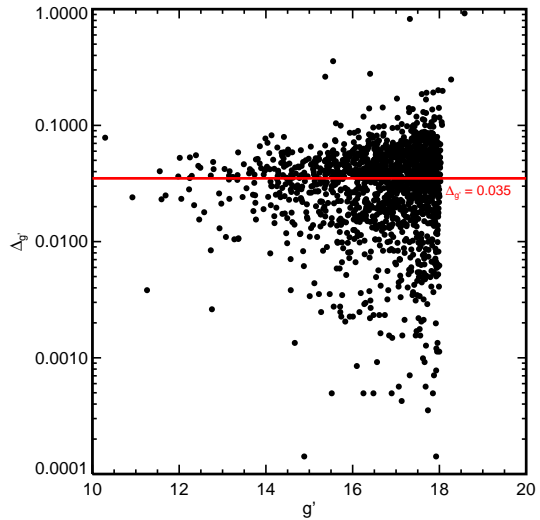}
    	\caption{Photometric error estimates for CFHT/MegaCam \griz\ photometry: 
        We imaged the R147 field with four separate but 
        partially overlapping pointings (see Figure \ref{f:2mass}). 
        We matched all stars in the overlap regions in our bright source catalog 
        (short exposures), and find a total of 1575 unique sources with 
        $g' < 18$. This figure plots the mean versus standard deviation of 
        $g'$ for the 2 -- 4 independent measurements, 
		depending on the number of overlapping regions containing the source.
		We find a typical value of $\Delta_{g'} = 0.035$, 
		but this is probably larger than what should be assumed for the 
		photometric precision across the field, 
		because one of the sources usually lies very close to the edge of 
		a field, where the mosaic dithering is incomplete, 
		and so the photometry is less reliable.
         \label{f:mcerr}}
\end{center}\end{figure}

We estimate the photometric error by making use of the overlapping 
regions between the four imaged fields (see Figure \ref{f:2mass}). 
We matched all stars in the overlap regions in our bright source catalog 
(short exposures), and find a total of 1575 unique sources with $g' < 18$. 
Figure \ref{f:mcerr} plots the mean versus standard deviation of 
$g'$ for the 2 -- 4 independent measurements, 
depending on the number of overlapping regions containing the source.
We find a typical value of $\Delta_{g'} = 0.035$ mag., 
but this is probably larger than what should be assumed for the 
photometric precision across the field, 
because one of the sources usually lies very close to the edge of 
one of the fields, where the mosaic dithering is incomplete
and the photometry less reliable.

We have identified a $\approx$0.15 mag. zero point error in the $z'$ band, 
and the persistence of a low-frequency mode that was not removed by the flat fielding. 
Given these issues with $z'$-band, we cautiously analyze the ($g' -i'$) CMD 
and provide photometry in Table \ref{t:mem}.
We will first analyze our spectroscopic results and 2MASS photometry, 
before fitting isochrones to our optical data. 
We will show in Section \ref{s:myisofit} that essentially identical 
cluster properties are determined from these 3 data sets. 
This consistency suggests the $g'$ and $i'$ zero points are likely accurate.
		
The Elixir and TERAPIX pipelines were developed and have been successfully used to 
reduce similar MegaCam imaging for the CFHT Legacy Survey
\footnote{For a list of publications, see http://www.cfht.hawaii.edu/Science/CFHTLS/cfhtlspublications.html}. 
Despite these quality assurances, we consider this photometry to be preliminary, 
and we are currently working to further validate the accuracy of the Elixir and TERAPIX calibration.
	
\section{Identifying the Ruprecht 147 membership}
\label{s:membership}
We identify stars as R147 members based on their common space motion 
and placement on a color -- magnitude diagram. 
Our initial membership list is drawn from the NOMAD and UCAC-3 
astrometric catalogs and subjected to radial velocity vetting.
We queried NOMAD and UCAC-3 for stars within a 
radius of 2\degree\ of the cluster center.  
Stars were accepted as candidates if their proper motions 
(\S \ref{s:pm}) were within 8 mas/yr of the cluster mean 
(see Figure \ref{f:dias} for a proper motion 
vector point diagram).
We adopted the values of \citet{khar2005}
for the cluster center and mean proper motion: 
$(\alpha$ [h:m:s], $\delta$ [d:m:s]) = (19:16:40, -16:17:59) and 
$(\mu_\alpha , \mu_\delta)$ [mas/yr] = (-0.6, -27.7) -- 
after we identified the highest confidence members, 
we recalculated these locations and note no significant change 
(Table \ref{t:cluster}).

Table 6 gathers data for 108 stars of interest. 
The first column provides a designation internal to this paper: 
CWW \# (CWW = Curtis, Wolfgang and Wright). 
The stars are ordered according to increasing $V$ magnitude 
(provided by NOMAD). 
The table also includes the 2MASS ID (and therefore RA and Dec position);  
PPMXL proper motions in mas/yr; 
the preliminary CFHT/MegaCam $g'$ and $g' - i'$ optical photometry; 
2MASS $J$ and $J - K_S$ NIR photometry; 
and radial velocities RV$_{\rm LP}$ and RV$_{\rm H}$ (Lick/Palomar and Hectochelle, respectively).  
A membership probability is assigned to each of these data, 
according to criteria discussed below and summarized in Table \ref{t:criteria}, 
and is listed in the order: 
(1) radial distance in proper motion space from the cluster mean, 
(2) RV$_{\rm LP}$, (3) RV$_{\rm H}$, 
(4) proximity to cluster locus on the 2MASS ($J - K_S$) CMD and 
(5) CFHT/MegaCam ($g' - i'$) CMD. 

The derivation of quantitative membership probabilities is precluded by 
the large uncertainties in proper motion and our Lick/Palomar velocities, 
combined with the intrinsic spread in the R147 main sequence due 
to unresolved stellar multiplicity and 
the possibility of differential reddening, 
along with non-negligible photometric error.  

Instead we designate three confidence levels: 
`Y' for \textit{yes this is consistent with cluster membership}, 
`P' for \textit{possible / probable member}, and 
`N' for \textit{not likely / non-member}. 
Each membership criterion is independently assessed and assigned a confidence level designation 
(whenever data are unavailable, a `-' is assigned instead). 
The following sections address each criterion, and establish the ranges for each level 
(summarized in Table \ref{t:criteria}). 
The results from all fields are then reviewed and an overall membership confidence level is 
assigned to each star according to the same `Y', `P', `N' scheme. 

We find 81 stars of highest confidence, 
21 stars with `P' possible member status, 
and 6 stars with little to no probability of membership - at least as single star members
(multiple star systems could show RVs and photometry inconsistent with membership 
as we have defined it, while still being gravitationally bound members of R147). 

\subsection{Proper motion}
We now primarily use PPMXL proper motions to assess membership, 
although our original membership list was derived from NOMAD and UCAC-3.
The typical PPMXL errors range from 1 to 5 mas/yr in proper motion, 
and so we designate stars with proper motion within this 5 mas/yr of the cluster mean, 
as `Y' members. 
Only stars within $\sim$8 mas/yr were considered in our initial candidate list.
and we found that the majority of bright candidates 
had velocities consistent with cluster membership.
There are six stars that we had originally classified as `Y' or `P' according to 
their NOMAD proper motions, but which have PPMXL proper motions more than 8 mas/yr 
different than the cluster's mean motion.
Despite this large discrepancy from the PPMXL data, 
we classify these stars with conflicting proper motion data as `P', 
and will consider their velocities and photometry when assigning their final probability.

\subsection{Radial velocity}
\label{s:rv}
The General Catalogue of Stellar Radial Velocities \citep{wilsonRV} 
contains a single entry for a cluster member:
HD 180015 (HIP 94635, classified as K0III, CWW 1). 
\citeauthor{wilsonRV} reported RV = 41 \kms, 
with quality designation `C', corresponding to 
a typical uncertainty = 2.5 \kms\ and maximum uncertainty = 5 \kms.
This was the first and only available RV until \citet{redgiants}
observed three other cluster red giants:  
HD 179691 (CWW 9) at 46.7 \kms, 
HD 180112 (CWW 10) at 40.1 \kms, and 
HD 180795 (CWW 7) at 40.8 \kms, 
with S/N $>100$, and precision estimated at 0.5 -- 0.8 \kms.
The RV for HD 179691 is 6 \kms\ larger than the other two stars,   
too large to be explained by the cluster velocity dispersion,
which implies this star is either a SB1 binary or a non-member.
We observed these stars at Lick/Palomar and measure 
RV$_{\rm LP}$ = 42.1, 41.4, 42.4 \kms, respectively. 
While our measurements for the second two stars are in basic agreement with 
\citet{redgiants}, 
the velocity for HD 179691 now appears consistent with the cluster mean, 
supporting its membership and corroborating its SB1 status. 

\citet{chubak2012}
have also measured RVs with rms errors $\sim$50 \mps\ for 
our 5 single-lined Keck/HIRES spectra (Table \ref{t:sme}).
We list these here in \kms, with our Lick/Palomar velocities in parenthesis for comparison:
CWW 44: 41.41 (42.2), CWW 91: 41.50 (42.8), CWW 21: 40.35 (41.9), 
CWW 78: 41.02 (40.8), and CWW 22: 46.63 (51.9).
We followed up CWW 22 with Hectochelle and measure RV = 38 \kms\ on both nights, 
and classify it a SB1.

\begin{figure}\begin{center}
\plotone{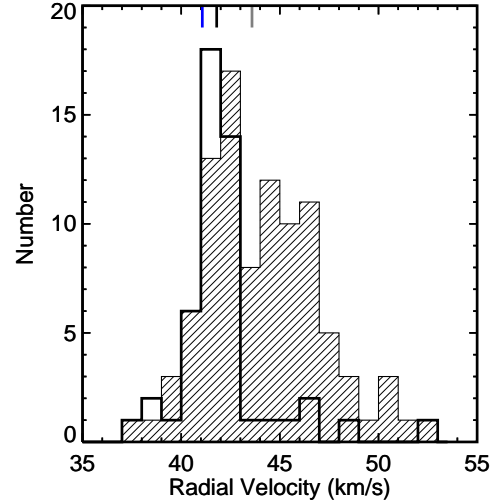}
    \caption{R147 radial velocity distribution. 
      The gray hash designates RVs measured from Lick/Palomar (98 stars),
      with RV$_{\rm LP} = 44 \pm 3$ \kms. 
      The black line plots 45 stars with Hectochelle velocities, 
	  RV$_{\rm H} = 41.6 \pm 1.5$ \kms.
      Tick marks on top indicate median values, color-coded to RV source: 
      blue shows the cluster RV = $41.07 \pm 0.52$ \kms from 
      the 4 Keck stars (described in \S \ref{s:rv}), 
      gray and black represent Lick/Palomar and Hectochelle respectively.
      The width of each distribution is as expected from the RV precision
      of each survey, and should not be interpreted as a resolved 
      cluster velocity dispersion. 
      According to the quoted RV precision of \citet{chubak2012}, 
      the 0.5 \kms\ dispersion in our 4 Keck velocities might 
      actually be the intrinsic cluster velocity dispersion.
      \label{f:rvhist}}
\end{center}\end{figure}

Selecting the six stars above showing no evidence of binarity, 
we find a typical cluster radial velocity of $40.86 \pm 0.56$ \kms 
(if we take only the 4 Keck stars analyzed by \citeauthor{chubak2012}, 
then we find the cluster RV is $41.07 \pm 0.52$ \kms).
We take stars with RVs consistent with this value as high-probability cluster members. 
Figure \ref{f:rvhist} plots 98 stars with Lick/Palomar velocities (shown in gray hash) 
with RV$_{\rm LP} = 43.8 \pm 3.2$ \kms. 
Also shown are 45 stars with Hectochelle velocities (black line) 
with RV$_{\rm H} = 41.6 \pm 1.5$ \kms. 
The blue tick mark at the top shows the typical cluster velocity from above at 40.57 \kms.
The width of each distribution is consistent with the RV precision
of each survey, and should not be interpreted as a resolved cluster velocity dispersion
\footnote{M67, a much richer cluster, has a measured velocity dispersion of 0.5 \kms\
from radial velocities measured by \citet{Mathieu1983}. 
Assuming virial equilibrium, the cluster velocity dispersion can be 
approximated as $\sigma_v ({\rm km s^{-1}}) \sim \sqrt{\frac{G M}{R}}$.
With $\sim$500 known members and similar size, we expect the M67 velocity
dispersion to be about twice that of R147. 
If the 50 \mps\ RV precision estimate by \citet{chubak2012} is valid, 
then the 0.5 \kms\ RV dispersion in our Keck RVs might actually be the intrinsic 
velocity dispersion for Ruprecht 147, on par with M67 despite the lower number of members.}
 
We have RVs from Lick/Palomar (RV$_{\rm LP}$) for nearly all stars listed in 
Table 6, except the 6 putative blue stragglers and 4 double-lined spectroscopic binaries (SB2s); 
and RVs from Hectochelle (RV$_{\rm H}$) for 50 stars in the central square degree. 
The Hectochelle velocities are more precise (Figure \ref{f:rvhist}), 
so whenever available, RV$_{\rm H}$ is used to determine 
the confidence in membership. 
Some stars have RV$_{\rm LP}$ within the highest confidence interval, 
and RV$_{\rm H}$ in a lower level. 
In these cases, if the star has `Y' confidence level proper motions, RV$_{\rm LP}$ and 
photometry, we set the overall confidence to `P', 
and consider the star a candidate SB1
(e.g. CWW 92 has RV$_{\rm LP}$ = 45 \kms\ and RV$_{\rm H}$ = 25 \kms). 

\subsection{The Color - Magnitude Diagram and Stellar Populations}
\label{s:cmd}
Before assiging membership confidence designations, we check that 
the stars are confined to the region of color -- magnitude space 
expected for a coeval stellar population with the properties we 
determine best describe R147.

\begin{figure*}\begin{center}
\plottwo{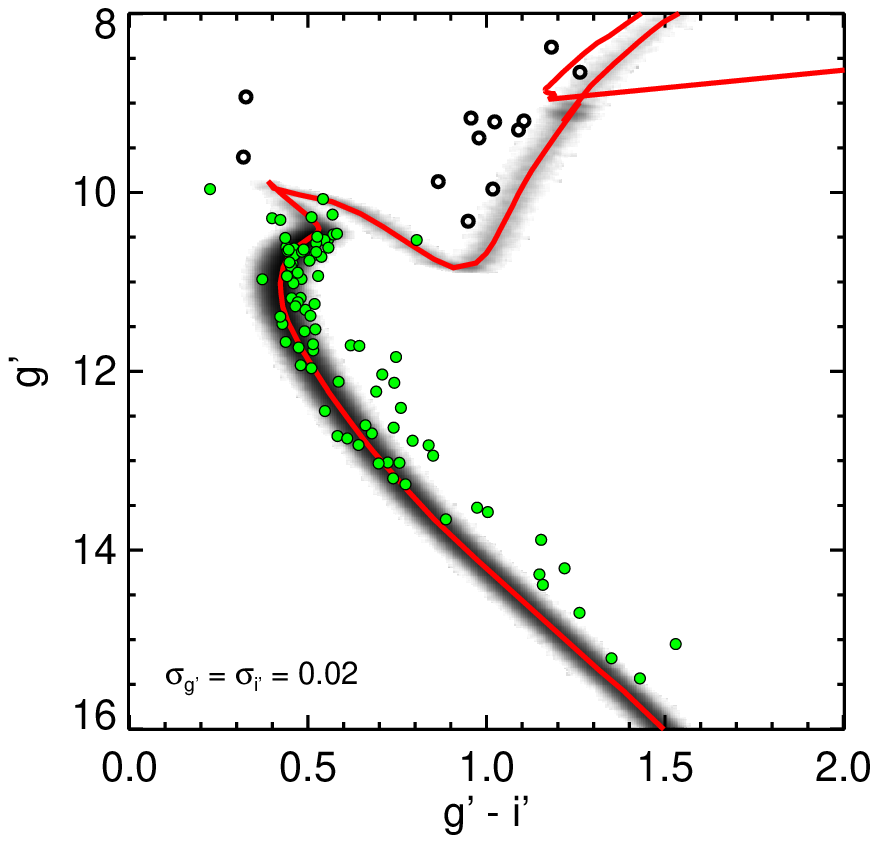}{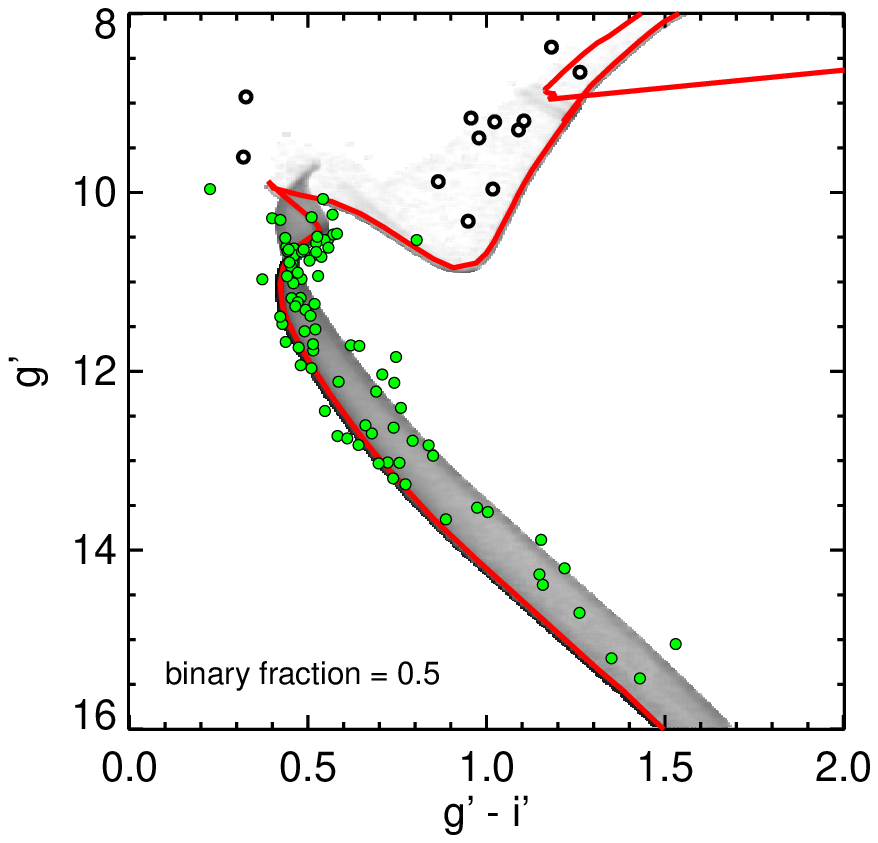}
\plottwo{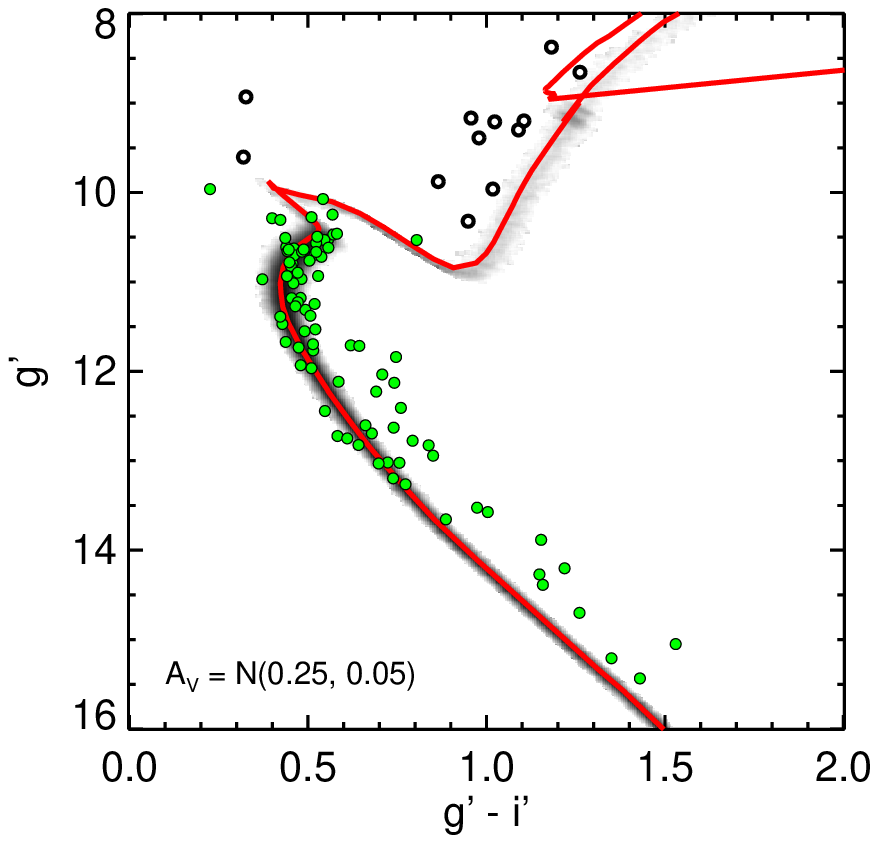}{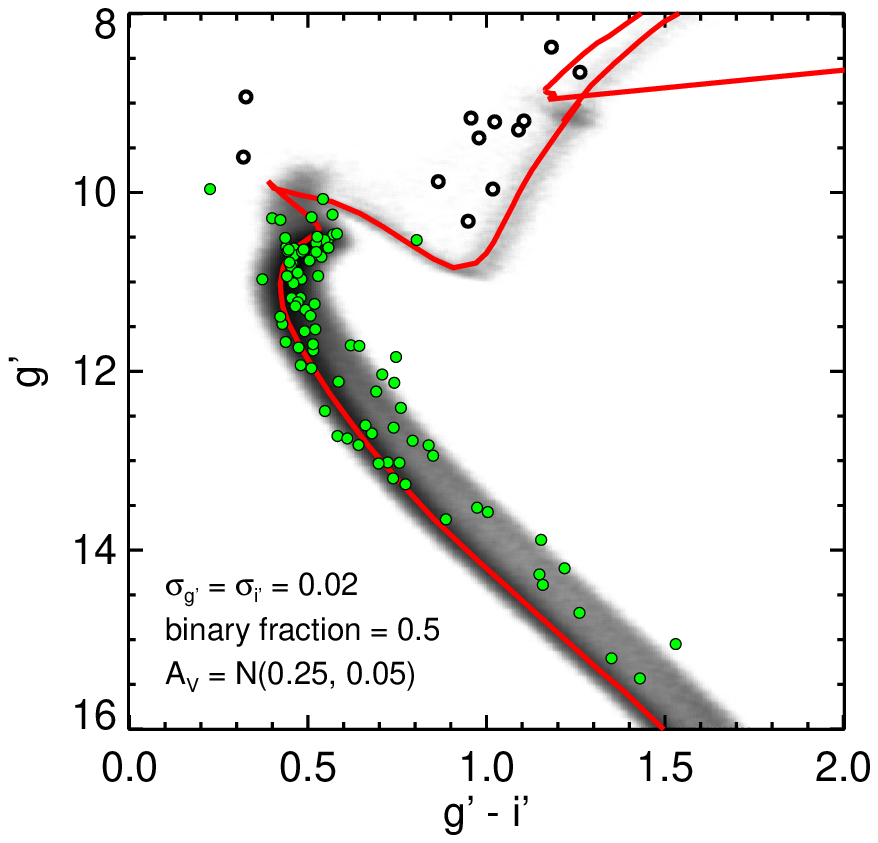}	
    \caption{Each panel illustrates a source of main sequence broadening, 
    demonstrating why the R147 main sequence 
    might appear thicker than a textbook ``beads on a wire" CMD. 
    The R147 stars are plotted in green and saturated stars are plotted with an open circle. 
	A Padova isochrone is overlaid in red with age = 2.51 Gyr, $m - M$ = 7.32 ($d = 291$ pc), 
    $A_V = 0.23$, and [M/H] = +0.065. 
    The shaded regions in each panel represent simluations of $10^6$ stars, 
    with masses uniformly distributed between 0.06 and 1.6 \msun, 
    and photometry queried from the previously quoted Padova model.  
    The simulated photmetry has been binned (0.005 mag. in $g' - i'$, 0.01 mag. in $g'$), 
    log scaled, and smoothed with a 5 pixel boxcar, 
    to highlight possible regions of color-magnitude space occupied by R147 members. 
    The top-left panel only includes photometric error, set at $\sigma_{g'i'} = 0.02$  
    and assuming normally distributed errors. 
    The top-right panel only includes binaries, with the binary fraction set at 50\%, 
    and the secondary masses uniformly distributed between zero and the primary mass. 
    The bottom-left panel only includes differential extinction, 
	normally distributed about the typical cluster value of $A_V$ = 0.25, 
	with $\delta$$A_V = N(0, 0.05)$. 
   	\label{f:sim}}
\end{center}\end{figure*}

We mapped out this locus by simulating a rich cluster with the properties 
we find for R147 from isochrone fitting (\S \ref{s:isofit}). 
Figure \ref{f:sim} shows CMDs for such simulated clusters. 
In this case, we simulated a cluster with $10^6$ stars, 
with masses uniformly distributed between 0.6 and 1.6 \msun. 
We set the binary fraction to 50\%, with companion masses uniformly 
distributed between zero and the primary mass. 
Differential extinction is introduced according to a Gaussian with 
$\mu = 0.25$ and $\sigma = 0.05$ mag., 
and photometric precision is set at 0.02 mag. for \griz\ and 0.025 mag. for $JHK_S$.   
The simulated photometry is drawn from a Padova isochrone with 
$\log t = 9.4$ (2.5 Gyr), [M/H] = +0.065, and $m - M$ = 7.35 mag.
We bin, log-scale, and smooth the synthetic photometry to highlight the R147 locus in 
color -- magnitude space.

Stars overlying the shaded region (basically, the region bound by the single star and 
equal mass binary sequences) are given the highest confidence designation. 
The simulation demonstrates that atypical differential reddening along a 
particular line of sight or relatively high photometric error 
can place stars outside the locus. 
Stars in these regions are assigned `P'. 
These could also be triple systems or exotic products of stellar mergers.
CWW 67 is the only star existing beyond the equal mass triple sequence, 
and we assign it the lowest designation, `N'. 

Confidence assignment is an iterative process, 
since we identify high-confidence members using isochrone fits, 
and these fits require a list of high-confidence members so that 
unlikely or non-members do not throw off the fit. 

\subsubsection{Stellar Populations}
\textbf{Blue Stragglers:} 
In addition to the potential triple systems, 5 -- 6 
stars occupy a space of the CMD outside the cluster locus 
beyond the main sequence turnoff (MSTO): 
6 clearly separate in the 2MASS CMD, but only 5 in \griz.
These five stars have proper motions consistent with the cluster, 
but lack RV measurements due to rotational line broadening 
(CWW 24, the sixth outlier in 2MASS, does have a measured 
RV$_{\rm LP}$ = 41.8 \kms, and so we assume that 2MASS photometric 
error is responsible for scattering it out of the cluster locus).
We classify these 5 stars as blue stragglers 
(see Table 6, blue stragglers are listed 
as `BS' in the Notes column). 
For the photometric probabilities, instead of the `Y/P/N' scheme, 
we assign a `B' for blue straggler.

\textbf{Red Giants:}  
We find 11 red giants in the cluster.  
The TERAPIX photometric errors suggest that only the four brightest red giants 
are saturated in \griz\, even with 1 second exposures. 
Other stars down to $\approx$9.5 are quoted as saturated in each band across the 4 fields. 
After consulting the raw frames, reduced images, and considering the 9.5 magnitude 
saturation limit found for other stars, we conclude that the entire red giant branch 
has unreliable optical photometry. 
This explains the apparent mismatch between our best isochrone fit and the optical RGB.

CWW 14 is fainter and has reliable photometry, although 
it is 0.15 mag. blueward of the red giant branch in ($g' - i'$), 
and 0.08 mag. (4-$\sigma$) blueward in ($J - K_S$).
\citet{Mathieu1990} identify a SB1 system in M67, S1040, 
which lies 0.2 mag. to the blue of the red giant branch in ($B-V$).
This system was previously suggested to consist of a star further 
down the giant branch with a companion star near the MSTO. 
\citet{Landsman1997} identified broad Lyman absorption features, demonstrating 
that the companion is actually a hot white dwarf, and that the 
system likely underwent a period of mass transfer. 
CWW 14 is an outlier in NIR, making it less likely to be a ``red straggler", 
and is probably a MSTO -- RGB binary.

\textbf{Main Sequence dwarfs:}  
We use our best isochrone fit to determine approximate spectral types 
for the R147 membership. We assume masses of 1.1 \msun\ for G0 and 0.8 \msun\
for K0 dwarfs \citep{handbook}, then locate the boundaries in the CMD 
from the isochrone. 
We find that the MSTO is located around mid-F. 
The subgiant branch down to F8 on the main sequence is well populated with $\approx$52 stars.
We also identify $\approx$27 G dwarfs and $\approx$8 K dwarfs down to mid-K 
(We quote approximate numbers because of the approximate nature of our spectral typing). 
The 9 stars lacking \griz\ photometry appear in the 2MASS CMD 
as follows: 2 red giants, 3 MSTO F stars, 2 G dwarfs and 2 K dwarfs.

This method ignores the existence of binaries but illustrates the 
top-heavy nature of our membership list. 
This is likely due to a combination of observational bias and cluster evaporation 
(star clusters tend to lose their lowest mass members first, 
and `evaporate' from the bottom up). 
The typical NOMAD proper motion error is $\sim$10 mas/yr by $V \sim$ 12.
The K dwarfs have $V > 13$, making candidate identification from 
proper motions difficult. 
Therefore we are almost certainly missing significant numbers of low mass dwarfs.

\subsection{Notes on Particular Stars}
\label{s:members}

\subsubsection{Apparent Non-members}
CWW 77 has RVs inconsisent with R147 (RV$_{\rm LP}$ = 51 \kms, RV$_{\rm H}$ = 52 \kms), 
but its ($g' - i'$) and ($J - K_S$) photometry place it near the equal mass binary sequence, 
so it could be an SB1. 

CWW 67 has a low RV$_{\rm LP}$ = 34 \kms\ and is 1.3 mag. above the ($g' - i'$) 
main sequence and 1.4 mag. above in ($J - K_S$), 
but an equal mass triple would sit 1.2 mag. above the main sequence, 
so membership seems very unlikely (although a fourth lower mass companion could 
theoretically explain these discrepancies, so membership is difficult to 
definitively rule out).  

CWW 72  sits $\approx$1 mag. above the main sequence in ($g' - i'$) and $\approx$0.75 mag. in 
($J - K_S$) -- if a member, this could be a triple or an equal mass binary with 
inaccurate optical photometry. 
CWW 72 is a SB2, as seen in the Keck/HIRES spectrum. 
This star was observed previously at Lick and the spectrum exhibited no sign of binarity 
(otherwise we would not have selected it for observation with Keck). 
We also observed CWW 72 on two consecutive nights with MMT/Hectochelle. 
The CCF from the first night exhibits a tall and sharp peak with RV$_{\rm H1}$ = 46.6 \kms. 
The CCF from the second night is lopsided, suggesting that the signature of the companion was 
beginning to manifest and that the period of this system could be on the order of days. 
The CCF shape and resulting radial velocity from the first night point to a systematic velocity 
$\approx$5 \kms\ greater than the R147 bulk motion, which cannot be explained by 
the cluster's velocity dispersion or RV precision. 
If CWW 72 is a member, then it is (at least) a triple, perhaps with two approximately equal mass 
primary components orbiting with a period of days, 
and a fainter companion modulating the RV on a longer timescale (needed to explain the 5 \kms\ 
systematic offset).

Finally, CWW 50 sits 0.05 mag. blueward of the ($g' - i'$) main sequence, 
but is on the ($J - K_S$) main sequence. 
In Section \ref{s:av} we discuss the possibility that this star 
is less extinguished and reddened than the rest of the cluster. 
If this is not the case, 
perhaps a hot white dwarf is pulling it blueward while not introducing much 
NIR flux, or there is an atypically large photometric error in one of the optical 
bands (2-$\sigma$), 
or else CWW 50 is not a member. 
We list it as `P' because it is only inconsistent in ($g' - i'$), 
and while RV$_{\rm LP}$ = 47.6 \kms, the Lick/Palomar velocity 
precision does not rule out membership.

\begin{deluxetable*}{lllll}
\tablecaption{Criteria for membership \label{t:criteria}}
\tablewidth{0pt}
\tabletypesize{\scriptsize}
\tablehead{\colhead{Data} & \colhead{Source} & 
\colhead{Highest: Y} & \colhead{Probable: P} & \colhead{Low or Non-member: N}}
\startdata
$r_\mu$\tablenotemark{a}	&	NOMAD, UCAC-3, Adam Kraus	&	$< 5$	&	5 -- 8  &	$>$ 8 \\
Radial Velocity\tablenotemark{b} & Lick \& Palomar & 39 -- 47 & 36 -- 39, 47 -- 50 & 33 -- 36, 50 -- 53 \\
Radial Velocity & Hectochelle &	40 -- 43 & 38.5 -- 40, 43 -- 44.5 & else \\
($J - K_S$) CMD	&	2MASS	&	Overlaps with simulation\tablenotemark{c}	& $\pm$ 0.2 mag.	&	beyond equal mass triples	\\
($g'-i'$) CMD & CFHT/MegaCam & Overlaps with simulation & $\pm$ 0.2 mag. & beyond equal mass triples \\
\enddata
\tablecomments{See Section \ref{s:properties} for a discussion of membership criteria. 
Values in parenthesis denote acceptable ranges. Values equal to endpoints are assigned to the higher level}
\tablenotetext{a}{Radial distance in proper motion space from the mean value for R147, 
with units of mas/yr.}
\tablenotetext{b}{Radial velocities measured in \kms.}
\tablenotetext{c}{See Section \ref{s:cmd} and/or Figure \ref{f:sim} for discussion}
\end{deluxetable*}

\subsubsection{Notes on 2MASS photometry}
\label{s:2mass}
Our \griz\ imaging shows CWW 51 is an optical double, 
with a star 1.65 arcseconds away with a similar \griz\ SED 
(the magnitude difference in each band is 0.02 -- 0.05 mag. between the two stars). 
The components are separated by 1.65'', which translates into 
a minimum physical separation of 495 AU, assuming a cluster distance of 300 pc. 
This suggests that the pair actually form a wide binary, 
although their angular proximity could also be explained by a chance alignment. 
This double was not resolved in the 2MASS Point Source Catalog. 
Adding 0.75 mag. to the $J$ band magnitude (halving the brightness, to reflect just the one star) 
moves CWW 51 next to 
the stars it neighbors in the ($g' - i'$) CMD. 
Table 6 quotes, and the figures in this work plot, the 2MASS 
photometry for CWW 51, despite this realization, 
although we do include a footnote referencing this in the Table.  

\begin{figure}\begin{center}
\plotone{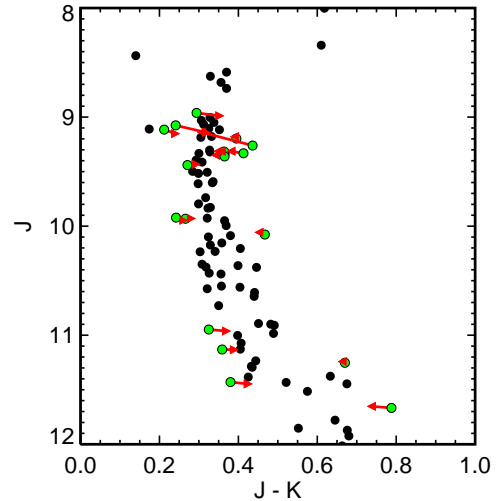}
\caption{PSF photometry (green circles) vs aperture photometry (red triangles) 
	for 18 outliers on the 2MASS ($J - K_S$) CMD. 
	All shift closer to the locus when the aperture 
	photometry is used instead of the default PSF photometry. 
	No star already in the locus shifts appreciably outside when aperture photometry is used instead.
	One star slides from blueward from the main sequence by 0.037 mag., or approximately equal to 
	the $J$ and $K_S$ errors added in quadrature (not shown). 
	\label{f:ap2mass}}
\end{center}\end{figure}

The 2MASS Point Source Catalog provides PSF photometry by default in most cases. 
Figure \ref{f:ap2mass} shows 23 outlier stars on either side of the 
($J - K_S$) main sequence, out of the 80 stars `Y/P' stars with 
aperture photometry that are not blue stragglers. 
If aperture photometry is used instead, each of these stars 
moves towards the cluster locus. 
No star already in the locus shifts appreciably outside when aperture photometry is used instead. 
One star slides from blueward from the main sequence by 0.037 mag., or approximately equal to 
the $J$ and $K_S$ errors added in quadrature.
The fact that the majority of outliers' photometry systematically moves 
toward the cluster locus suggests to us that for many stars in these fields 
and at these magnitudes, the aperture photometry is superior. 
We do not assign lower confidence levels to PSF photometry 
outliers, if their aperture photometry is consistent with membership. 
We include the aperture photometry for 18 stars in 
Table 6 and all other figures 
(CWW 22, 24, 27, 28, 30, 37, 38, 43, 48, 49, 57, 59, 67, 68, 90, 91, 94, and 100).

\subsubsection{SB2 systems} 
\label{s:sb2}
\begin{figure*}\begin{center}
	\plottwo{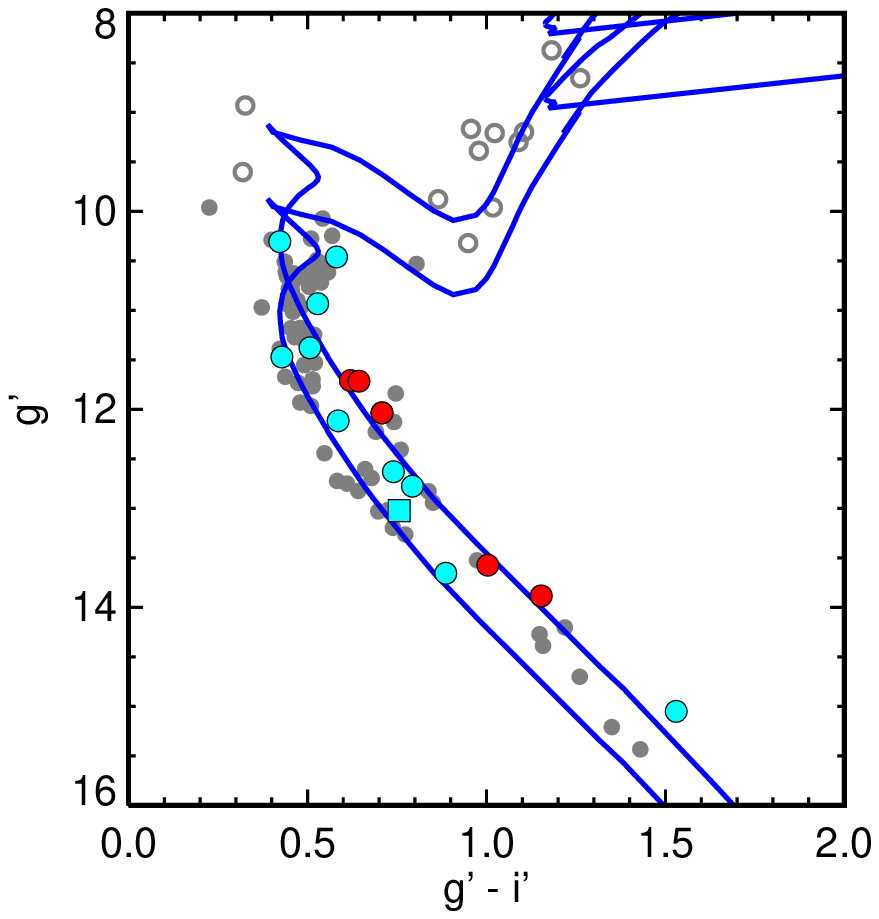}{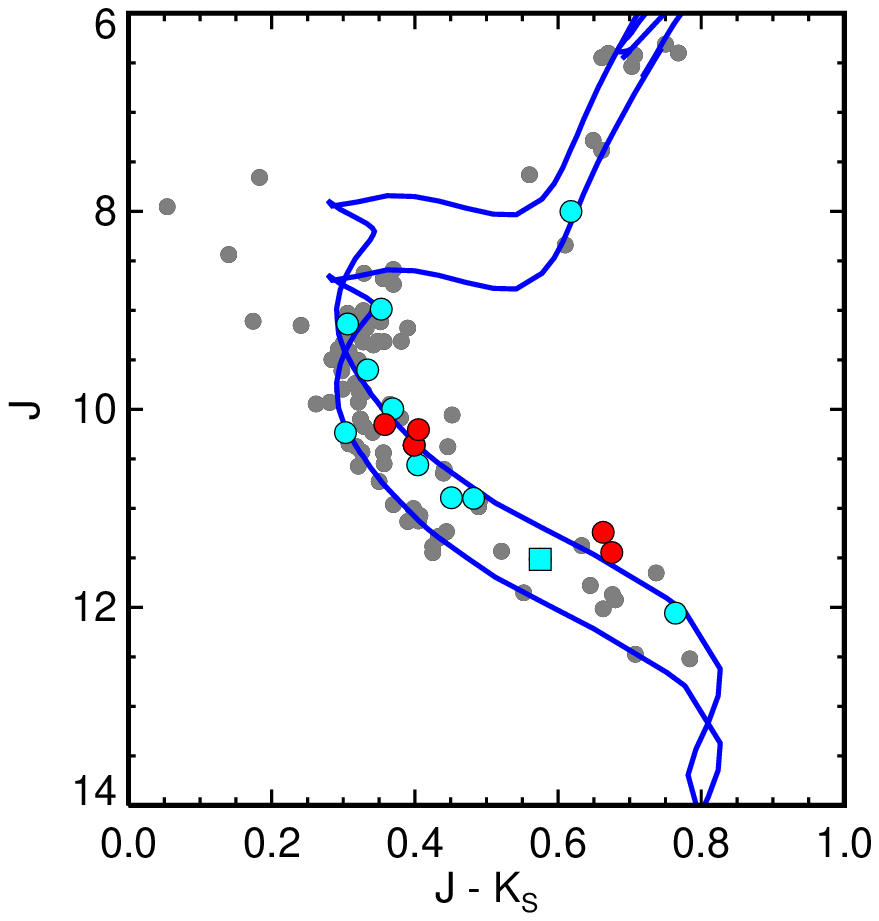}
	\caption{Ten stars with discrepant Hectochelle RVs are plotted in cyan 
	(RV more than 1.5 \kms\ away from cluster average).
	Five SB2s are plotted in red. 
	The rest of the cluster is shown in gray, with our best fit isochrone (Padova model) 
	overplotted in dark blue showing the single and binary sequences 
	($\log t = 9.4$, [M/H] = +0.064, $m - M = 7.32$, $A_V = 0.23$).  
	All of the SB2s were identified by a double-peaked cross-correlation function, 
	which indicates these should be nearly equal mass ratio systems. 
	The fact that they are all clustered around the binary sequence (shifted ``up" 0.75 mag.) 
	coroborates their equal mass status and the validity of our isochrone fit.
	All but 3 SB1 candidates are also shown near or on the binary sequence. 
	The square shows CWW 99 on the ($g' - i'$) single star sequence, but in ($J - K_S$) the 
	star is 0.5 mag above this sequence. 
	The other two stars might be high-mass ratio systems (and therefore do not manifest in 
	shifts on the CMD), stars that have received gravitational kicks (so their RVs 
	are no longer consistent with the cluster), or are non-members.}
	\label{f:binary}
\end{center}\end{figure*}

CWW 64, 65, 66, 68, and 72 showed double-peaked cross-correlation functions in one of the RV epochs, 
indicating these systems to be nearly equal mass binaries (the case of CWW 72 is discussed above). 
Figure \ref{f:binary} plots the ($g' - i'$) and ($J - K_S$) CMDs, 
with the five SB2s highlighted red. The single star and equal mass binary sequences 
from our best isochrone fit (Padova model) are plotted in green. 
All five SB2s are clustered around the equal mass binary sequence, 
corroborating their equal mass status and the validity of our isochrone fit.
\footnote{CWW 68 actually sits on the equal mass triple sequence. 
Both CWW 64 and 65 are midway between the 
equal mass binary (-0.75 mag.) and triple (-1.2 mag.) sequences, 
at 1.0 mag. brighter than the main sequence, 
which can occur when an equal mass binary system, which manifests as the SB2, 
has a third companion with 50\% the luminosity of each primary.}

\subsubsection{SB1 binary candidates from discrepant RVs}
\label{s:sb1}
Figure \ref{f:binary} also plots in cyan 10 stars with Hectochelle RVs 
inconsistent with the cluster: 
CWW 19, 22, 27, 53, 69, 70, 77, 92, 95, 99, and 106 
(CWW 19 is a red giant. We do not include it in the optical plot because 
it is saturated).
In Section \ref{s:rv} we suggested CWW 92 is a SB1 from the 
20 \kms\ difference between RV epochs. 
All other SB1 candidates have similar RVs from Lick/Palomar and Hectochelle, 
and are considered candidate SB1s because RV$_{\rm H}$ is at least a few standard 
deviations away from the cluster average, although this discrepancy 
forces a `P' classification 
(except CWW 77, which we classify as `N' as noted above). 

Figure \ref{f:binary} shows all but CWW 70 are clustered around the binary sequence. 
CWW 70 has RV$_{\rm LP}$ = 38 \kms\ and RV$_{\rm H}$ = 39 \kms. 
Despite this 2-$\sigma$ discrepancy, we cannot rule out membership. 
The star might have a low mass companion or received a gravitational ``kick". 
For example, 
a G2 V (1 \msun) -- M0 V (0.5 \msun) binary system with semi-amplitude velocity $K_1$ = 5 \kms, 
zero eccentricity and zero inclination, will have a period $P \approx 12$ years. 
The large luminosity difference in a high mass ratio binary means the system will not 
stand out in either photometry (it will lie on the single star sequence) 
or in spectra (the secondary is too faint to manifest as a SB2).
This means although the RVs for CWW 70 are inconsistent with single star membership, 
we cannot rule out the possibility of a low mass companion at large separation, 
which would induce a measureable velocity offset, 
but modulated at a period much longer than our 2 year baseline. 

CWW 99 is plotted as a blue square in Figure \ref{f:binary}. 
It sits on the single star ($g' - i'$) sequence, 
but is 0.5 mag. above the ($J - K_S$) sequence. 
This can be explained by a low mass companion, 
which would show up more prominently in NIR than optical. 

\section{Inferring Cluster Properties}
\label{s:properties}
The properties of stellar clusters (age, composition, distance and interstellar extinction) 
are commonly estimated by fitting isochrones to broadband photometric 
color -- magnitude diagrams (CMD). 
Often a ``chi-by-eye" technique is employed, 
where sets of isochrones representing varying cluster parameters are overlaid 
on a CMD, and the apparent best fit or series of best fits are selected 
to establish acceptable ranges for these fundamental parameters. 
This technique can be successful when one or more of these properties 
can be well constrained. For example, clusters might be nearby 
or sit above/below the galactic plane and suffer little extinction, 
and the closest benchmarks have parallax distances from Hipparcos and/or HST/FGS. 
These independent constraints break the high degree of degeneracy 
between each variable (e.g. metallicity works in a similar direction 
to the interstellar reddening vector).

None of Ruprecht 147's properties have been previously well measured.
At a distance of over 200 pc, the cluster has a HIP2 
distance measurement from 3 stars that appears
unreliable and is apparently too close by a significant fraction (\S \ref{s:distance}).
We first describe our efforts to independently constrain the 
interstellar extinction with the \citet{dustmap} dust map (\S \ref{s:av}), 
and the composition from spectroscopic analysis (\S \ref{s:sme}), 
then we will use isochrone models to determine the cluster's age and distance. 
Specifically, we fit a spectroscopically derived \teff\ -- \logg\ diagram with 
Padova isochrones with abundances fixed by the spectroscopic metallicity 
(\S \ref{s:isosme}). 
We then query the resulting best fit Padova isochrone for a star with 
\teff\ and \logg\ closest to the values for the early G dwarf we derive with SME, 
and perform a brute force $\chi^2$ SED fit for distance and visual extinction 
to the resulting synthetic $g'r'i'JHK_S$ photometry. 

Next we develop an efficient and automated 2D cross-correlation 
isochrone fitting techinque, and fit Padova and Dartmouth models 
to the ($g' - i'$) and ($J - K_S$) CMDs to determine the age, distance and visual extinction. 
We will see that the NIR and optical fits agree very well with each other
(Figures \ref{f:myisojk}, \ref{f:myisogi}). 
We also fit the ($g' - i'$) CMD with the $\tau^2$ maximum-likelihood method, 
and derive consistent results, validating our isochrone fitting technique.
We find an age from both NIR and optical photometric isochrone fitting consistent with our spectroscopic results, 
which supports our earlier decision to break the degeneracy between these parameters with this age (\S \ref{s:isofit}).

We will compare results from Padova, Dartmouth and PARSEC models; 
and then discuss the differences between our photometric distance and 
the parallax distance inferred from 3 HIP2 cluster members.
Finally, we synthesize the results from these various isochrone fits 
and present our preferred set of parameters describing the 
age, composition, distance and visual extinction for Ruprecht 147 (\S \ref{s:result}).
\subsection{Interstellar Extinction}
\label{s:av}
\begin{figure}\begin{center}
\plotone{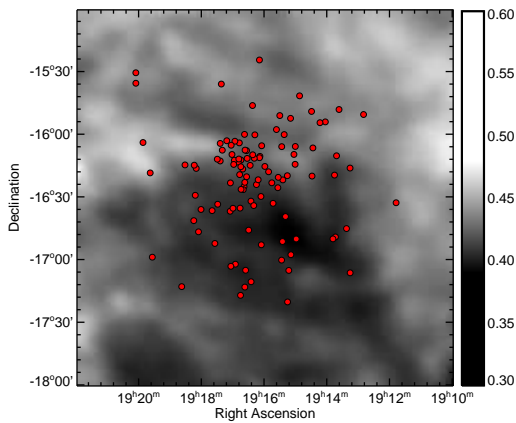}
	\caption{Visual extinction map with R147 members designated by red dots.
	$A_V$ calculated from the dust map of \citet{dustmap}, assuming $R_V$ = 3.1. 
	The map was smoothed with a 3 arcminute boxcar. \label{dust}}
\end{center}\end{figure}

The large apparent size of R147 on the sky introduces the possibility of 
differential extinction across the cluster. 
Figure \ref{dust} plots $A_V$ from the dust map of \citet{dustmap}, 
assuming $R_V = 3.1$\footnote{Many studies have demonstrated that on average,
$R_V = 3.1$ for a diffuse ISM. 
This relationship between extinction and reddening is of course dependent on the 
composition and physical conditions in the intervening ISM. 
Considering Ruprecht 147, there are no dense molecular clouds along the line of sight. 
We also do not expect any additional reddening intrinsic to the cluster or its stars, 
because this is an old cluster, and it and it’s stars are no longer enshrouded in dust and gas. 
Recently, \citet{Jones2011} used over 56,000 Sloan M dwarf 
spectra to map visual extinction and $R_V$ in the nearby Galaxy.
While their $R_V$ distribution peaks near 3.1, their median value is actually 3.38,
and ranges from 2 to 5.5. 
We will assume the canonical value of $R_V = 3.1$ for the diffuse ISM 
for this preliminary study and intend to return to this issue in a future work.}. 
Many cloud structures are apparent in the field, 
showing $A_V$ to vary from 0.3 -- 0.5 mag.

Considering the cluster's proximity, some of this dust undoubtedly 
lies beyond the cluster. 
\citet{drimmel2001} determine the Sun to lie 14.6 $\pm$ 2.3 pc above the Galactic midplane, 
and measure a dust scale height of $h_d = 188$ pc at the Solar Circle.
The Galactic latitude of R147 ranges from -12\degree\ to -14\degree.
At a distance of 250 -- 300 pc, this latitude places R147 50 -- 70 pc below the Sun, 
and 35 -- 55 pc below the midplane (less, if a larger $Z_\astrosun$ is assumed), 
or about 20-30\% of the dust scale height. 
The local bubble has very little dust in it out to $\sim 150$ pc \citep{bubble},
suggesting it is possible that most of the dust is behind the cluster. 

Section \ref{s:cmd} described star cluster simulations used 
to assess photometric membership probabilities, 
by introducing photometric scattering sources to explain the observed width of 
the R147 main sequence, including photometric error, binarity, and differential extinction. 
Figure \ref{f:sim} plots simulated cluster CMDs including 
each of these photometric scattering sources separately, and all combined. 
The binarity simulation demonstrates that the single star and 
equal mass binary sequences pinch together near the MSTO. 
Differential reddening smears the CMD along a negative-sloped diagonal 
(extinction plus reddening). 
If there is non-negligible differential reddening, 
the ``binary pinch" should be smeared out. 

\begin{figure}\begin{center}
\plotone{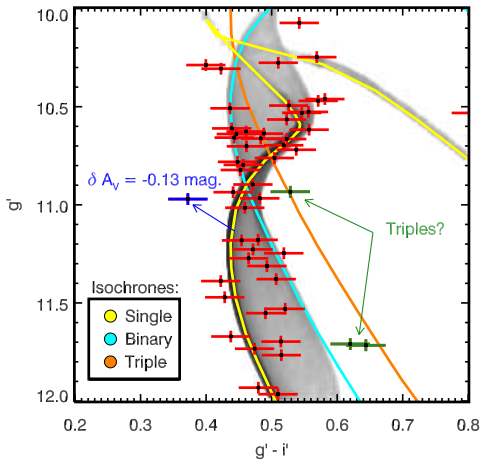}
    \caption{Looking for differential extinction. The `Y' and `P' members are plotted with 
    0.03 mag. error bars in color and magnitude. Single star (yellow) and 
    equal mass binary (cyan) sequence isochrones (Padova) are plotted with 
    $\log t = 9.4$, [M/H] = +0.1, $m - M = 7.48$, and $A_V = 0.23$. 
	The gray shading illustrates the results of our Monte Carlo cluster simulation, 
	including binaries and neglecting differential extinction and photometric error, 
	from Figure \ref{f:sim} (top -- right panel). 
	The single and binary sequences cross at $g' \approx 10.9$, forming a ``pinch" at the turnoff. 
	Differential extinction should smear this pinch along the reddening vector shown in blue. 
	The three outliers redward of the binary sequence could be triples. 
	CWW 50, the one outlier blueward of the single star main sequence, might either suffer 
	atypically large photometric error, or is not a member, or else 
	needs to be reddened by $\delta A_V = 0.13$ in order to place it on the main sequence. 
	Interestingly, the visual extinction from the \citet{dustmap} dust map at the position of 
	CWW 50 is the lowest in the field, and exactly 0.13 mag. lower than the median value for 
	the region of radius $r = $1\degree.2 centered on the cluster and encompassing all R147 members.
    \label{f:diffAv}}
\end{center}\end{figure}

Unfortunately, we have only identified 6 members at this pinch. 
Figure \ref{f:diffAv} shows four are confined within the pinch. 
CWW 50 sits 0.08 mag. blueward, and CWW 53 sits 0.06 mag. redward.
These values are 2 -- 3 times larger than the expected photometric error. 
CWW 53 has RVs from two epochs at 5 \kms\ greater than the 
cluster mean, and photometry placing it near the equal mass triple sequence. 
The uncertainty in membership and multiplicity means we cannot use CWW 53 to test for 
differential reddening. 
CWW 50 must be reddened by $\delta$$A_V = 0.13$ mag. in order to place it on the 
single star sequence. 
In Section \ref{s:isosme}, we find an optimal $A_V = 0.23$ mag., so 
a particular line of sight of $A_V = 0.10$ mag. is not impossible.
Ideally, we would like to check if the nearest R147 neighbors to CWW 50 also appear less 
extinguished than expected. Unfortunately, the nearest neighbor is $\sim$10' away. 
It is also noteworthy that the \citet{dustmap} dust map extinction along this line of sight is 
$A_V = 0.296$ mag., the lowest value in the entire field, and 0.13 mag. lower than the 
median $A_V$ for the region of radius $r = $1\degree.2 encompassing all R147 members. 

The R147 main sequence is thicker than 1 magnitude at various points.
Unfortunately in many cases, our radial velocities do not have 
sufficient precision to firmly establish these stars as cluster members 
(we still designate them `Y' members because the RVs are 
consistent with the R147 bulk motion, within the precision of our Lick/Palomar RV survey). 
We also only have one RV epoch for the majority of stars, 
and so stellar multiplicity is impossible to diagnose at this point. 
More precise velocities are required before we attribute these photometric outliers 
to differential extinction. 

There is also a strip of 7 main sequence dwarf stars 
blueward of our best isochrone fit in the ($g' - i'$) CMD. 
Adding $\delta$$A_V$ = 0.05 mag. to these stars shifts them onto the isochrone, 
This translates into a 0.03 mag. shift in color, 
which is within the photometric precision, 
and so differential extinction is not required to explain these stars 
apparent blueward offset in the optical CMD.
No net offset is seen in the NIR CMD.

We will postpone further investigation into differential extinction to a future study, 
and in this work will fit single $A_V$ models, with 
values constrained by the dust map at $A_V < 0.5$.

\subsection{Metallicity}
\label{s:sme}
We analyzed five Keck/HIRES spectra with Spectroscopy Made Easy 
\citep[SME,][]{sme}, using the procedure described in \cite{valenti2005}.
SME uses the Levenberg-Marquardt algorithm to fit observed echelle spectra 
with synthetic spectra generated assuming LTE and plane-parallel geometry, 
yielding effective temperature, surface gravity, metallicity, 
projected rotational velocity, 
and abundances of the elements Na, Si, Ti, Fe, and Ni 
\footnote{\citet{valenti2005} did not solve for magnesium abundance because 
of the degeneracy between [Mg/Fe] and \logg\ when fitting 
the synthetic spectra to the gravity-sensitive Mg b triplet \citep{firstMgb}. 
\citet{Fuhrmann1997} recommends fitting for [Mg/Fe] first, using 
the weak \mgi\ lines at $\lambda$4571 or $\lambda$5711, 
and then fixing the abundance and fitting the Mg b wings to 
derive \logg. 
\citet{valenti2005} decided to exclude from analysis wavelengths $\lambda < 6000$ \AA, 
except for the region encompassing the Mg b triplet, 
because of severe line blending in the blue in cool stars which 
would complicate the spectral synthesis fit.} 

\begin{figure}\begin{center}
\plotone{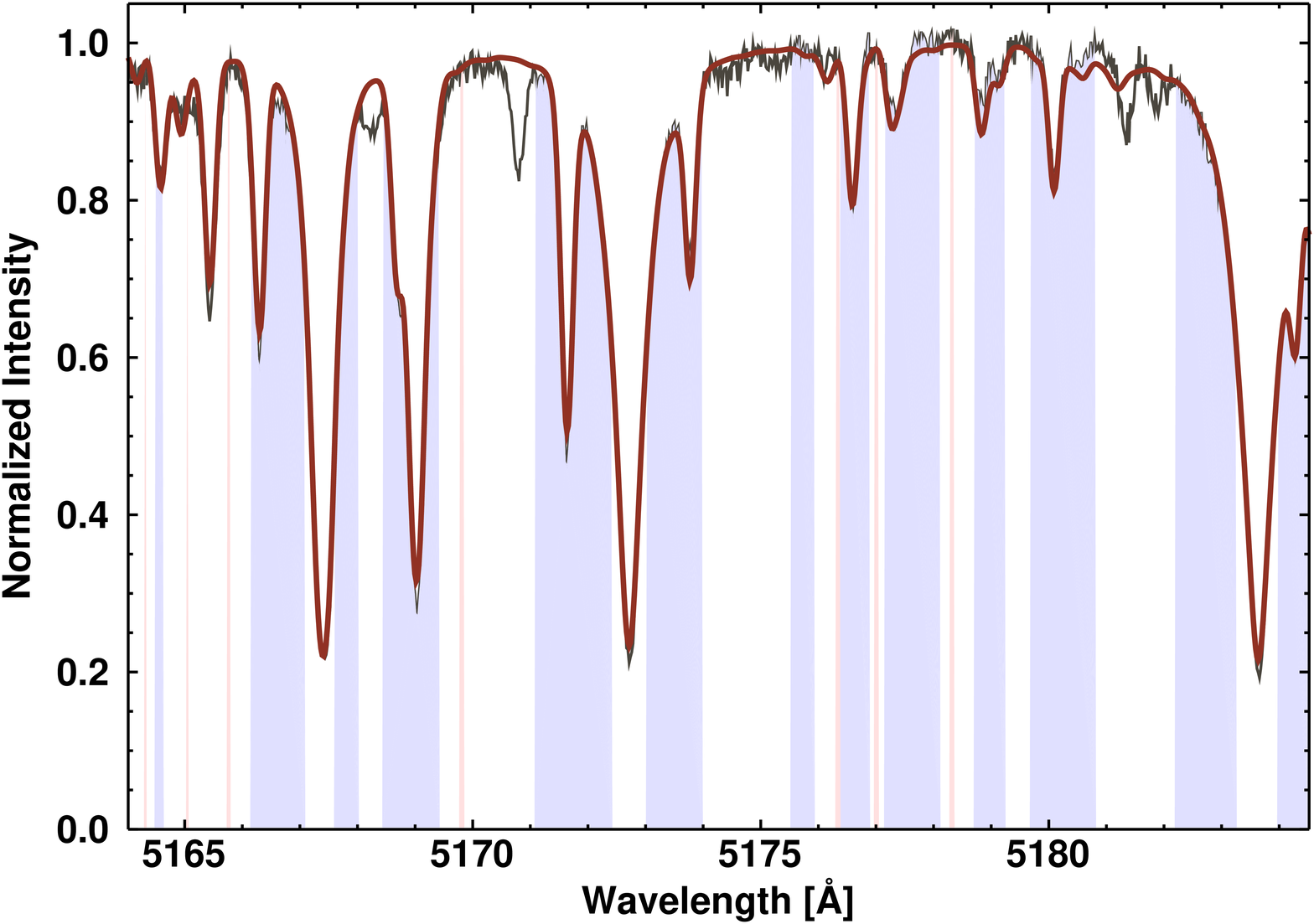}
    \caption{A Keck/HIRES spectrum of the order encompassing the Mg b triplet for 
 	CWW 44, an F MSTO star, is shown in black. 
    The synthetic spectrum resulting from our SME analysis is overlaid in red. 
    The spectrum segments included in the 
    fit are highlighted in purple, and the salmon stripes identify the continuum regions. 
    Table \ref{t:sme} lists our SME results for this and four other stars. 
    \label{f:smeg2}}
\end{center}\end{figure}

Upon obtaining an initial fit to a spectrum, 
\teff\ was perturbed $\pm$100K and run again. 
The three solutions were then averaged, with the standard 
deviation set as the parameter uncertainty, 
except in cases where this uncertainty is less than the statistical 
uncertainities measured in \citet{valenti2005}: 
44 K in effective temperature, 0.03 dex in metallicity, 
0.06 dex in the logarithm of gravity, 
and 0.5 \kms\ in projected rotational velocity. 
Additional corrections are applied to the final values based 
on the analysis in \citet{valenti2005} of Vesta and
abundance trends in binary pairs with \teff\ 
(see Figure \ref{f:smeg2} for the CWW 44 Keck/HIRES spectrum 
and SME synthetic spectrum fit in the order encompassing Mg b).

Our results for the five stars are presented in Table \ref{t:sme} and 
indicate that the cluster has a slightly super-Solar metallicity 
of [M/H] = +0.07 $\pm$ 0.03, from CWW 91, 44 and 21\footnote{
We will often use [M/H] = +0.064 throughout this work, 
because of the way the Padova isochrone webtool parameterizes metallicity: 
[M/H] = $\log (Z / Z\astrosun)$, with $Z = 0.019$. 
Setting $Z = 0.022$ gives [M/H] $\approx$ +0.064.}.
We neglected the results from CWW 22 because it is hottest 
(complicating the fit to gravity, described in the next subsection) and 
has the poorest $\chi_\nu^2$ fit. 
\citet{valenti2005} suggest using [Si/H] as a proxy for 
the $\alpha$-process abundance. 
We find [$\alpha$/Fe] = [Si/H] $-$ [Fe/H] $\approx$ 0.0 
([Si/Fe] = -0.03 and 0.0 for CWW 44 and 91). 

We find a much lower metallicity for CWW 78, [M/H] = -0.11 $\pm$ 0.03 
and [Fe/H] = 0.0 $\pm$ 0.02. 
This outlier has otherwise satisfied every 
criterion for membership, with proper motions, photometry and 
a precise RV all consistent with the cluster. 
For this work, we will assume that this peculiar metallicity 
can be explained by a complication in the SME analysis, 
and will look into this in a future study.

\citet{redgiants} analyzed high-resolution, high signal-to-noise spectra 
of three red giant members of R147, and their results are compiled in 
Table \ref{t:sme}. 
The first star, HD 179691, has a radial velocity 
inconsistent with the cluster, indicating it is either an SB1 or not a member.
The other two red giants show super-Solar iron abundance, 
consistent with our SME results.

In the future, we will more rigorously determine the cluster metallicity 
combining photometry, spectroscopy, and cluster properties. 
\begin{turnpage}
\begin{deluxetable*}{cccccc|ccc}
\tabletypesize{\scriptsize}
\tablecaption{Spectroscopic analysis of R147 stars \label{t:sme}}
\tablewidth{0pt}
\tablehead{
\colhead{Property} & \colhead{CWW 44\tablenotemark{a}} & 
\colhead{CWW 91\tablenotemark{a}} & \colhead{CWW 21\tablenotemark{a}} & 
\colhead{CWW 22\tablenotemark{a}} & \colhead{CWW 78\tablenotemark{a}} &
\colhead{HD 179691\tablenotemark{b}} & \colhead{HD 180112\tablenotemark{b}} & 
\colhead{HD 180795\tablenotemark{b}} 
}
\startdata
Type & mid-F MSTO     & G0/2 V         & subgiant & mid-F MSTO / SB1? & late-F V & 
	K1 III & K0 III & K0 III \\
\teff\ (K) & 6273 (5) & 5747 (62) & 6129 (25) &  6350 (80) & 6115 (52) & 
	4573 (80) & 4733 (80) &	4658 (80) \\
\logg\ (gm cm s$^{_-2}$) &  4.11 (0.02) & 4.35 (0.11) & 3.79 (0.07) & 3.6 (0.06) &  4.27 (0.08) &
	2.28 (0.15) & 2.53 (0.15) & 2.43 (0.15) \\
RV (\kms) & 41.41 & 41.50 & 40.35 & 46.63 & 41.02 &
	46.7 & 40.1 & 40.8 \\
\vsini\ (\kms) &  6.87 (0.69) & 0.32 (0.33) & 6.50 (0.61) & 6.91 (0.73) & 6.09 (0.65) &
	-	&	-		&	-	\\
$\left[ \rm{M/H} \right]$ & 0.07 (0.01)  & 0.06 (0.03) & 0.09 (0.03) & -0.01 (0.04) & -0.11 (0.03) &
	-	& - &	-	\\
$\left[ \rm{Na/H} \right]$ & 0.22 (0.02) & 0.23 (0.01) & 
0.12 (0.01) & -0.02 (0.08) & -0.14 (0.03) &
	0.24	&	0.16	&	0.24	\\
$\left[ \rm{Si/H} \right]$ & 0.14 (0.01) & 0.11 (0.02) & 
0.10 (0.01) &  0.00 (0.03) &  0.02 (0.02) &
0.08 $\pm$ 0.08		&	0.15 $\pm$ 0.06	&	0.25 $\pm$ 0.08	\\
$\left[ \rm{Ti/H} \right]$ & 0.28 (0.02) & 0.16 (0.02) & 
0.25 (0.03) &  0.03 (0.05) &  0.06 (0.04) &
-0.03 $\pm$ 0.07	&	0.04 $\pm$ 0.05 &	-0.02 $\pm$ 0.04 \\
$\left[ \rm{Fe/H} \right]$ & 0.17 (0.01) & 0.11 (0.02) & 0.22 (0.01) &  
0.08 (0.03) & -0.00 (0.02) &
	-	&	-	&	-	\\
\feih & - & - & - & - & - & 
	0.03 $\pm$ 0.06	& 0.14 $\pm$ 0.06 & 0.16 $\pm$ 0.04 \\
\feiih & - & - & - & - & - & 
	-0.02 $\pm$ 0.09 	&	0.07 $\pm$ 0.07	&	0.08 $\pm$ 0.04	\\
$\left[ \rm{Ni/H} \right]$ & 0.07 (0.01) & 0.05 (0.02) & 0.13 (0.01) & 
-0.03 (0.06) & -0.02 (0.03) & 
-0.04 $\pm$ 0.08	&	0.04 $\pm$ 0.06	&	0.12 $\pm$ 0.07	\\
$\chi^2_\nu$ & 2.90          & 2.86          & 4.07			 & 5.88 		  &  2.18 &
	-	&	-	&	-	\\
$\log t$ (years)	& - & - & - & - & - & 	
	9.2 $\pm$ 0.5	&	9.0 $\pm$ 0.4	&	9.0 $\pm$ 0.4 \\ 
\enddata
\tablecomments{Rows and SME statistical uncertainties: 
(Type) Rough Spectral Type,  
(\teff) Effective Temperature: $\sigma$ = 44 K,  
(\logg) Surface Gravity: $\sigma$ = 0.06 dex,
(RV) Radial Velocity (v $\sin i$) projected rotational velocity: $\sigma$ = 0.5 \kms,  
([M/H]) Metallicity = $\log_{10} Z / Z_{\astrosun}$ ([Na/H] .. [Ni/H])
Sodium, Silicon, Titanium, Iron and Nickel abundance: $\sigma$ = 0.03 dex,  
($\chi^2_\nu$) Reduced $\chi^2$ of the fit,  
($\log t$) age in years assuming d = 280 pc}
\tablenotetext{a}{Our SME analysis results of Keck/HIRES spectra}
\tablenotetext{b}{Red giants analyzed by \citet{redgiants}, reproduced here for comparison.}
\end{deluxetable*}
\end{turnpage}
\subsection{Fitting Isochrone Models to Spectroscopic Properties}
\label{s:isosme}

\begin{figure*}\begin{center}
\plottwo{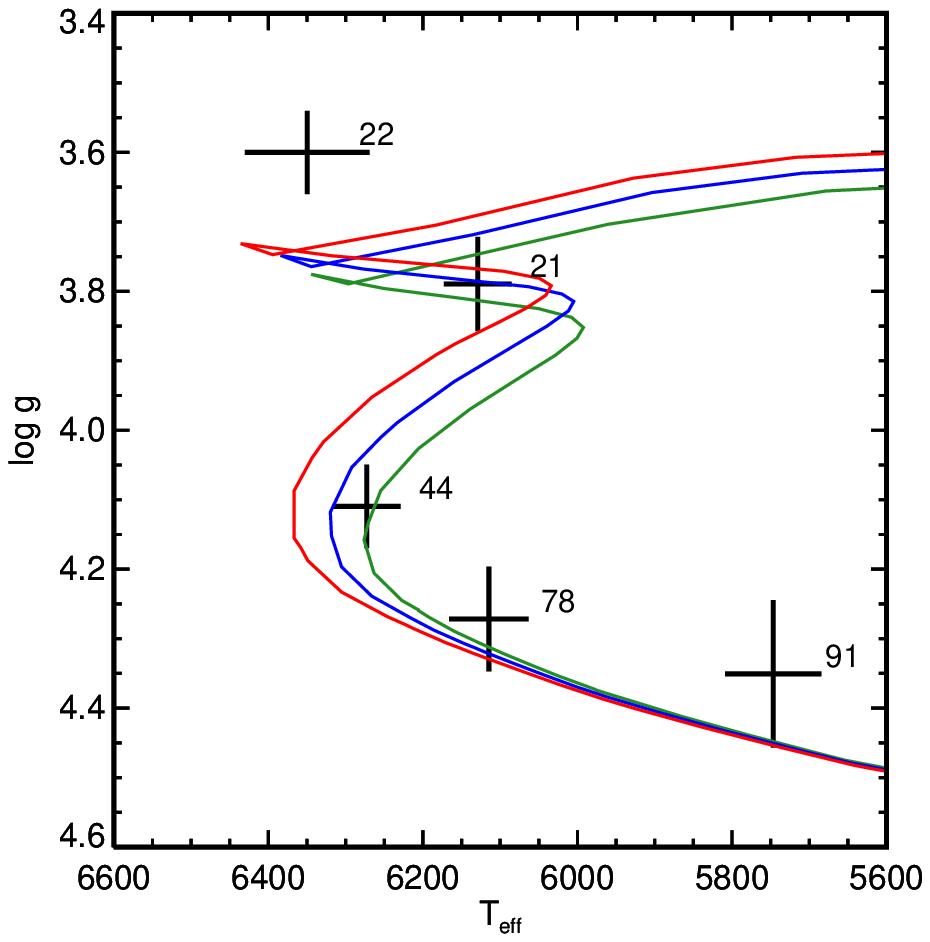}{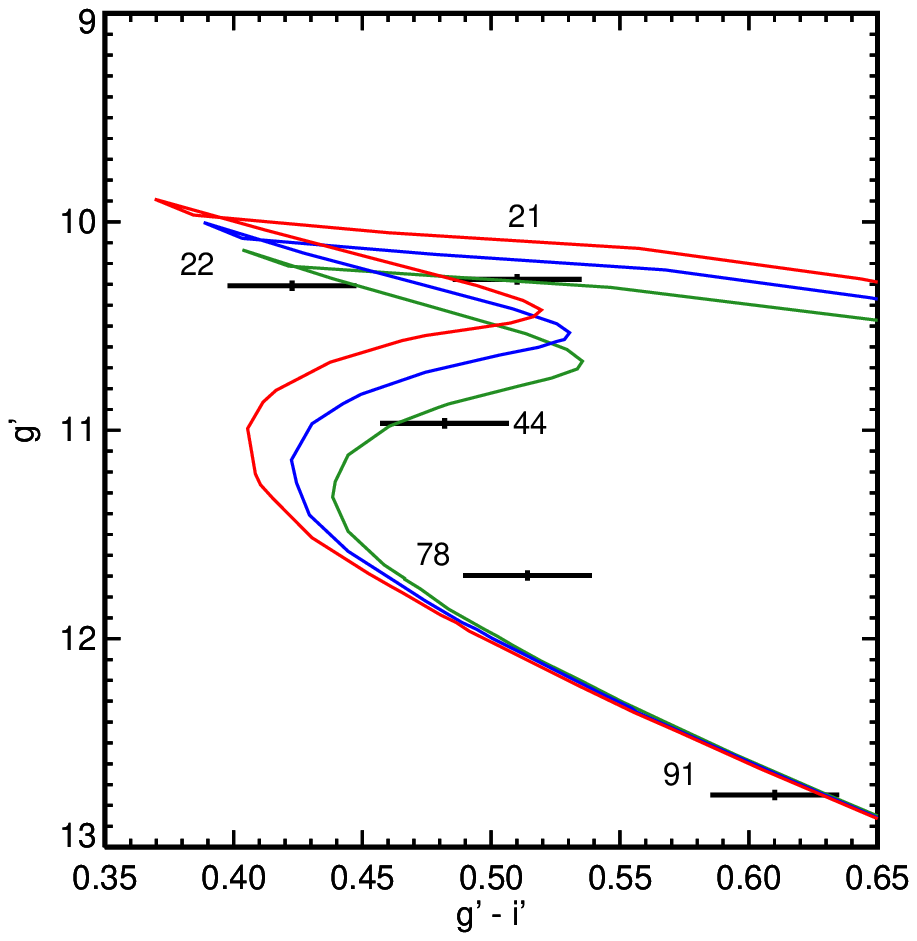}
    \caption{Left: \teff\ -- \logg\ diagram of five stars with Keck/HIRES spectra and
    properties derived with SME. 
    Numbers indicate CWW ID. 
    Padova isochrones overlaid with [M/H] = +0.064 and $\log t$ = 
    9.3 (blue, 2 Gyr), 9.4 (red, 2.51 Gyr), and 9.5 (green, 3.16 Gyr). 
	Isochrones with ages of 2.34 and 2.7 Gyr encompass the error bars 
	of CWW 44. 
	The red isochrone at 2.51 Gyr shows the best fit to these spectroscopic properties;  
	we also derive this age solution in our isochrone fits to NIR and optical photometric 
	color-magnitude diagrams.
	Right: CFHT/MegaCam ($g' - i'$) CMD for the same five stars, with 0.03 mag. error bars. 
	Padova isochrones of same age and color scheme are overlaid with $m - M = 7.35, A_V = 0.25$. 
	For a discussion, see \S \ref{s:isosme}.
	\label{f:tefflogg}}
\end{center}\end{figure*}

Figure \ref{f:tefflogg} shows the \teff\ -- \logg\ diagram resulting from our SME analysis, 
along with a ($g' - i'$) CMD for the 5 stars, along with their CWW IDs. 
The CMD shows that our SME results place the stars on the \teff\ -- \logg\ diagram 
with the correct relative positions. 
We have selected Padova isochrone models with [$\alpha$/Fe] = 0 and [M/H] = +0.1, 
and attempt a ``chi-by-eye" fit by overlaying models with ages at $\sim$2, 2.5, and 3 Gyr 
($\log t = 9.3, 9.4, 9.5$).
Fitting the \teff\ -- \logg\ diagram is powerful because it is 
independent of $m - M, A_V$, and color -- temperature transformations. 

The MSTO star CWW 44 in theory provides a tight constraint on age and metallicity, 
but we are cautious of the accuracy of \logg, 
because the broad wings of the Mg b lines provide the gravity constraint, 
and their sensitivity decreases at higher temperature and lower gravity.
Jeff Valenti has suggested that the Mg b wings provide useful constraints on 
gravity for dwarfs coller than $\sim$6200 K\footnote{http://www-int.stsci.edu/$\sim$valenti/sme.html}, 
which is approximately the temperature for 4 of 5 stars we analyze. 
\citet{Fuhrmann1997} is able to derive an accurate \logg\ for Procyon (F5V) from 
Mg b, so perhaps our concern is unwarranted. 
Assuming we have derived accurate stellar properties, 
we find that models fit best with $\log t = 9.4 \pm 0.05$ (2.25 -- 2.8 Gyr), 
which encompass the error bars of CWW 44. 

The models barely pass through the error bars for the early G dwarf, CWW 91, 
even though this should be the one star of the five we know has broad enough Mg b wings to 
provide adequate constraint on \logg, since it is most similar to the Sun. 
If we fix \logg\ according to the Padova isochrone ($\log t = 9.4$ (2.5 Gyr) and [M/H] = +0.064), 
we find \logg\ = 4.45 instead of 4.35, at mass $M = 1.03$ \msun. 

\begin{figure*}\begin{center}
\plottwo{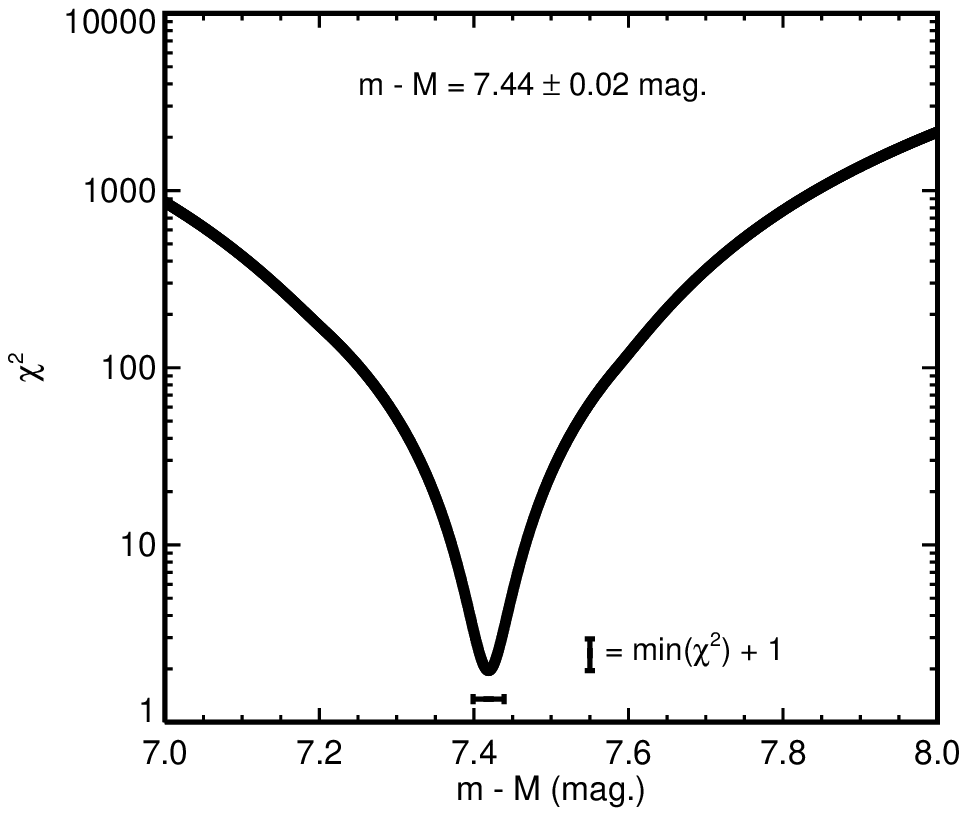}{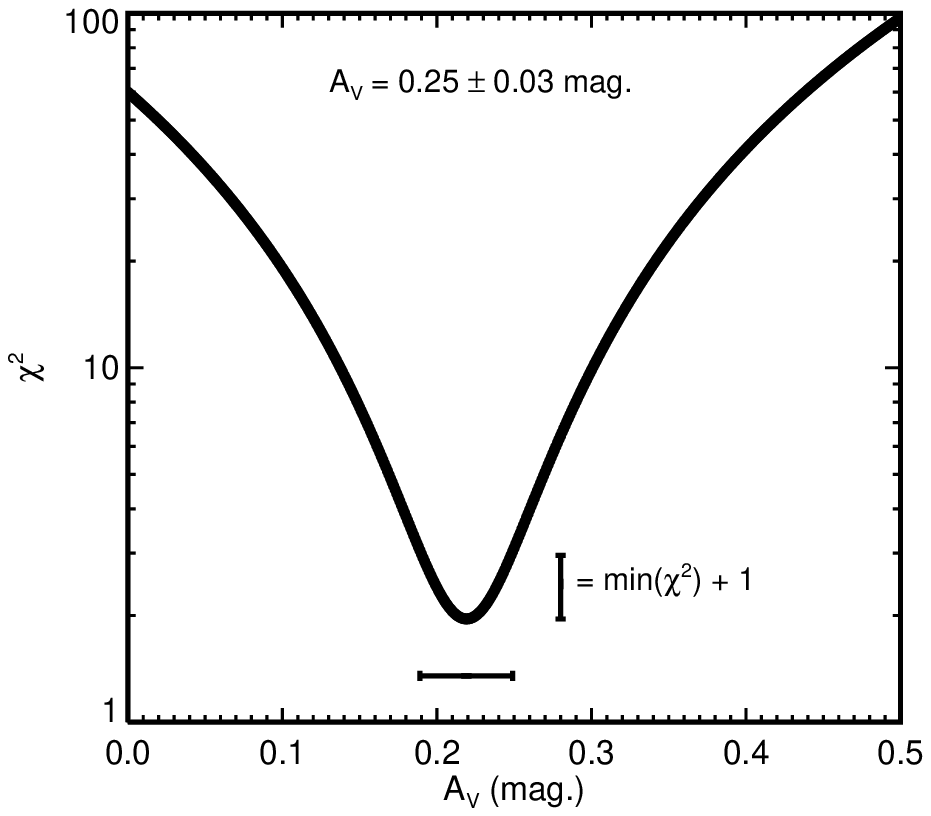}
    \caption{Inferring distance and extinction: 
    brute force least squares fit of $g'r'i'JHK_S$ photometry for CWW 91 
    to a 1.03 \msun\ star with SED drawn from a Padova isochrone with $\log t = 9.4$ and 
    [M/H] = +0.064 (mass, age and metallicity suggested by SME analysis of a Keck/HIRES spectrum). 
    We calculate $\chi^2 = \Sigma \left(x_i - \mu_i \right)^2 / \sigma_i^2$, 
    where $x_i$ is the measurement, $\mu_i$ is the model value, and 
    assuming $\sigma_{g'r'i'} = 0.03$ mag. and $\sigma_{JHK_S} = 0.025$ mag.
    $\chi^2$ is minimized at $m - M = 7.44 \pm 0.02$ and $A_V = 0.25 \pm 0.03$.
	See \S \ref{s:isosme} for a discussion.
    \label{f:chi}}
\end{center}\end{figure*}

We queried the $g'r'i'$ and $JHK_S$ photometry for a 1.03 \msun\ star 
from this isochrone and perform a 
brute force least-squares fit for distance modulus and visual extinction for CWW 91 
in the range $m - M = 7 - 8$ and $A_V = 0 - 0.5$, with 0.01 mag. step sizes, 
$JHK_S$ errors according to 2MASS ($\sigma$ = 0.023, 0.026, 0.021 mag.) and 
$g'r'i'$ errors set to $\sigma$ = 0.03 mag.
We find a minimum $\chi^2$ at $m - M = 7.44 \pm 0.02, A_V = 0.25 \pm 0.03$. 
(see Figures \ref{f:tefflogg}, \ref{f:chi}).
We perturbed [M/H] $\pm 0.02$ dex (our error bar from the SME analysis) 
and $\log t \pm 0.5$ (our error bar from fitting isochrones to the \teff\ -- \logg\ diagram), 
then re-fit and find uncertainties of 0.04 and 0.01 mag. for $m - M$ and $A_V$.
We then perturbed \teff\ by $\pm 50$K (the SME statistcal error bar) and re-fit, 
and find uncertainties of 0.06 and 0.04 mag. for each parameter. 

Adopting these conservative errors, we find 
$m - M = 7.44 \pm 0.06$ and $A_V = 0.25 \pm 0.04$.
We repeated this analysis with solely the optical $g'r'i'$ photometry, 
and then with just the 2MASS NIR $JHK_S$ photometry, 
and find $m - M$ = 7.44 and 7.41 respectively, and $A_V = 0.18$ for both cases.
This give us confidence in the $g' > 10$ CFHT photometry.

We introduced reddening to the synthetic SED using the relationships provided by 
the Padova CMD website\footnote{for a G2 dwarf, 
using a \citet{Cardelli1989} extinction curve with $R_V = 3.1$}: 
$A_{g'}/A_V = 1.167$, $A_{r'}/A_V = 0.860$, $A_{i'}/A_V = 0.656$, 
$A_{J}/A_V = 0.290$, $A_{H}/A_V = 0.183$ , $A_{K_S}/A_V = 0.118$.

\subsection{Fitting isochrone models to broadband photometry}
\label{s:isofit}
\label{s:myisofit}
Up to this point, we have fit isochrones to a \teff\ -- \logg\ diagram 
with values derived with SME for 5 stars with Keck/HIRES spectra.  
Now we perform a more traditional fit to the broadband optical and NIR 
photometry of all cluster members, and find results 
consistent with our spectroscopic solution. 
We initially worked with Padova models 
\citep{padova1, padova2}\footnote{http://stev.oapd.inaf.it/cgi-bin/cmd} 
because they were the only group, 
to our knowledge, that provided isochrones in the CFHT/MegaCam \griz\ filter set. 
Aaron Dotter has since released Dartmouth models in this set, 
and we will briefly compare results bewteen these two models in Section \ref{s:dart}.

Given the high degree of degeneracy between age, distance and visual extinction 
(we fix metallicity according to our SME result, [M/H] = +0.064, [Fe/H] = +0.1), 
we developed an automated isochrone fitting technique that 
can efficiently cover a large parameter space.
Inspired by the $\tau^2$ method of \citet{naylor2006}, 
described and utilized in Section \ref{s:tau2}, 
our method simulates a star cluster with a particular age and composition, 
and computes 2D cross-correlations between the resulting synthetic and 
actual CMD density distributions in order to find the distance and visual extinction 
that best aligns the model to the data.
The age and composition control the morphology of the stellar locus on a CMD, 
while changes in distance modulus and extinction simply translate the locus 
across the CMD plane.

Using our cluster simuator (\S \ref{s:cmd}), we map the stellar locus with $10^4$ stars,
the binary fraction set at 70\%, and no differential extinction, 
for ages ranging from 1 to 4 Gyr. 
The simulated cluster CMD is binned by 0.005 magnitudes in each dimension, 
as is the actual R147 CMD.
Photometric error is not introduced to the synthetic photometry. 
Instead, the position for each star on the real CMD is broadened by a 2D gaussian 
according to the assumed photometric error for each band:  
we use 0.02 mag. for $g'$ and $i'$, and the errors provided by the 2MASS Point Source Catalog for $J$ and $K_S$.

For the NIR ($J - K_S$) fit, all `Y' members were binned except for the blue stragglers, 
and the giants that appear to be undergoing core helium fusion according to their optical 
and NIR photometry offset from the main red giant branch.
\citet{dartmouth} note that the Padova isochrones are hotter in the lower main sequence, 
starting at $\approx$ 0.8 \msun, 
than other commonly used models including the DSEP (Dartmouth Stellar Evolution Program) 
and Yale-Yonsei (Y$^2$). 
The left and central panels of Figure \ref{f:PadDart} illustrates this 
difference in \teff\ -- \logg\ and ($g' - i'$), 
but the right panel shows that the discrepancy does not significantly affect the NIR isochrone.
We do remove the K dwarfs from our optical isochrone fits. 
We also discard the optical red giant branch because of the saturated $i'$ band photometry.

We compute the 2D cross-correlation between the real and synthetic CMD distributions 
with the IDL function \texttt{CONVOLVE}. 
Locations on the resulting image corresponding to negative extinction are set to zero.
The point of peak signal provides the shift 
required to best align the isochrone model to the data. 

\begin{figure*}\begin{center}
\plottwo{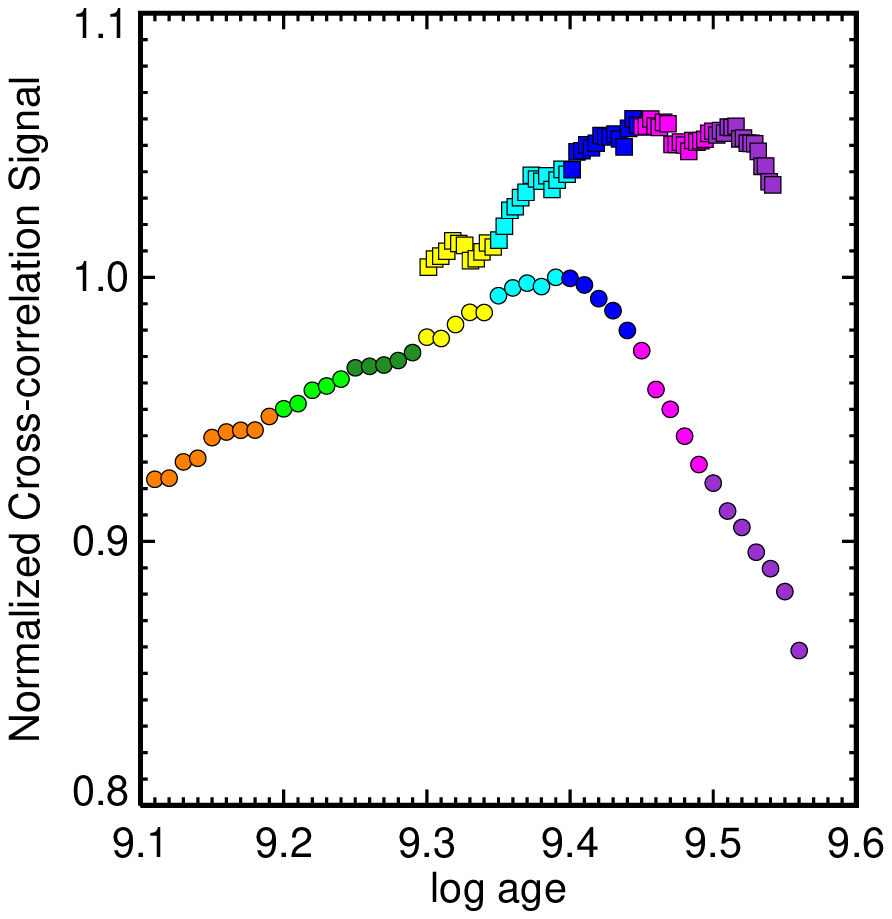}{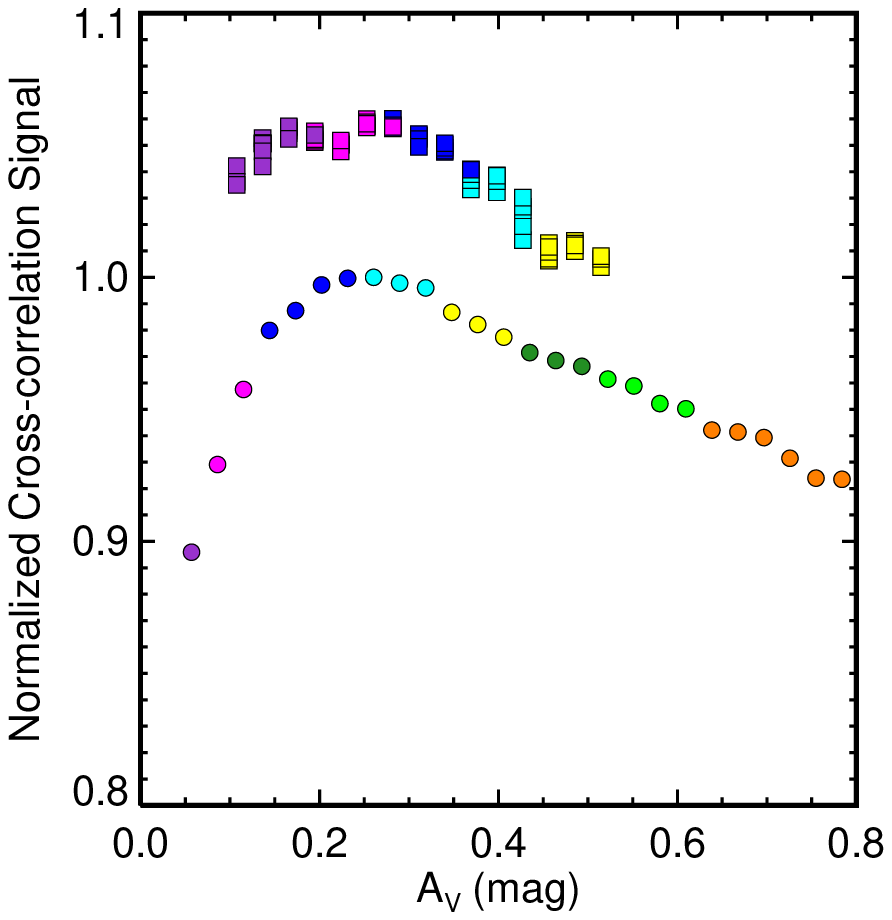}
\plottwo{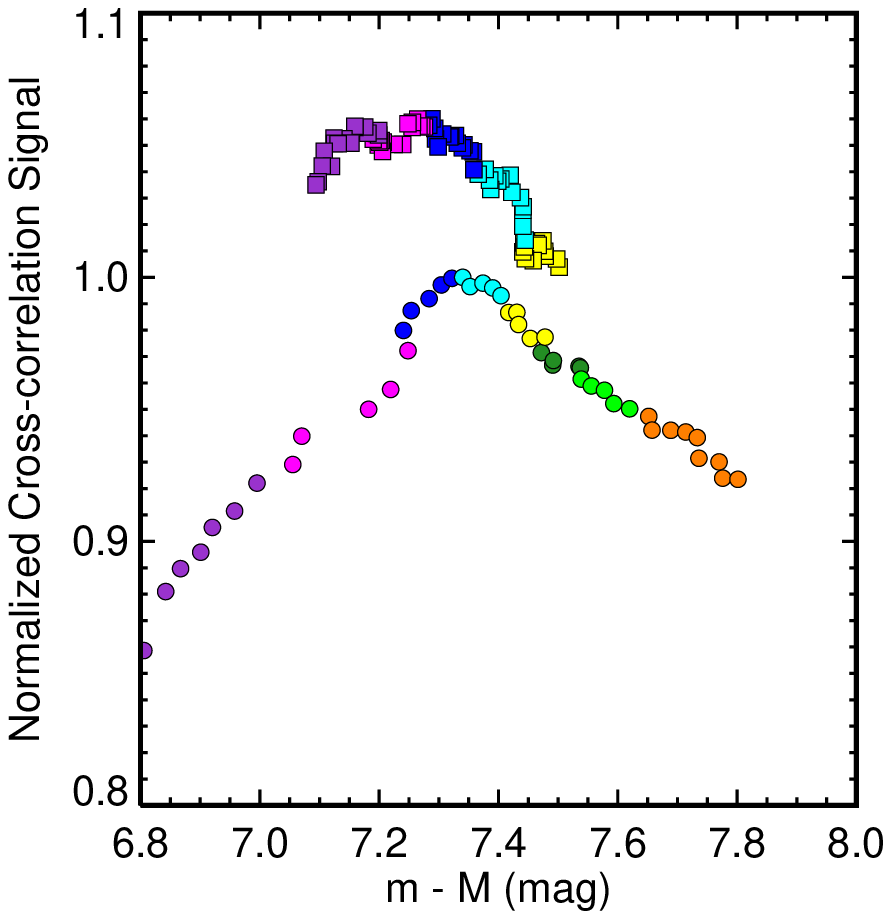}{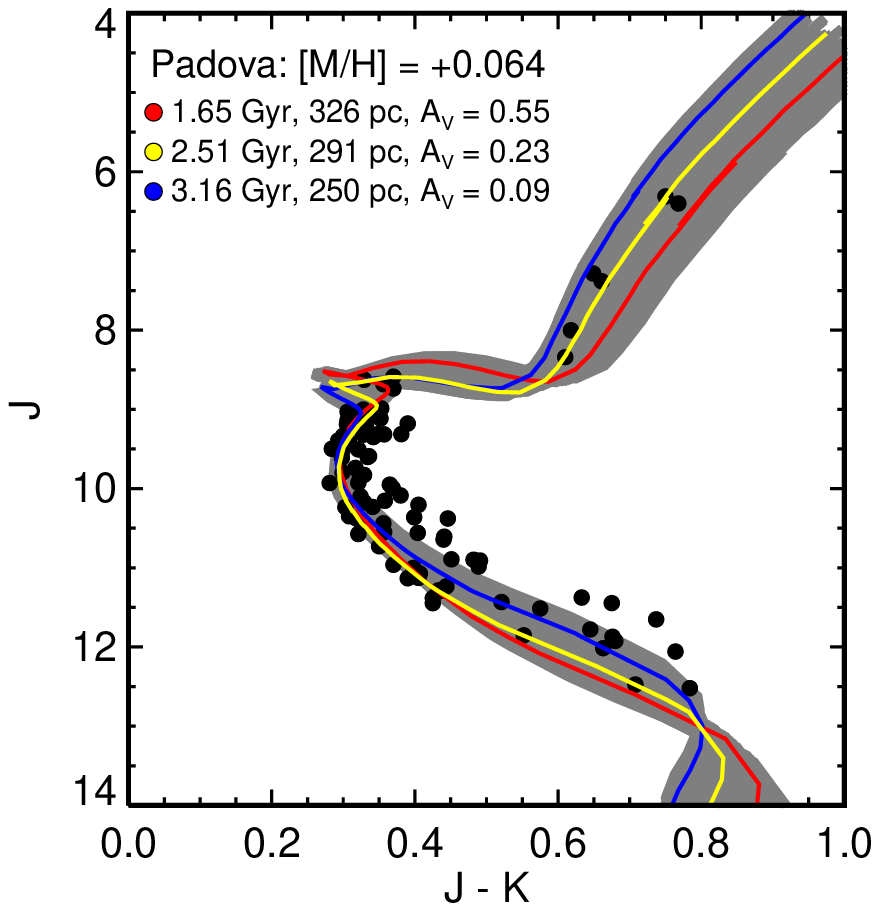}
    \caption{Isochrone fitting to NIR CMD with our 2D cross-correlation method:  
	We tested Padova models (circles) with $\log t = 9.1$ to 9.6 (0.01 step size, $\approx$1.25 to 4 Gyr) 
	and [M/H] = +0.064, and Dartmouth models (squares) with [Fe/H] = +0.1 with 
	ages running from 2 to 3.5 Gyr.
	Our technique simulates stars clusters with $10^4$ stars, 
	including binaries, 
	and computes the $m - M$ and $A_V$ that best matches the R147 photometry. 
	The model with maximum signal in each age / visual extinction / distance modulus 
	bin (0.01, 0.05 mag., 0.025 mag.) is plotted in each panel, 
	color coded according to age. 
	We find that the model with the maximum cross-correlation signal has 
	an age of 2.45 Gyr, $m - M$ = 7.34 and $A_V = 0.26$. 
	\textbf{Bottom -- Right: } All solutions are plotted in gray
	along with R147 photometry as black points.
	Three models are highlighted: 
	the best fit model (yellow), along with an older and younger model 
	for reference.
	See \S \ref{s:myisofit} for more discussion.
	\label{f:myisojk}}
\end{center}\end{figure*}

Although this is not a statistically rigorous method for isochrone fitting, 
this cross-correlation method is conceptually straightforward, simple to code, 
it can efficiently test hundreds of models in an automated fashion to more 
quickly cover the age -- composition parameter space (a few minutes), 
and it provides a diagnostic for model selection:  the model with 
the maximum cross-correlation signal. 
In the next section, we will demonstrate that this method provides 
solutions essentially identical to the $\tau^2$ maximum likelihood method.
Three panels in Figure \ref{f:myisojk} plot age, distance and extinction versus the 
cross-correlation signal, normalized to the maximum value, 
with metallicity fixed at [M/H] = +0.064. 
The results from fitting Dartmouth models are also plotted, and 
offset by +0.06 for clarity: see \S \ref{s:dart}.
Although broad, there are clear peaks in each diagram
at $\log t = 9.4$ (2.51 Gyr), $m - M$ = 7.32 (291 pc), and $A_V = 0.23$ 
or $E(B - V) = 0.075$ assuming $R_V = 3.1$.

The fourth panel of Figure \ref{f:myisojk} plots the isochrones (gray) 
of each solution for $\log t = $ 9.1 to 9.56, in steps of 0.01. 
The R147 members used in the fit are plotted in black.
The best model quoted above is overlaid in yellow, 
and two additional models are also included at younger and older ages, 
providing points of reference for how the fits appear to 
rotate across the CMD counter clockwise with increasing age.  
This rotation is primarily caused by the overabundance of 
cluster members at the MSTO, which serves to anchor 
each fit to the MSTO location. 
Those models which also pass through the handful of K dwarfs and red giant branch 
are rewarded with a higher cross-correlation signal, 
creating the broad peaks in the other panels of Figure \ref{f:myisojk}. 
Perhaps if the K dwarfs and red giants were weighted more heavily than 
the turnoff stars we would see more strongly peaked results.

\begin{figure*}\begin{center}
\plottwo{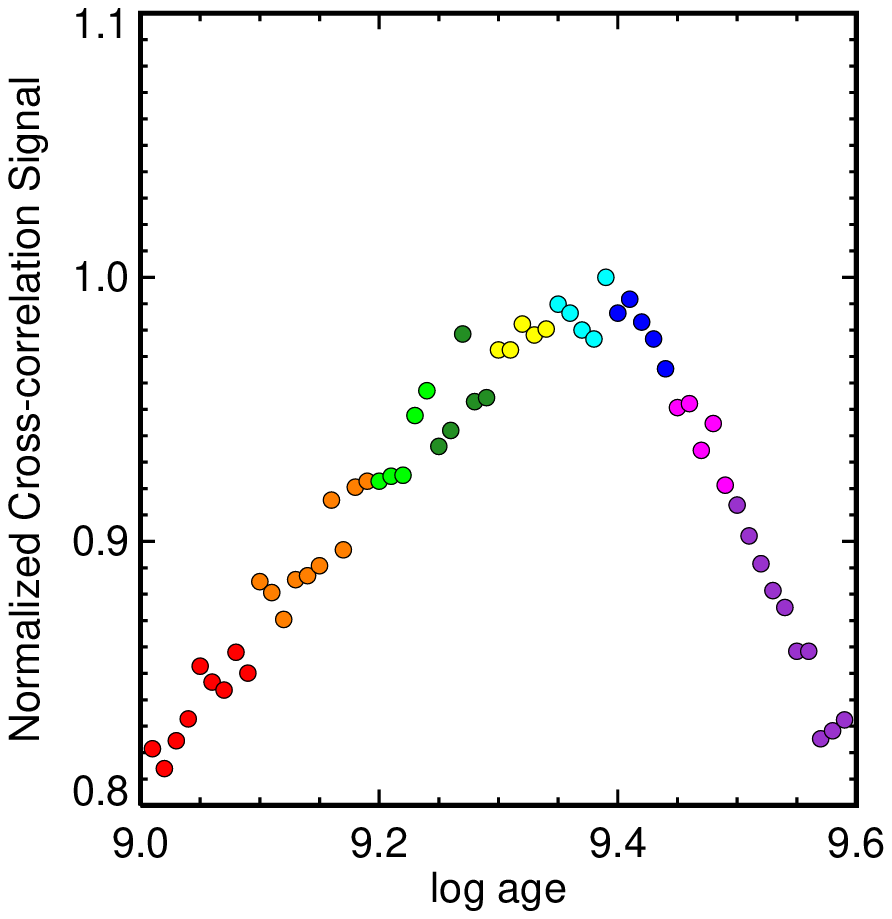}{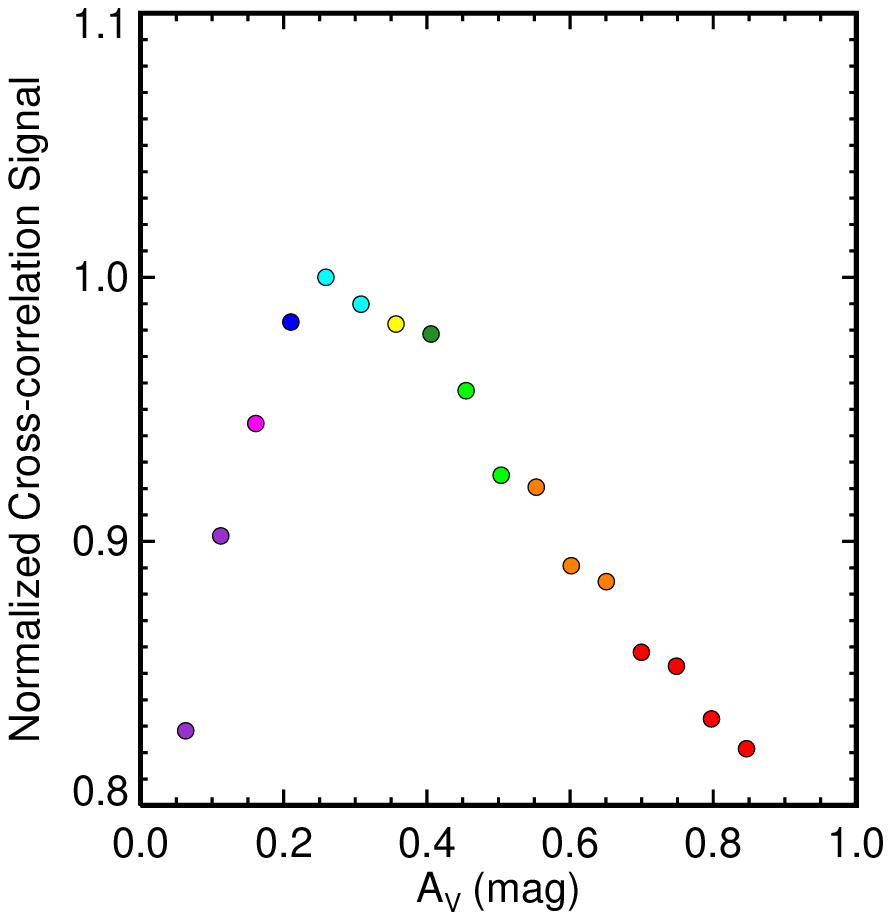}
\plottwo{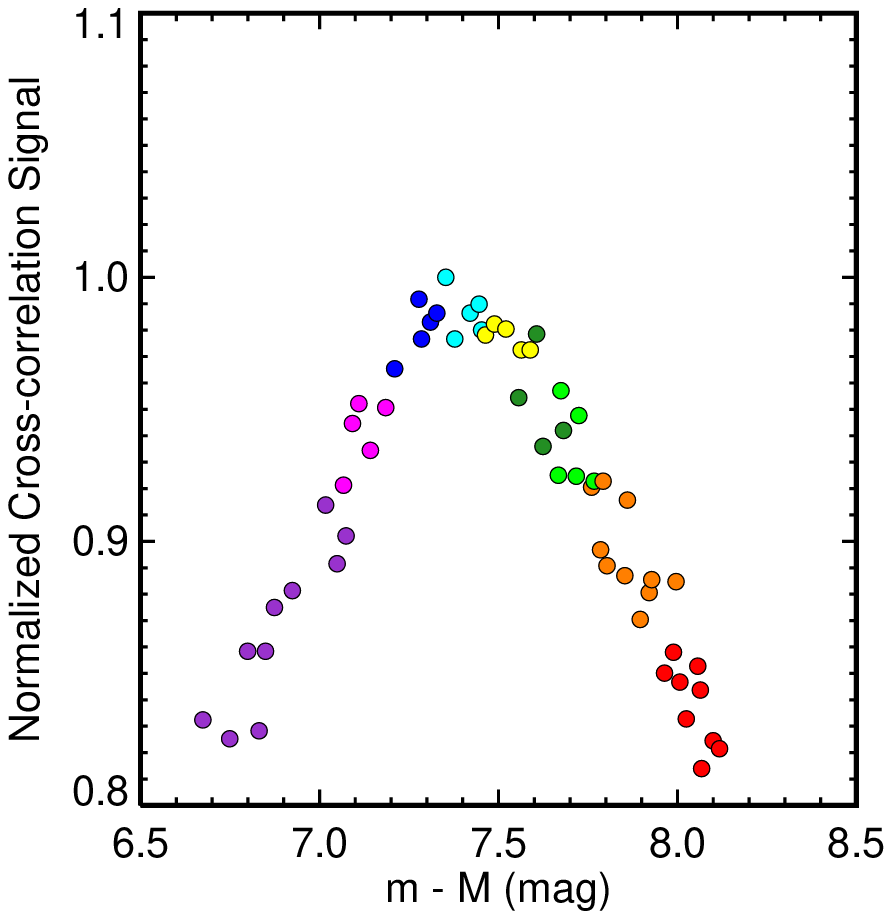}{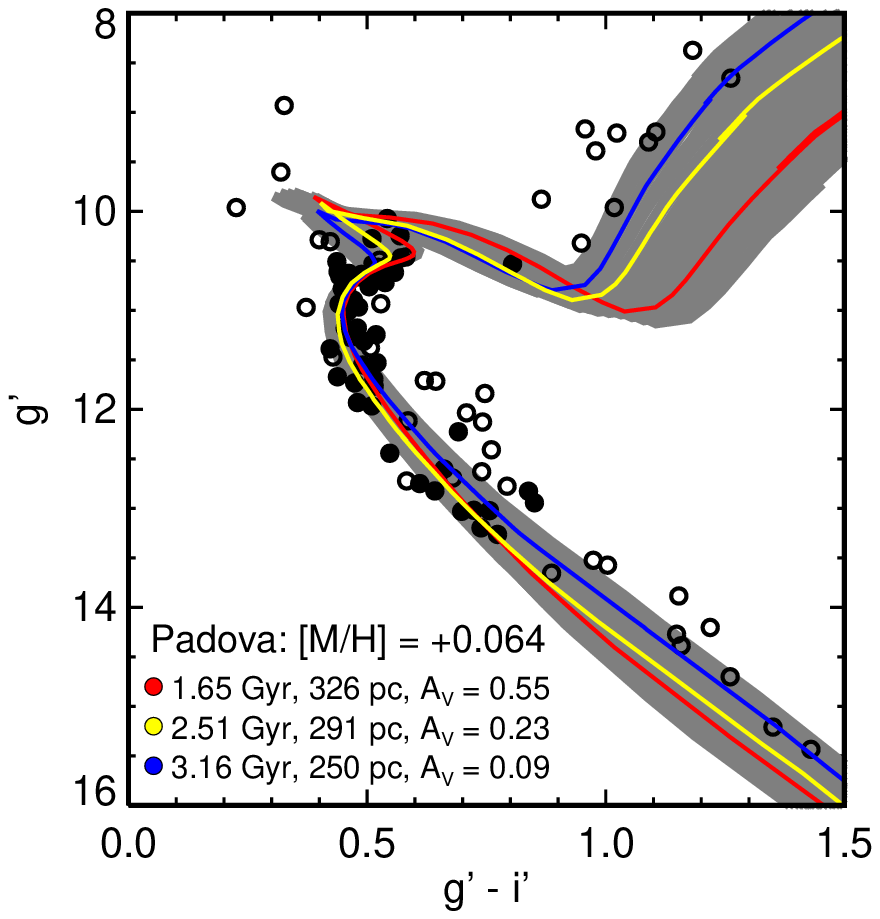}
    \caption{Similar to Figure \ref{f:myisojk}, except we fit 
    the optical CFHT/MegaCam ($g' - i'$) color -- magnitude diagram, 
    with the saturated red giants discarded (open circles). 
    The lower K dwarfs were also removed (also open circles), because the Padova 
    models are known to run hotter than the main sequence at $\sim$0.8 \msun, 
    as illustrated in Figure \ref{f:PadDart}. 
    We find that the model with the maximum cross-correlation signal has 
	an age of 2.45 Gyr, $m - M$ = 7.35 and $A_V = 0.26$. 
	The isochrones plotted in the bottom -- right panel have the following 
	properties:  
	Red -- $\log t$ = 9.4, [M/H] = +0.08, $m - M = 7.35$ and $A_V = 0.25$.
	Blue -- $\log t$ = 9.53, [M/H] = +0.08, $m - M = 6.92$ and $A_V = 0.11$.
	See \S \ref{s:myisofit} for more discussion.
	The fact that we derive essentially identical properties from our spectroscopic 
	\teff\ - \logg\ diagram, and our NIR and optical CMDs validates the 
	accuracy of the CFHT Elixir and TERAPIX photometric reduction 
	(\S \ref{s:griz}).
	\label{f:myisogi}}
\end{center}\end{figure*}

We ran the optical fit with [M/H] = +0.08 and find the 
best model is $\log t = 9.39$, $m - M$ = 7.35, $A_V = 0.26$. 
The optical results are presented in Figure \ref{f:myisogi}.
The consistency between the optical and NIR CMD fits and the parameters 
obtained with the \teff\ -- \logg\ diagram 
further justifies our use of these prelimnary optical data.

\subsection{Validating our method with $\tau^2$}
\label{s:tau2}
\citet{naylor2006} have developed a maximum-likelihood method 
called $\tau^2$ for fitting model isochrones to color -- magnitude diagrams 
\citep[see also][]{naylor2009}\footnote{code available at http://www.astro.ex.ac.uk/people/timn/tau-squared/}. 
This method simulates a cluster CMD from an isochrone, with a user-defined 
binary fraction. The user supplies a star catalog including the color, magnitude, and 
photometric errors which the $\tau^2$ code assumes are normally distributed. 
This method is powerful because it naturally accomodates errors in 
both color and magnitude, and accounts for the binary sequence. 
The code performs a grid search across a specified range of distances and ages,
for isochrones of a given metallicity and reddening, 
and identifies best values, confidence intervals, and returns two diagnostics for 
assessing how well the model describes the data: a
reduced-$\tau^2$ (analogous to $\chi^2_\nu$) and a probability value, $Pr$. 

While the $\tau^2$ code includes an isochrone library, we make use of the 
user-supplied isochrone feature and pass it the Padova grids, 
with [M/H] = +0.064.
The $\tau^2$ code does not currently solve for $A_V$, so we de-redden our catalog before 
running $\tau^2$, and iterate for a range of reddening values.
We selected the 56 R147 members of highest confidence (`Y'), excluding the later K dwarfs and red giants. 
We ran $\tau^2$ for $A_V = 0.0 $to 0.5, and 
allowed $\tau^2$ to search distances ranging from 200 to 400 pc 
(range suggested by the HIP1 ($\sim$270 pc) and HIP2 ($\sim$200 pc) parallaxes, plus 
a little extra on the far side),
and ages 1 to 4 Gyr (step size is 0.01 in $\log t$, encompassing the 2.5 Gyr value suggested by 
our fit to the \teff\ -- \logg\ diagram in Figure \ref{f:tefflogg}).

\begin{figure}\begin{center}
\plotone{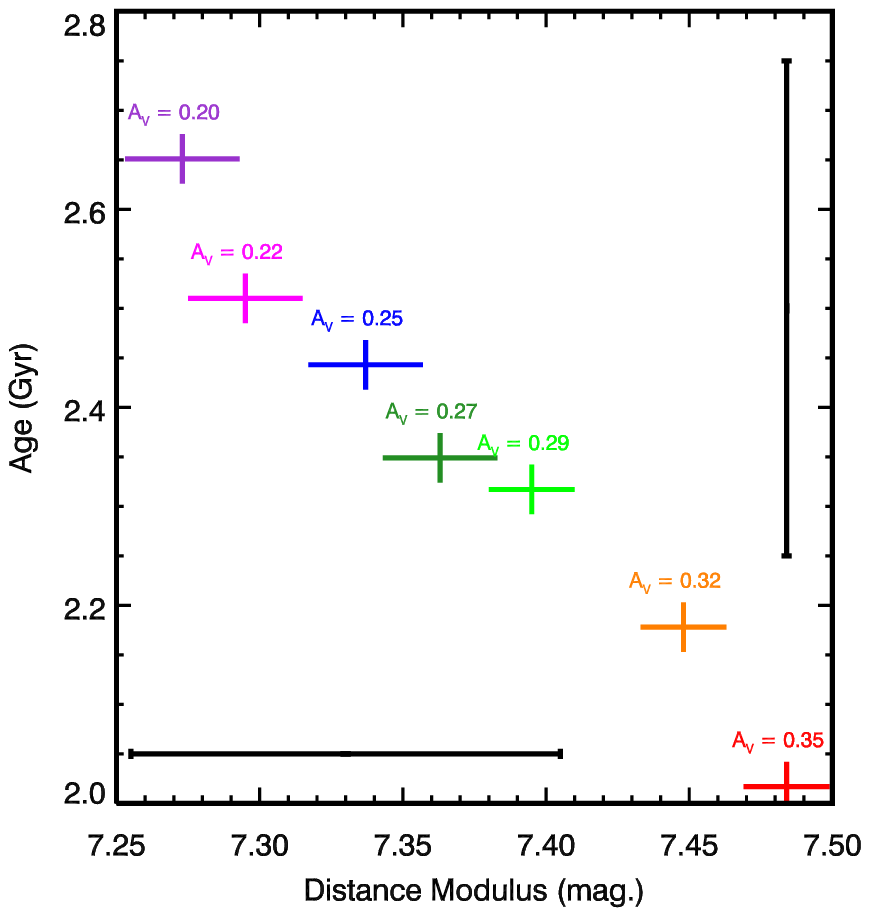}
    \caption{This plot illustrates the degeneracy between extinction, 
      age and distance in isochrone fits to broadband photometry. 
	  Results are plotted from fitting 7 different $A_V$ values with $\tau^2$, 
	  with [M/H] fixed at +0.064 according to our SME analysis. 
	  All fits returned high $\tau^2$ probabilities, 
	  and so model selection is only possible when we place additional 
	  constraints from our spectroscopic analysis. 
	  In Figure \ref{f:tefflogg}, we showed that a 5 star 
	  \teff\ -- \logg\ diagram was best fit by a Padova isochrone 
	  with age = 2.5 $\pm$ 0.25 Gyr, illustrated by the error bar 
	  on the right side of this figure. 
	  This corresponds to a distance modulus of $m - M = 7.35 + 0.05 - 0.1$,
	  shown by the error bar at the bottom of the figure, 
	  and $A_V = 0.25 + 0.08 - 0.05$.
	  These properties are identical to the values we derived when fitting both 
	  the optical and NIR CMDs with our 2D cross-correlation techinque, 
	  validating both our fitting method and the optical photometric reduction.
	  See \S \ref{s:tau2} for more discussion.
      \label{f:degen}}
\end{center}\end{figure}

We find high probabilities for a large suite of models 
demonstrating the flatness of the $\tau^2$ space and high degree of 
degeneracy between age, metallicity, distance and extinction. 
The degeneracy is accentuated in this particular case because we 
had to remove the saturated red giant branch from the fit.
Figure \ref{f:degen} plots the best distance and age values for the range of extinctions.
If we apply the age constraint from the \teff\ -- \logg\ diagram fit 
(2.25 to 2.7 Gyr, illustrated by the black error bar on the right side of Figure \ref{f:degen}), 
then this restricts distance and extinction to $m - M = 7.25 - 7.4$ and $A_V = 0.2 - 0.3$. 
At 2.5 Gyr, $A_V = 0.25$ and $m - M = 7.35$, the $\tau^2$ maximum-likelihood method 
derives values equal to those we determined with our 2D cross-correlation method.

\subsection{Comparision between Padova, PARSEC and Dartmouth isochrone models}
\label{s:dart}
Many groups are developing stellar evolution models. 
While there is remarkable agreement between these models, 
there are noticeable departures especially around the turnoff and red giant branch 
due to choice of input physics and Solar composition \citep{dartmouth}.
In fact, the model choice is the greatest source of uncertainty when 
determining the fundamental properties of star clusters 
via isochrone fitting.

When we began our analysis, the Padova group provided the only 
models with synthetic MegaCam photometry. 
Since then, Dartmouth models have become available. 
Most recently, Padova has issued updated isochrones, 
which are now referred to as PARSEC models 
\citep{parsec}:  
they have revised the major input physics, 
lowered the Solar metallicity from $Z = 0.019$ to $Z \approxeq 0.015$, 
and now include the pre-main sequence (irrelevant for this work).

\begin{figure*}\begin{center}
\plotone{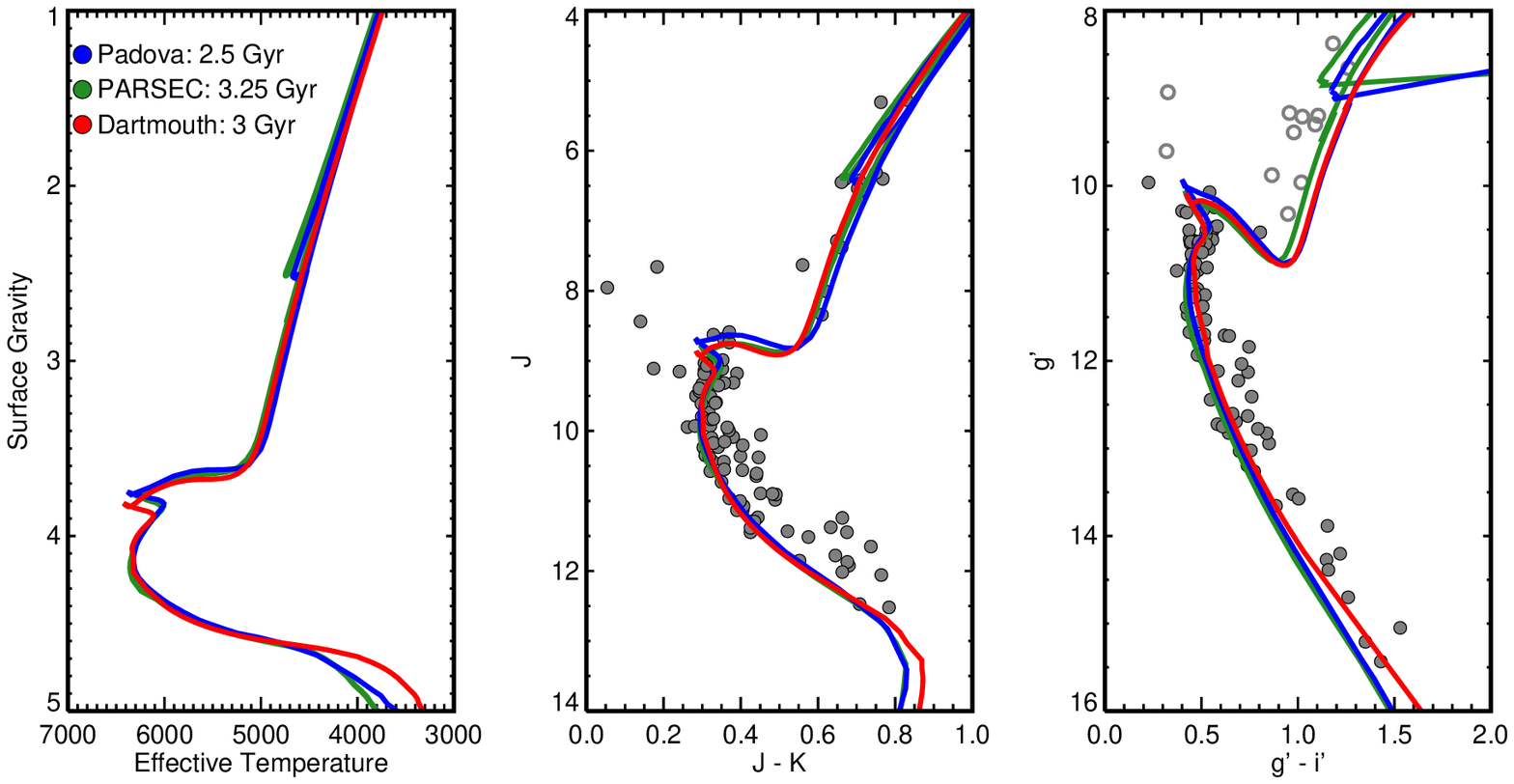}
    \caption{A Padova isochrone (blue, 2.5 Gyr, [M/H] = +0.064) is 
    compared to the newly released PARSEC model(green, 3.25 Gyr) 
    and a Dartmouth model (red, 3 Gyr), with metallicity set at [M/H] $\approx +0.06$.
    The Dartmouth age parameter must be increased by 500 Myr to match 
    the Padova main sequence turnoff. 
	The Padova lower main sequences are known to run hotter than many other 
	stellar evolution models, diverging at approximately 0.8 \msun.
	While this causes the Padova optical main sequence to undershoot the K dwarfs, 
	the NIR sequence well matches the Dartmouth model until 0.7 \msun. 
	No stars below this mass were surveyed in this initial study, 
	so the Padova models can be used to fit the entire NIR main sequence.    
	\label{f:PadDart}}
\end{center}\end{figure*}

The Padova (now PARSEC) webtool parameterizes composition with $Z$, 
where [M/H] $= \log (Z / Z\astrosun)$. 
We used $Z = 0.022$ and 0.017 to query Padova and PARSEC models at [M/H] $\approx$ +0.065.
The Dartmouth webtool\footnote{http://stellar.dartmouth.edu/models/}
uses [Fe/H] instead of $Z$, 
and we set [Fe/H] = +0.1, according to the SME result for CWW 91.
The left panel of Figure \ref{f:PadDart} shows that isochrones from 
Padova at 2.51 Gyr, PARSEC at 3.25 Gyr and Dartmouth at 3 Gyr 
map out the same sequence on the \teff\ -- \logg\ diagram, 
except for the low mass departure at $\sim$0.8 \msun\ already noted.
The middle and right panels plot the optical and NIR isochrones, 
with $A_V = 0.25$ and $m - M = 7.35$. 
The differences between models introduces an additional age uncertainty of 
at least 750 Myr.
Assuming PARSEC more accurately models stellar evolution than their Padova predecessor, 
the difference between PARSEC and Dartmouth reduces the age 
uncertainty to only $\sim 250$ Myr.
Figure \ref{f:myisojk} displays the results from fitting the 2MASS ($J - K_S$) CMD 
with Dartmouth [Fe/H = +0.1 models, illustrated with the square symbols, and offset 
vertically from the Padova results by +0.06 for clarity.
The peak is shifted to older ages relative to Padova, while the 
distance and extinction remain basically consistent -- a consequence 
of the similarity between the 2.5 Gyr Padova model and 3 Gyr Dartmouth model 
in \teff\ -- \logg\ space, 
and each group's color -- temperature transformations.

We will explore additional models
(starting with BaSTI: Bag of Stellar Tracks and Isochrones, 
available at http://albione.oa-teramo.inaf.it/)
in a more detailed analysis in a future work, 
where we intend to perform a simultaneous 7-band isochrone fit using 
our optical \griz\, 2MASS $JHK_S$, and our UKIRT $JK$ photometry.

\subsection{Distance:  photometric versus astrometric}
\label{s:distance}
\citet{dias2001} identified HIP 94635 (CWW 1) and HIP 94803 (CWW 2) 
as kinematic members of Ruprecht 147, and used their HIP1 parallaxes 
to derive a distance of 270 pc to the cluster. 
HIP1 lists $\pi = 3.57 \pm 1.01$ and $3.75 \pm 1.04$ mas for 
these stars respectively, 
while HIP2 provides larger values of 
$5.48 \pm 0.65$ and $4.92 \pm 0.79$. 
Disregarding the advice of \citet{vanLeeuwenbook} against deriving 
distances and distance moduli from parallaxes when the relative error is 
greater than 10\%, 
one would determine a distance from HIP1 of 270 pc, and 193 pc from HIP2.

\begin{figure*}\begin{center}
\plottwo{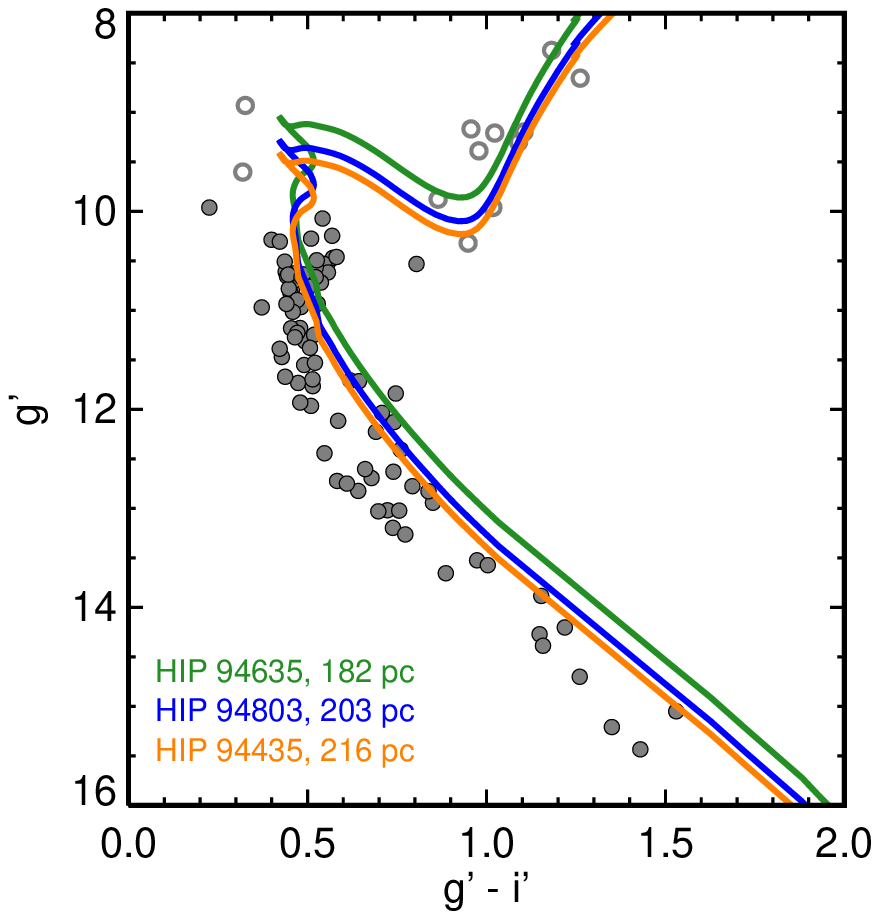}{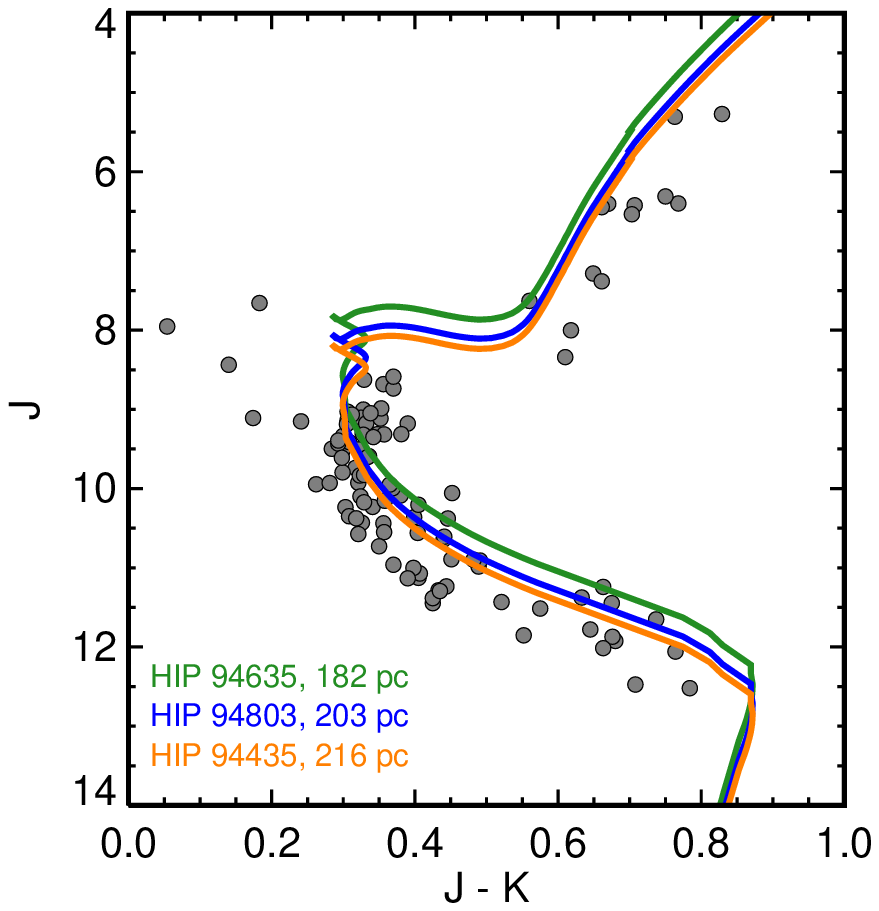}
    \caption{The R147 optical and NIR CMDs are plotted, 
    along with Padova isochrones: 2.5 Gyr, [M/H] = +0.064, $A_V = 0.25$, 
    and distance moduli corresponding to the HIP2 parallaxes for the 3 Hipparcos members: 
    HIP 94635, 94803, and 94435 (CWW 1, 2, and 13).
	While the HIP2 parallaxes are all approximately in agreement, 
	there is simply no way to place an isochrone at the distances implied by these parallaxes. 
	We determine a photometric distance from optical and NIR isochrone fitting 
	that is about 100 pc further than the HIP2 parallax distances, 
	reminiscent of the Pleiades distance problem \citep[][see \S \ref{s:distance}]{FGS:pleiades}.
	\label{f:hip}}
\end{center}\end{figure*}

We also identify HIP 94435 (CWW 13) as a member of R147. 
Although HIP1 lists a discrepant parallax $\pi = 2.42 \pm 1.25$ (413 pc), 
HIP2 gives $4.64 \pm 1.19$ (216 pc) consistent with the other 2 HIP stars. 
While the HIP2 results are self-consistent, 
Figure \ref{f:hip} demonstrates there is simply no way an isochrone can be drawn through either 
the optical ($g' - i'$) or NIR ($J - K_S$) CMDs at the HIP2 distance moduli. 
The HIP2 parallax distances are all too close by $\approx$100 pc, 
compared to the photometric distances. 
This is reminiscent of the Pleiades problem, 
where the HIP1 parallax placed the cluster 0.23 mag. closer than 
distances inferred from main sequence fitting and other methods.
\citet{FGS:pleiades} utilized the HST Fine Guidance Sensor 
to measure a new parallax distance consistent with the other non-HIP results.

\section{Final Synthesis and Sources of Uncertainty}
\label{s:result}
We have fit Padova isochrone models to three separate datasets: 
a \teff\ -- \logg\ diagram consisting of 5 stars with values derived from SME, 
and both NIR ($J - K_S$) and optical ($g' - i'$) CMDs. 
We have chosen to use the Padova isochrones for our preliminary investigation 
into the propeties of R147 because they provide colors in the MegaCam filter set.
Other models (e.g. Dartmouth and Yonsei-Yale) show differences in 
the main sequence turnoff region, which provides the primary age constraint; 
and in the lower main sequence, where Padova runs bluer than 
Dartmouth and Yonsei-Yale. 
Our results, especially for age, therefore depend heavily on our chosen model.

Our solution is subject to additional sources of uncertainty, 
including photometric error, 
unresolved multiple star systems, 
and the possibility of differential extinction. 
Main sequence fitting, both to the single star and equal mass binary sequences, 
provides the primary constraint on the sum of distance and extinction. 
Ideally, these sequences would be vertically offset by 0.75 magnitudes (i.e. double the brightness), 
but for R147, there are separations of 1 magnitude or more. 
This unexpected offset could be explained by differential extinction, 
which would widen the sequence in both directions, 
by a population of triple systems, or if the stars are not actually cluster members. 
In this preliminary investigation, we have not yet untangled 
this $\delta$$A_V$ -- multiplicity -- membership degeneracy on a star by star basis, 
but such an analysis would improve the precision of the cluster parameters.

At $\sim$300 pc and 1\degree.25 in angular radius, we expect a physical radius of $\sim$5 pc, 
which introduces a differential distance modulus of $\delta(m - M) = 0.02$ across the cluster. 
This is comprable to photometric error, 
and should be part of a more comprehensive error analysis. 

While fitting the fitting the \teff\ -- \logg\ diagram, 
assuming [M/H] = +0.064,
we find $\log t = 9.4 \pm 0.03$, or $t = 2.5 \pm 0.02$ Gyr. 
If we increase or decrease the metallicity by $\pm 0.02$ dex, 
the age error bars remain similar and the best value for 
$\log t$ shifts by 0.02. 
When we performed the brute force SED fit to the early G dwarf 
for distance and extinction, we perturbed [M/H] by $\pm$0.02 dex, 
$\log t$ by $\pm$0.5, and \teff\ by $\pm$50K, 
and found $m - M = 7.44 \pm 0.06$ and $A_V = 0.25 \pm 0.04$.

For a given metallicity and extinciton, $\tau^2$ returns 
typical uncertainties of 40 -- 100 Myr in age 
and 0.03 -- 0.05 in distance modulus. 
Although the $\tau^2$ code does calculate two diagnostics useful 
for model selection, the reduced $\tau^2$ and a probability value, 
for the range of parameters we searched, the high degree of degeneracy 
between the four cluster parameters enabled $\tau^2$ to find 
solutions which delivered high probabilities ($>$60\%) 
and reduced $\tau^2$ values all $\approx$1. 
Instead, we will select our preferred parameter set by 
fixing the metallicity according to our spectroscopic SME results: 
[M/H] = +0.065, and the age according to the \teff\ -- \logg\ result 
at 2.5 Gyr. 
This breaks the degeneracy and we can then accept the corresponding 
distance and and visual extinction from $\tau^2$ as best values: 
$m - M$ = 7.35 ($d$ = 295 pc) and $A_V$ = 0.25. 

This is essentially identical to our 2D cross-correlation results, 
where we found peak values at 
an age of 2.45 Gyr, $m - M = 7.34$ and $A_V = 0.26$, 
for the [M/H] = +0.065 case. 

We set our preferred values with generous error bars at 
age = 2.5 $\pm$ 0.25 Gyr ($\log t = 9.4 \pm 0.03$), 
[M/H] = 0.07 $\pm$ 0.03, $m - M$ = 7.25 to 7.45 
($d = $ 280 to 310 pc), 
and $A_V = $ 0.20 to 0.30. 
We set the metallicity error bars according to our SME analysis, 
ignoring the one anomalously low metallicity result, and the 
hottest star with a poor $\chi_\nu^2$ fit. 
We set the age and error bars according to our fit to the 
\teff\ -- \logg\ diagram (Figure \ref{f:tefflogg}), 
which is corroborated by the 2D cross-correlation fit (Figure \ref{f:myisogi}). 
We set the $A_V$ range and lower bound $m - M$ value 
by placing these metallicity and age constraints 
on our $\tau^2$ results (Figure \ref{f:degen}). 
We extend the upper bound on $m - M$ past 7.4 to 7.45 mag. 
to encompass the result of our SED fit to the G0/2 dwarf. 
These results are summarized in Table \ref{t:cluster}, 
with the values of \citet{khar2005} and \citet{redgiants} 
provided for comparison. 

\begin{deluxetable*}{cccccccl}
\tabletypesize{\scriptsize}
\tablecaption{Table of R147 cluster properties.\label{t:cluster}}
\tablewidth{0pt}
\tablehead{
\colhead{$\mu_\alpha$} & \colhead{$\mu_\delta$} &
\colhead{RV\tablenotemark{a}} & \colhead{Age\tablenotemark{b}} & \colhead{Distance} & 
\colhead{$A_V$\tablenotemark{c}} & \colhead{Metallicity} &
\colhead{Reference} \\
\colhead{mas/yr} & \colhead{mas/yr} & \colhead{\kms} & 
\colhead{Gyr} & \colhead{pc} & \colhead{mag.} & & 
}
\startdata
-1.1 & -27.3 & 41.1 & 2.51 $\pm$ 0.25 &  295 $\pm$ 15   & 0.25 $\pm$ 0.05  & [M/H] = +0.07 $\pm$ 0.03  & This work \\ 
-0.6 & -27.7 & 41 & 2.45           &  175            & 0.47         & ...            & \citet{khar2005} \\
... & ... & 40.5 & 1.26 $\pm$ 1.16 &  280 $\pm$ 100  & 0.34         & \fei\ = 0.16, \feii\ = 0.08 & \citet{redgiants} \\
\enddata

\tablecomments{Our proper motions are median values for `Y' and `P' members.}
\tablenotetext{a}{Our RV is the average of our 4 Keck velocities. 
\citeauthor{khar2005}'s velocity is from \citet{wilsonRV}. 
\citeauthor{redgiants}'s velocity is the average of the two apparent single stars. 
See Section \ref{s:rv}.}
\tablenotetext{b}{All three groups determined ages from Padova isochrones. 
Our age is the best fit parameter from a fit 
to a \teff\ - \logg\ diagram (\S \ref{s:isosme}) and 
the NIR and optical CMDs and  (\S \ref{s:myisofit}).
This parameter is heavily model-dependent: using PARSEC and Dartmouth models, 
we instead find 3.2 and 3 Gyr respectively (\S \ref{s:dart}).
\citeauthor{khar2005} fit an optical $(B - V)$ CMD with a Solar composition Padova model 
(Figure \ref{f:khar}, \S \ref{s:litrev}). 
\citeauthor{redgiants} determined composition, \teff, and \logg\ for 3 red giants from 
their high resolution spectroscopy, interpolated Padova models to the metallicities 
for each star, and fit for mass and age using their derived \teff\ and \logg\ values.}
\tablenotetext{c}{$A_V$ assumes a $R_V = 3.1$ reddening law.}
\end{deluxetable*}

\section{Summary, Discussion and Upcoming Work}
\label{s:summary}

Over 170 years passed since Herschel first cataloged Ruprecht 147 
before astronomers finally investigated its properties and membership. 
\citeauthor{dias2001} and \citeauthor{khar2005} demonstrated that a group of 
20 -- 40 stars at the location of R147 were in fact moving together in 
the plane of the sky, 
and estimated this group's properties, although their analysis 
was hindered by ($B - V$) photometry with a limiting magnitude near
the main sequence turnoff. 
While \citet{khar2005} was able to determine an age (2.45 Gyr) from 
the MSTO consistent with the results of our analysis, 
the ($B - V$) main sequence is dominated by photometric error and 
therefore provides a weak constraint on the distance, 
which their isochrone fitting has apparently placed 125 pc too close, 
at 175 pc compared to the 300 pc distance we find here.
Nevertheless, these works by \citeauthor{dias2001} and \citeauthor{khar2005} 
are significant because they essentially re-discovered R147. 
\citet{redgiants} first spectroscopically determined the composition for 3 red giant members, 
showing the cluster to be super-Solar (\S \ref{s:litrev}).

We queried the NOMAD catalog for stars within 1\degree .5 of the cluster center, 
and out of the 750,000 stars, we find 1348 with proper motion within 5 mas/yr of 
the R147 value \citep[astrometric values from ][]{khar2005}.
We conducted an initial radial velocity survey at Lick and Palomar Observatories
and for the first time confirm that over 100 stars are likely members of Ruprecht 147 and 
they are indeed moving together in three dimensions through the Galaxy (\S \ref{s:angrv}, \ref{s:rv}). 
We followed up this initial survey with high resolution and signal-to-noise \caii\ spectra 
with MMT/Hectochelle, and used these second epoch RVs, at higher precision, 
to investigate binarity (\S \ref{s:mmt}, \ref{s:sb2}, \ref{s:sb1}). 
We have imaged the cluster in four optical bands (\S \ref{s:griz}), 
and combined with 2MASS NIR photometry (\S \ref{s:2mass}), 
used the resulting CMDs to establish a membership list with 81 high-confidence members, 
21 possible members, and 6 unlikely members (\S \ref{s:membership}).

We have obtained high-resolution, high-SNR spectra of five members (\S \ref{s:keck}), and 
determine the metallicity to be super-Solar using the SME spectral synthesis code, 
and find [M/H] = +0.07 $\pm$ 0.03 and zero $\alpha$-enhancement (\S \ref{s:sme}).

We have fit Padova isochrones to the \teff\ -- \logg\ diagram resulting from our 
spectroscopic analysis, 
and find that the age of R147 is best fit by a Padova isochrone with age of 2.5 Gyr (\S \ref{s:isosme}).
\teff\ -- \logg\ diagram isochrone fitting is independent of distance and visual extinction, 
which makes it a powerful tool, but this also means it does not directly provide any 
information on these parameters. 
We queried the best fit Padova isochrone for a star with \teff\ and \logg\ closest to 
the values for the early G dwarf (CWW 91) we derived with SME, 
then performed a brute force SED fit to the resulting synthetic $g'r'i'JHK_S$ photometry, 
and find a minimum $\chi^2$ at $m - M$ = 7.44 (308 pc) and $A_V$ = 0.25. 

We consulted the dust map of \citet{dustmap} to set an upper limit on the amount of visual 
extinction toward R147, $A_V < 0.5$ (\S \ref{s:av}), 
then fit Padova isochrones to the ($g' - i'$) and ($J - K_S$) CMDs using a 
2D cross-correlation technique developed here, which was inspired by 
and validated with the \citet{naylor2009} $\tau^2$ maximum-likelihood method (\S \ref{s:isofit}). 
We find that without additional constraints from spectroscopy or additional 
photometric bands, just fitting a single CMD with isochrone models yields 
a suite of solutions all with high $\tau^2$ probabilities, due to the 
high degree of degeneracy between age, composition, distance and visual extinction. 
If we break this degeneracy with the spectroscopic metallicity and age, 
we find $m - M$ = 7.33 and $A_V$ = 0.23 from $\tau^2$.

We found that the best models derived from fitting both ($J - K_S$) and 
($g' - i'$) CMDs with this technique corroborates 
the age we determined from the \teff\ -- \logg\ fit, as well as the 
distance and extinction corresponding to this age in our $\tau^2$ fits. 

We recognize significant uncertainty in our solution from the unresolved binary population 
and possibility of differential extinction across this large cluster.
These results are also heavily model-dependent (\S \ref{s:dart}).

\subsection{Discussion and Future Work} 
The R147 single star main sequence is not well defined, but blends smoothly into what 
is apparently the binary population. 
Evidently, R147 has a large binary fraction. 
The stellar population has encountered approximately 3 Gyr of 
Galactic gravitational tidal forces.
Evaporation of the lightest-mass members should proceed first, 
both because low mass stars are preferentially ejected in 3-body encounters, 
and mass segregation gives them a larger effective radius 
and more susceptibility to Galactic tides.
This process also preferentially ejects single stars from the cluster,  
as multiple star systems of similar spectral type have a greater bound mass. 
The ongoing dynamical evolution of an open cluster thus
tends to increase the binary fraction, and we may be seeing this effect in the R147 main sequence. 

Our membership list is top heavy, dominated by F stars (dwarfs, MSTO and subgiants) and 
red giants, with fewer numbers of G dwarfs, and only a handful of early K dwarfs. 
This dynamical evolution and evaporation could also 
explain the paucity of low mass members, 
and perhaps if any are left, they exist predominantly in multiples. 
But this can also be explained by observational bias: 
the NOMAD and UCAC-3 proper motion errors increase for the fainter members. 
These stars also begin to blend into the Galactic background in the color -- magnitude 
diagrams, further complicating candidate identification. 
We are currently conducting a radial velocity survey for lower mass members, 
but this question will only finally be settled by deriving precise proper motions for the 
faint stars in the R147 field, which we intend to do by re-imaging the cluster 
with MegaCam in the near future. 

If we are able to identify single M dwarfs, these will be the only old ($>$1 Gyr), single 
cool dwarfs with known ages and compositions bright enough to admit close 
spectroscopic study. 
Once the white dwarf population is identified, it will provide an
independent age estimate for the cluster and inform studies of white dwarf cooling curves.
At 300 pc, chromospheric activity diagnostics are measureable, as is $L_{\rm X}$, 
and R147 should prove useful for studying the evolution of angular momentum and 
magnetic activity at intermediate ages.

For these reasons and more, we will continue our efforts to characterize Ruprecht 147 and  
establish it as a new and important benchmark for stellar astrophysics.


\acknowledgments
J.L.C.\ acknowledges support from the National Science Foundation 
Graduate Research Fellowship Program supported by NSF grant No. DGE1255832, 
the Stephen B.\ Brumbach graduate fellowship, 
and the Zaccheus Daniel travel grant program.
A.W.\ acknowledges a Hunter R.\ Rawlings III Cornell Presidential Research Scholarship.

J.T.W.\ conceived of and oversaw the project, 
and collected all of the candidate stellar spectra. 
A.W.\ and J.T.W.\ conducted the radial velocity survey at Palomar and Lick 
and produced the preliminary R147 membership list. 
J.M.B.\ conducted the SME analysis. 
J.A.J.\ was PI on the program that obtained the optical photometry from CFHT/MegaCam, 
and also conducted preliminary SME analysis of Lick and Keck spectra 
to determine the composition of R147 
(results are not reported in this work, though they informed this study). 
J.L.C.\ and J.T.W.\ performed the Hectochelle RV survey. 
J.L.C.\ was primarily responsible for the writing of the manuscript 
and the final analysis and synthesis of all data. 

We would like to thank \gabor\ and Andrew Szentgyorgyi for assisting the Hectochelle data reduction; 
Tim Naylor for assisting us with his $\tau^2$ code; 
Aaron Dotter for providing access to the Dartmouth isochrones with the CFHT/MegaCam filters; 
David Monet and Stephen Levine for providing the NOMAD catalog on HDD; 
James Graham and James Lloyd for supporting this research; 
Matthew Muterspaugh for sharing and swapping Lick 3-m time; 
and Kevin Covey for providing extensive comments on an early draft of this paper.
We would also like to thank all observing staff and telescope operators 
at MMT/FLWO, CFHT, Lick, Palomar, and Keck; 
and Geoff Marcy, Andrew Howard and the California Planet Survey observing team for 
their assistance in acquiring the Keck spectra.
We also appreciate our anonymous referee for offering thorough and constructive 
suggestions that improved this paper.
Finally, we thank Debra Fischer, Jeff Valenti, Adam Kraus, Steve Saar, S\o ren Meibom, 
Andrew West, Kevin Covey, Marcel Ag\"{u}eros, Suzanne Hawley, Ivan King, 
Jay Anderson, Bob Mathieu, Ken Janes, and Eric Mamajek 
for helpful conversations, suggestions, and support. 

This work is based on observations obtained with MegaCam, a joint project of CFHT and CEA/DAPNIA, 
at the Canada-France-Hawaii Telescope (CFHT) which is operated by the 
National Research Council (NRC) of Canada, 
the Institute National des Sciences de l'Univers of 
the Centre National de la Recherche Scientifique of France, 
and the University of Hawaii.
Observing time was granted by the University of Hawaii Institute for Astronomy TAC.
These data were reduced at the TERAPIX data center located at the Institut d'Astrophysique de Paris.

Observations reported here were obtained at the MMT Observatory, 
a joint facility of the Smithsonian Institution and the University of Arizona. 
MMT telescope time was granted by NOAO 
(Project PA-10A-0378, PI J. Wright), 
through the Telescope System Instrumentation Program (TSIP). 
TSIP is funded by NSF. 

This publication makes use of data products from the Two Micron All Sky 
Survey, which is a joint project of the University of Massachusetts and 
the Infrared Processing and Analysis Center/California Institute of 
Technology, funded by NASA and the NSF.

Some of the data presented herein were obtained at the W.M. Keck 
Observatory, which is operated as a scientific partnership among the 
California Institute of Technology, the University of California and 
the National Aeronautics and Space Administration. The Observatory was 
made possible by the generous financial support of the W.M. Keck 
Foundation.  The authors wish to recognize and acknowledge the very 
significant cultural role and reverence that the summit of Mauna Kea 
has always had within the indigenous Hawaiian community.  We are most 
fortunate to have the opportunity to conduct observations from this 
mountain.

This research made use of Montage, 
funded by NASA's Earth Science Technology Office, 
Computation Technologies Project, under Cooperative Agreement 
Number NCC5-626 between NASA and the California Institute of Technology. 
Montage is maintained by the NASA/IPAC Infrared Science Archive.

This research made use of the WEBDA database operated at the Institute for Astronomy of the University of Vienna, 
NASA’s Astrophysics Data System Bibliographic Services, 
and the SIMBAD database and the VizieR catalogue access tool 
operated at CDS, Strasbourg, France. 

Any opinions, findings, and conclusions or recommendations expressed in this material are those of the authors 
and do not necessarily reflect the views of the National Science Foundation or any other 
institute cited above.


{\it Facilities:} \facility{Shane}, \facility{Hale}, \facility{MMT}, \facility{CFHT}, \facility{Keck:I}





\bibliographystyle{apj}

\clearpage
\pagenumbering{gobble}
\LongTables
\begin{landscape}
\begin{deluxetable}{ccccccccccccc}
\tabletypesize{\scriptsize}
\tablecaption{Membership list \label{t:mem}}
\tablewidth{0pt}
\tablehead{
\colhead{CWW ID} & \colhead{2MASS ID} & 
\colhead{$\mu_{\rm RA} (\sigma_\mu)$} & \colhead{$\mu_{\rm Dec} (\sigma_\mu$)} & 
\colhead{$g'$} & \colhead{$g' - i'$} & \colhead{$J$}\tablenotemark{a} & \colhead{$J - K_S$}\tablenotemark{a} & 
\colhead{RV$_{\rm LP}$} & \colhead{RV$_{\rm H}$} & 
\colhead{Mem.}\tablenotemark{b} & \colhead{MemFlag}\tablenotemark{c} & \colhead{Notes}\tablenotemark{d}
}
\startdata
1 & 19152612-1605571 & -1.0 (0.7) & -27.4 (0.6) & 8.37 & 1.18 & 5.31 & 0.76 & 38.5 & - & Y & YP-YY & RG \\
2 & 19172384-1604243 & -2.2 (0.8) & -27.6 (0.6) & 8.66 & 1.26 & 5.27 & 0.83 & 43.4 & - & Y & YY-YY & RG \\
3 & 19161966-1634094 & -1.5 (1.6) & -25.4 (1.5) & 8.65 & -0.10 & 8.05 & -0.02 & - & - & Y & Y- -BB & BS \\
4 & 19171130-1603082 & -0.9 (1.5) & -29.1 (1.5) & 9.20 & 1.10 & 6.40 & 0.77 & 42.7 & 41.1 & Y & YYYYY & RG \\
5 & 19164073-1616411 & -0.3 (1.6) & -25.4 (1.5) & 8.81 & -0.04 & 7.95 & 0.05 & - & - & Y & Y- -BB & BS \\
6 & 19170343-1703138 & -0.7 (1.9) & -30.1 (1.8) & 9.21 & 1.02 & 6.42 & 0.71 & 46.2 & - & Y & YY-YY & RG \\
7 & 19183747-1712575 & -0.4 (1.9) & -26.7 (1.8) & - & - & 6.41 & 0.67 & 42.4 & - & Y & YY-Y- &  \\
8 & 19181439-1641226 & 1.8 (1.5) & -25.3 (1.5) & 8.93 & 0.33 & 7.66 & 0.18 & - & - & Y & Y- -BB & BS \\
9 & 19140272-1554055 & -0.4 (1.5) & -26.1 (1.4) & 9.30 & 1.09 & 6.31 & 0.75 & 42.1 & - & Y & YY-YY & RG/SB1? \\
10 & 19155129-1617591 & -2.7 (1.6) & -26.8 (1.5) & 9.17 & 0.96 & 6.45 & 0.66 & 41.4 & 40.6 & Y & YYYYY & RG \\
11 & 19180978-1616222 & -1.4 (1.5) & -27.3 (1.5) & 9.39 & 0.98 & 6.54 & 0.70 & 44.2 & - & Y & YY-YY & RG \\
12 & 19164388-1626239 & -1.2 (1.6) & -24.6 (1.5) & 9.60 & 0.32 & 8.44 & 0.14 & - & - & Y & Y- -BB & BS \\
13 & 19131526-1706210 & -0.1 (1.1) & -27.1 (0.8) & - & - & 7.28 & 0.65 & 46.4 & - & Y & YY-Y- &  \\
14 & 19134817-1650059 & -1.5 (1.4) & -26.2 (1.3) & 9.88 & 0.87 & 7.63 & 0.56 & 43.6 & - & Y & YY-YY & RG \\
15 & 19164574-1635226 & -1.6 (1.6) & -27.1 (1.5) & 9.96 & 1.02 & 7.38 & 0.66 & 46.1 & 41.4 & Y & YYYYY & RG \\
16 & 19164823-1611522 & 1.9 (1.6) & -26.1 (1.7) & 9.96 & 0.23 & 9.11 & 0.17 & - & - & Y & Y- -BB & BS \\
17 & 19165670-1612265 & -1.7 (1.7) & -26.8 (1.8) & 10.07 & 0.54 & 8.59 & 0.37 & 44.2 & 40.6 & Y & YYYYY &  \\
18 & 19193373-1658514 & 0.3 (4.2) & -26.2 (4.7) & - & - & 8.68 & 0.36 & 47.2 & - & Y & YP-Y- &  \\
19 & 19161456-1624071 & -0.2 (2.0) & -28.6 (2.0) & 10.32 & 0.95 & 8.00 & 0.62 & 43.8 & 43.9 & P & YYPYY & RG \\	 	
20 & 19160865-1611148 & -3.4 (1.6) & -29.3 (1.8) & 10.25 & 0.57 & 8.63 & 0.33 & 41.8 & 41.7 & Y & YYYYY &  \\
21 & 19132220-1645096 & -5.4 (1.5) & -29.1 (1.6) & 10.28 & 0.51 & 8.74 & 0.37 & 41.9 & - & Y & YY-YY &  \\
22 & 19172382-1612488 & -1.7 (1.7) & -30.8 (1.8) & 10.31 & 0.42 & 9.26 & 0.44 & 51.9 & 38.2 & P & YNNYY & SB1 \\	
23 & 19154269-1633050 & -1.8 (2.2) & -30.3 (2.3) & 11.67 & 0.44 & 10.38 & 0.32 & 34.7 & 41.2 & Y & YNYYY & SB1? \\
24 & 19172865-1633313 & 1.1 (1.7) & -27.6 (1.8) & 10.29 & 0.40 & 9.11 & 0.21 & 41.8 & - & Y & YY-YB & BS? \\
25 & 19133648-1548104 & -1.4 (1.4) & -28.5 (1.5) & 10.53 & 0.80 & 8.34 & 0.61 & 40.7 & - & Y & YY-YY & RG \\
26 & 19153282-1620388 & 0.2 (1.8) & -27.1 (2.0) & 10.51 & 0.44 & 9.03 & 0.31 & 46.1 & 42.0 & Y & YYYYY &  \\
27 & 19171984-1607383 & -2.0 (1.9) & -30.8 (2.0) & 10.46 & 0.58 & 8.96 & 0.29 & 48.3 & 48.4 & P & YPNYY &  \\		
28 & 19152638-1700159 & -2.1 (2.0) & -29.6 (2.3) & 10.53 & 0.56 & 9.08 & 0.24 & 43.9 & - & Y & YY-YY &  \\
29 & 19173931-1636348 & 1.2 (1.6) & -25.4 (1.6) & 10.47 & 0.57 & 9.00 & 0.33 & 41.6 & - & Y & YY-YY &  \\
30 & 19155841-1615258 & -2.8 (1.9) & -28.9 (2.1) & 10.57 & 0.52 & 9.20 & 0.39 & 41.4 & 40.7 & Y & YYYPY &  \\
31 & 19195154-1603583 & -2.4 (2.0) & -27.3 (2.2) & - & - & 9.05 & 0.34 & 41.7 & - & Y & YY-Y- &  \\
32 & 19151540-1619517 & -1.8 (1.8) & -26.8 (2.1) & 10.70 & 0.52 & 9.30 & 0.33 & 45.9 & 42.2 & Y & YYYYY &  \\
33 & 19181155-1629141 & 1.4 (1.9) & -27.0 (1.9) & 10.66 & 0.44 & 9.39 & 0.29 & 41.2 & - & Y & YY-YY &  \\
34 & 19165477-1702129 & 3.2 (3.2) & -28.7 (3.5) & 10.63 & 0.46 & 9.41 & 0.31 & 45.8 & - & Y & YY-YY &  \\
35 & 19163976-1626316 & 0.1 (1.9) & -24.8 (2.1) & 10.53 & 0.55 & 9.10 & 0.32 & 45.1 & 41.1 & Y & YYYYY &  \\
36 & 19153626-1557460 & -0.7 (1.9) & -27.2 (2.1) & 10.49 & 0.53 & 9.07 & 0.31 & 46.4 & 41.8 & Y & YYYYY &  \\
37 & 19163344-1607515 & -2.5 (2.1) & -27.6 (2.2) & 10.64 & 0.49 & 9.32 & 0.36 & 34.6 & 41.0 & Y & YNYYY &  \\
38 & 19142651-1606340 & -2.5 (2.0) & -27.2 (2.1) & 10.66 & 0.48 & 9.33 & 0.41 & 45.1 & 41.2 & Y & YYYPY &  \\
39 & 19150275-1609405 & -5.2 (1.9) & -28.3 (2.1) & 10.66 & 0.52 & 9.19 & 0.31 & 42.7 & 41.7 & Y & YYYYY &  \\
40 & 19163339-1620215 & -3.0 (1.9) & -24.9 (2.1) & 10.62 & 0.56 & 9.18 & 0.33 & 42.1 & 41.5 & Y & YYYYY &  \\
41 & 19170481-1636526 & 4.3 (2.0) & -28.6 (2.2) & 10.61 & 0.44 & 9.33 & 0.30 & 45.9 & - & Y & YY-YY &  \\
42 & 19183120-1614421 & -1.8 (2.2) & -26.9 (2.2) & 10.64 & 0.45 & 9.32 & 0.33 & 44.6 & - & Y & YY-YY &  \\
43 & 19180054-1636016 & 1.3 (1.9) & -26.3 (2.0) & 10.70 & 0.46 & 9.44 & 0.27 & 44.9 & - & Y & YY-YY &  \\
44 & 19164495-1717074 & 2.7 (2.3) & -27.2 (2.5) & 10.97 & 0.48 & 9.74 & 0.32 & 42.2 & - & Y & YY-YY &  \\
45 & 19150860-1657412 & -2.0 (2.0) & -30.0 (2.3) & 10.82 & 0.45 & 9.50 & 0.28 & 45.7 & - & Y & YY-YY &  \\
46 & 19163525-1705075 & 3.7 (2.3) & -28.0 (2.6) & 10.80 & 0.46 & 9.52 & 0.30 & 45.5 & - & Y & YY-YY &  \\
47 & 19131541-1616123 & -5.6 (1.7) & -30.0 (1.8) & 10.78 & 0.45 & 9.51 & 0.32 & 44.5 & - & Y & YY-YY &  \\
48 & 19164662-1619208 & 0.0 (2.1) & -26.4 (2.2) & 10.72 & 0.54 & 9.34 & 0.37 & 46.3 & 42.3 & Y & YYYYY &  \\
49 & 19142907-1549056 & 0.3 (1.9) & -25.3 (2.1) & 10.76 & 0.50 & 9.36 & 0.36 & 46.2 & - & Y & YY-YY &  \\
50 & 19162934-1645544 & -2.4 (2.0) & -27.4 (2.3) & 10.97 & 0.37 & 9.80 & 0.30 & 47.6 & - & P & YP-YP &  \\			
51 & 19163620-1607363 & -3.0 (2.5) & -29.5 (2.6) & 11.02 & 0.46 & 9.12\tablenotemark{e} & 0.35 & 44.9 & - & Y & YY-YY &  \\
52 & 19162169-1609510 & -0.9 (2.1) & -27.7 (2.3) & 10.90 & 0.47 & 9.59 & 0.34 & 39.2 & 40.1 & Y & YYYYY &  \\
53 & 19163231-1611346 & -2.8 (2.2) & -29.3 (2.3) & 10.93 & 0.53 & 9.60 & 0.33 & 46.9 & 45.2 & P & YYNYP &  \\		
54 & 19165573-1603220 & -0.8 (2.2) & -30.2 (2.4) & 11.39 & 0.42 & 10.10 & 0.32 & 48.9 & 42.8 & Y & YPYYY &  \\
55 & 19160452-1605313 & -0.6 (2.1) & -32.1 (2.2) & 10.94 & 0.44 & 9.61 & 0.30 & 45.4 & 41.8 & Y & YYYYY &  \\
56 & 19200522-1535360 & 1.1 (1.9) & -25.8 (2.0) & - & - & 9.83 & 0.33 & 40.7 & - & Y & YY-Y- &  \\
57 & 19170433-1623185 & -0.2 (2.2) & -28.6 (2.3) & 11.23 & 0.47 & 9.93 & 0.27 & 41.9 & 41.8 & Y & YYYYY &  \\
58 & 19172172-1535592 & -0.4 (1.7) & -28.1 (1.7) & 11.27 & 0.46 & 9.95 & 0.36 & 44.6 & - & Y & YY-PY &  \\
59 & 19151260-1705121 & 2.9 (2.5) & -30.6 (2.7) & 11.18 & 0.48 & 9.92 & 0.24 & 50.1 & - & P & YP-PY &  \\			
60 & 19114731-1632485 & 0.8 (2.2) & -27.1 (2.3) & 11.31 & 0.49 & 10.09 & 0.38 & 43.5 & - & Y & YY-PY &  \\
61 & 19145840-1650089 & -4.5 (2.2) & -29.1 (2.3) & 11.18 & 0.45 & 9.93 & 0.32 & 42.2 & - & Y & YY-YY &  \\
62 & 19164922-1613222 & -0.5 (2.2) & -24.9 (2.3) & 11.25 & 0.52 & 9.84 & 0.32 & 45.0 & 42.5 & Y & YYYYY &  \\
63 & 19152981-1551047 & -1.8 (1.7) & -26.5 (1.8) & 11.53 & 0.52 & 10.17 & 0.33 & 39.7 & 41.4 & Y & YYYYY &  \\
64 & 19152465-1651222 & -1.5 (2.4) & -26.1 (2.5) & 11.72 & 0.64 & 10.20 & 0.40 & - & - & P & Y--PP & SB2 \\			
65 & 19164440-1615338 & -3.6 (3.7) & -23.7 (3.7) & 13.88 & 1.15 & 11.45 & 0.68 & - & - & Y & Y--PY & SB2 \\
66 & 19150050-1614245 & -1.5 (2.2) & -27.5 (2.3) & 11.71 & 0.62 & 10.15 & 0.36 & - & - & P & Y--YP & SB2 \\			
67 & 19151498-1720177 & -0.1 (2.5) & -29.1 (2.8) & 11.84 & 0.75 & 10.08 & 0.47 & 34.7 & - & N & YN-NN &  \\					
68 & 19180536-1646438 & 3.4 (3.7) & -29.8 (3.7) & 13.57 & 1.00 & 11.25 & 0.67 & - & - & Y & Y--PY & SB2 \\
69 & 19161864-1611305 & -6.8 (2.2) & -30.6 (2.3) & 11.38 & 0.51 & 9.99 & 0.37 & 42.4 & 37.2 & P & PYNPY &  \\		
70 & 19163827-1625039 & 1.1 (2.2) & -29.6 (2.3) & 11.47 & 0.43 & 10.23 & 0.30 & 37.9 & 38.9 & P & YPPYY &  \\		
71 & 19154511-1623157 & -0.1 (2.2) & -27.0 (2.4) & 11.93 & 0.48 & 10.57 & 0.32 & 41.4 & 41.1 & Y & YYYYY &  \\
72 & 19165800-1614277 & -2.6 (2.3) & -30.8 (2.4) & 12.04 & 0.71 & 10.36 & 0.40 & 47.9 & 46.6 & P & YPNYP & SB2 \\ 	
73 & 19160523-1652561 & -3.5 (2.4) & -26.8 (2.4) & 11.73 & 0.47 & 10.43 & 0.33 & 42.5 & - & Y & YY-YY &  \\
74 & 19150925-1552241 & -0.1 (1.8) & -28.6 (1.8) & 11.55 & 0.49 & 10.23 & 0.34 & 42.1 & 42.2 & Y & YYYYY &  \\
75 & 19161121-1621485 & 4.2 (3.6) & -25.5 (3.6) & 13.20 & 0.74 & 11.43 & 0.52 & 39.5 & 42.2 & Y & YYYYY &  \\
76 & 19134334-1649109 & 6.6 (3.8) & -77.7 (3.8) & 12.69 & 0.68 & 11.13 & 0.41 & 43.5 & - & P & NY-YY &  \\			
77 & 19150012-1605517 & -0.4 (3.2) & -25.0 (3.2) & 12.12 & 0.59 & 10.56 & 0.40 & 50.5 & 52.6 & N & YNNYY &  \\				
78 & 19160879-1524279 & -3.8 (2.0) & -30.5 (2.1) & 11.70 & 0.51 & 10.35 & 0.31 & 40.8 & - & Y & YY-YY &  \\
79 & 19142816-1620023 & -2.6 (3.0) & -28.6 (3.1) & 12.60 & 0.66 & 11.00 & 0.40 & 42.3 & 42.1 & Y & YYYYY &  \\
80 & 19162501-1632018 & -2.0 (2.4) & -29.5 (2.5) & 12.13 & 0.74 & 10.38 & 0.45 & 41.5 & - & P & YY-PP &  \\			
81 & 19151897-1639244 & -2.1 (2.2) & -25.5 (2.4) & 11.77 & 0.51 & 10.44 & 0.36 & 41.2 & - & Y & YY-YY &  \\
82 & 19152406-1621519 & -1.3 (2.4) & -29.6 (2.4) & 11.97 & 0.51 & 10.55 & 0.36 & 47.9 & 42.5 & Y & YPYYY &  \\
83 & 19134126-1610201 & -5.3 (3.2) & -29.5 (3.3) & 12.23 & 0.69 & 10.64 & 0.44 & 42.0 & - & Y & YY-YY &  \\
84 & 19141294-1554291 & -1.8 (3.1) & -26.6 (2.9) & 12.41 & 0.76 & 10.61 & 0.44 & 46.7 & - & Y & YY-YY &  \\
85 & 19165940-1635271 & -2.8 (3.1) & -27.7 (3.1) & 12.82 & 0.64 & 11.23 & 0.44 & 42.6 & 42.9 & Y & YYYYY &  \\
86 & 19160589-1629481 & -0.3 (2.9) & -28.7 (2.9) & 12.94 & 0.85 & 10.98 & 0.49 & 44.7 & 41.6 & Y & YYYYY &  \\
87 & 19160785-1610360 & -4.5 (2.9) & -26.8 (3.0) & 12.83 & 0.84 & 10.91 & 0.49 & 45.0 & 42.4 & Y & YYYYY &  \\
88 & 19162477-1710375 & -4.3 (3.7) & -42.8 (3.7) & 13.02 & 0.72 & 11.28 & 0.43 & 47.0 & - & P & PY-YY &  \\   		
89 & 19173402-1652177 & -1.6 (2.5) & -31.9 (2.5) & 12.72 & 0.58 & 11.07 & 0.41 & 47.3 & - & P & YP-YY &  \\			
90 & 19163672-1713101 & -0.1 (2.9) & -31.8 (2.9) & 12.44 & 0.55 & 10.95 & 0.32 & 42.0 & - & Y & YY-YY &  \\
91 & 19164725-1604093 & -2.2 (3.1) & -29.5 (3.0) & 12.75 & 0.61 & 11.13 & 0.36 & 42.8 & 42.2 & Y & YYYYY &  \\
92 & 19164417-1612222 & -1.5 (3.6) & -29.0 (3.6) & 12.63 & 0.74 & 10.89 & 0.45 & 45.1 & 25.2 & P & YYNYY & SB1 \\	
93 & 19162203-1546159 & 1.5 (2.9) & -27.6 (2.9) & 13.02 & 0.76 & 11.29 & 0.43 & 41.8 & 41.7 & Y & YYYYY &  \\
94 & 19152141-1600107 & -2.6 (3.2) & -27.1 (3.2) & 13.26 & 0.77 & 11.43 & 0.38 & 43.5 & 42.8 & Y & YYYPY &  \\
95 & 19170128-1609423 & -1.5 (2.9) & -30.1 (2.9) & 12.78 & 0.79 & 10.90 & 0.48 & 40.6 & 39.1 & P & YYPYY &  \\		
96 & 19151156-1726308 & -0.5 (2.1) & -27.0 (2.1) & - & - & 10.73 & 0.35 & 40.8 & - & Y & YY-Y- &  \\
97 & 19170285-1605166 & -1.7 (3.6) & -27.9 (3.6) & 13.03 & 0.70 & 11.38 & 0.43 & 40.9 & 41.1 & Y & YYYYY &  \\
98 & 19162656-1614545 & 0.4 (2.9) & -29.5 (2.9) & 13.52 & 0.97 & 11.38 & 0.63 & 43.2 & 40.7 & Y & YYYPY &  \\
99 & 19161757-1600177 & -2.4 (3.0) & -29.4 (3.0) & 13.66 & 0.89 & 11.52 & 0.58 & 44.8 & 44.6 & P & YYNYY &  \\		
100 & 19145199-1541379 & 0.2 (3.7) & -3.7 (3.7) & 14.20 & 1.22 & 11.67 & 0.79 & 42.2 & - & N & NY-PY &  \\				
101 & 19153354-1625368 & -9.0 (3.7) & -31.6 (3.7) & 15.43 & 1.43 & 12.52 & 0.78 & 46.3 & 40.5 & Y & PYYYY &  \\ 
102 & 19124958-1550340 & 3.4 (3.8) & -34.1 (3.8) & - & - & 11.85 & 0.55 & 44.3 & - & P & PY-Y- &  \\				
103 & 19134512-1619340 & 19.9 (4.9) & -11.1 (4.9) & 14.27 & 1.15 & 11.92 & 0.68 & 45.0 & - & N & NY-YY &  \\			
104 & 19193779-1618312 & 11.4 (4.8) & -15.8 (4.8) & - & - & 11.78 & 0.65 & 52.4 & - & N & NN-Y- &  \\					
105 & 19181352-1614496 & 5.8 (3.7) & -58.2 (3.7) & 14.39 & 1.16 & 11.87 & 0.68 & 50.5 & - & N & NN-YY &  \\				
106 & 19163680-1623032 & -3.0 (3.7) & -31.9 (3.7) & 15.05 & 1.53 & 12.06 & 0.76 & 49.3 & 46.5 & P & YPNYY &  \\ 	
107 & 19163732-1600050 & -2.9 (3.7) & -34.8 (3.7) & 14.70 & 1.26 & 12.02 & 0.66 & 42.5 & 42.1 & Y & PYYYY &  \\ 	
108 & 19172940-1611577 & 9.1 (3.7) & -32.3 (3.7) & 15.21 & 1.35 & 12.47 & 0.71 & 48.6 & 42.8 & P & NPYYY &  \\  	
\enddata
\tablecomments{Column Notes: (1) CWW ID -- This work's star identification scheme, sorted by $V$ magnitude. 
CWW = Curtis, Wolfgand and Wright.  (2) 2MASS ID, also provides RA and Dec positions
(3, 4) RA and Dec proper motions in mas/yr from PPMXL catalog
 (5) CFHT/MegaCam $g'$ mag. (6) $g' - i'$ mag. (7) 2MASS $J$ mag. (8) 2MASS $J - K_S$ mag.
 (9) Lick / Palomar RV in \kms\ (10) Hectochelle RV in \kms\ (11) Membership probabilities
 (12) Membership probabilities for each criterion (13) Notes for individual stars.
 Vaues in parenthesis are measurement errors.}
\tablenotetext{a}{We use 2MASS aperture photometry instead of the default PSF photometry for 18
stars, based on our analysis that these stars, and only these stars, shift position on the ($J-K_S$) CMD
and that they all move toward the cluster locus. No neighbors are resolved in our optical imaging within 5".
The stars are CWW 22, 24, 27, 28, 30, 37, 38, 43, 48, 49, 57, 59, 67, 68, 90, 91, 94, and 100.}
\tablenotetext{b}{Membership Probability: Y = yes, highest confidence member,
P = possible / probable member, N = not likely / non-member}
\tablenotetext{c}{Membership Criteria: proper motion radial distance from cluster value;
 Lick / Palomar RV, Hectochelle RV, 2MASS ($J - K_S$) CMD, CFHT/MegaCam ($g' - i'$) CMD.
 Confidence intervals defined in Table \ref{t:criteria}. 
 A `B' flag indicates photometry consistent with blue stragglers. A dash `-' indicates no data.}
\tablenotetext{d}{Notes: BS = blue straggler, RG = red giant, SB2 = spectroscopic
 double line binary, SB1? = inconsistent RVs between multiple epochs}
\tablenotetext{e}{Our MegaCam imaging shows CWW 51 is an optical double, with a star 1.65$''$ away 
with a similar \griz\ SED. 
Assuming a cluster distance of 300 pc, this angular separation translates into 
a minimum physical separation of 495 AU.
This suggests that the pair actually form a wide binary, 
although their angular proximity could also be explained by a chance alignment. 
This double was not resolved in the 2MASS Point Source Catalog. 
Adding 0.75 mag. to the $J$ band magnitude (halving the brightness, to reflect just the one star) 
moves CWW 51 in the ($J - K_S$) CMD to its neighbors in the ($g' - i'$) CMD. 
Despite this realization, we quote the 2MASS PSC photometry here. See \S \ref{s:2mass}.}
\end{deluxetable}
\clearpage
\end{landscape}
\end{document}